\documentclass[11pt, a4paper]{article}
\pdfoutput=1
\usepackage{jheppub}
\usepackage{tikz}
\usepackage{arydshln}
\usepackage{amsmath}
\usepackage{relsize}
\usepackage[labelformat=simple]{subcaption}

\newcommand{\circled}[1]{\mathbin{
  \mathchoice
    {\mbox{\normalsize\textcircled{$#1$}}}
    {\mbox{\normalsize\textcircled{$#1$}}}
    {\tikz{
        \node[draw, shape=circle, minimum size=7pt, inner sep=0pt, font=\scriptsize] (a) {$#1$};}}
    {\mbox{\scriptsize\textcircled{$#1$}}}
  }
}

\newcommand{\qto}{\mathbin{
  \mathchoice
    {\to}
    {\to}
    {\tikz{
        \draw[->] (0,0) -- (0.3,0);}}
    {\to}
  }
}

\newcommand{\dqto}{\mathbin{
  \mathchoice
    {\to}
    {\to}
    {\tikz{
        \draw[->, densely dotted] (0,0) -- (0.3,0);}}
    {\to}
  }
}

\newcommand{\RM}{\check{R}}
\newcommand{\BW}[6]{W\biggl(\begin{array}{cc} #4 & #3 \\ #1 &
   #2\end{array} \biggm| #5, #6\biggr)}

\newlength{\qsep}
\usetikzlibrary{decorations.markings, calc, decorations.pathmorphing}
\tikzset{
  x=\qsep, y=\qsep,
  >=latex,
  font=\smaller,
  baseline={(current bounding box.center)},
  snake/.style={decorate, decoration={snake, segment length=6pt,
      amplitude=1pt, post length=4pt}},
  ->-/.style={decoration={
      markings, mark=at position #1 with
      {\arrow{>}}},postaction={decorate}},
  -<-/.style={decoration={
      markings, mark=at position #1 with
      {\arrow{<}}},postaction={decorate}},
  node/.style={draw, shape=circle, minimum size=11pt, inner
    sep=0pt},
  wnode/.style={node},
  tnode/.style={node, thick},
  bnode/.style={node, fill=black, text=white},
  dnode/.style={node, densely dashed},
  snode/.style={node, shape=rectangle},
  inode/.style={node, draw=none},
  q->/.style={->, shorten >=1pt, font=\smaller[2]},
  q<-/.style={q->, <-, shorten >=0pt, shorten <=1pt},
  dq->/.style={q->, densely dotted},
  dq<-/.style={dq->, <-, shorten >=0pt, shorten <=1pt},
  r->/.style={->, thick},
  r->-/.style={->-=#1, thick},
  dr->/.style={r->, densely dashed, thin},
  ddr->/.style={r->, densely dashdotted, thin},
  r<-/.style={r->, <-},
  dr<-/.style={dr->, <-},
  ddr<-/.style={ddr->, <-},
}

\DeclareFontFamily{U}{MnSymbolC}{}
\DeclareSymbolFont{MnSyC}{U}{MnSymbolC}{m}{n}
\DeclareFontShape{U}{MnSymbolC}{m}{n}{
    <-6>  MnSymbolC5
   <6-7>  MnSymbolC6
   <7-8>  MnSymbolC7
   <8-9>  MnSymbolC8
   <9-10> MnSymbolC9
  <10-12> MnSymbolC10
  <12->   MnSymbolC12}{}
\DeclareMathSymbol{\intprod}{\mathbin}{MnSyC}{'270}

\newcommand{\vev}[1]{\langle #1 \rangle}

\newcommand{\biggvev}[1]{\biggl\langle #1 \biggr\rangle}

\newcommand{\Tr}{\mathop{\mathrm{Tr}}\nolimits}

\newcommand{\SU}{\mathrm{SU}}
\newcommand{\SO}{\mathrm{SO}}

\newcommand{\SL}{\mathrm{SL}}

\newcommand{\U}{\mathrm{U}}

\newcommand{\longto}{\longrightarrow}

\newcommand{\Z}{\mathbb{Z}}

\newcommand{\R}{\mathbb{R}}
\newcommand{\C}{\mathbb{C}}

\let\nc\newcommand
\let\renc\renewcommand
\nc{\wbar}{\overline}
\let\td\tilde
\let\wtd\widetilde
\let\wht\widehat
\let\mcl\mathcal

\nc{\ab}{{\bar{a}}} \nc{\at}{\tilde{a}} \nc{\ah}{\hat{a}}
\nc{\bb}{{\bar{b}}} \nc{\bt}{\tilde{b}} \nc{\bh}{\hat{b}}
\nc{\cb}{{\bar{c}}} \nc{\ct}{\tilde{c}} 
\nc{\db}{{\bar{d}}} \nc{\dt}{\tilde{d}} \renc{\dh}{\hat{d}}
\nc{\eb}{{\bar{e}}} \nc{\et}{\tilde{e}} \nc{\eh}{\hat{e}}
\nc{\fb}{{\bar{f}}} \nc{\ft}{\tilde{f}} \nc{\fh}{\hat{f}}
\nc{\gb}{{\bar{g}}} \nc{\gt}{\tilde{g}} \nc{\gh}{\hat{g}}
\nc{\hb}{{\bar{h}}} \nc{\hh}{\hat{h}} 
\nc{\ib}{{\bar{\imath}}} \nc{\ih}{\hat{\imath}} 
\nc{\jb}{{\bar{\jmath}}} \nc{\jt}{\tilde{\jmath}} \nc{\jh}{\hat{\jmath}}
\nc{\kb}{{\bar{k}}} \nc{\kt}{\tilde{k}} \nc{\kh}{\hat{k}}
\nc{\lb}{{\bar{l}}} \nc{\lt}{\tilde{l}} \nc{\lh}{\hat{l}}
\nc{\mb}{{\bar{m}}} \nc{\mt}{\tilde{m}} \nc{\mh}{\hat{m}}
\nc{\nb}{{\bar{n}}} \nc{\nt}{\tilde{n}} \nc{\nh}{\hat{n}}
\nc{\ob}{{\bar{o}}} \nc{\ot}{\tilde{o}} \nc{\oh}{\hat{o}}
\nc{\pb}{{\bar{p}}} \nc{\pt}{\tilde{p}} \nc{\ph}{\hat{p}}
\nc{\qb}{{\bar{q}}} \nc{\qt}{\tilde{q}} \nc{\qh}{\hat{q}}
\nc{\rb}{{\bar{r}}} \nc{\rt}{\tilde{r}} \nc{\rh}{\hat{r}}
\renc{\sb}{{\bar{s}}} \nc{\st}{\tilde{s}} \nc{\sh}{\hat{s}}
\nc{\tb}{{\bar{t}}} \renc{\th}{\hat{t}} 
\nc{\ub}{{\bar{u}}} \nc{\ut}{\tilde{u}} \nc{\uh}{\hat{u}}
\nc{\vb}{{\bar{v}}} \nc{\vt}{\tilde{v}} \nc{\vh}{\hat{v}}
\nc{\wb}{{\bar{w}}} \nc{\wt}{\tilde{w}} \nc{\wh}{\hat{w}}
\nc{\xb}{{\bar{x}}} \nc{\xt}{\tilde{x}} \nc{\xh}{\hat{x}}
\nc{\yb}{{\bar{y}}} \nc{\yt}{\tilde{y}} \nc{\yh}{\hat{y}}
\nc{\zb}{{\bar{z}}} \nc{\zt}{\tilde{z}} \nc{\zh}{\hat{z}}

\nc{\Ab}{{\wbar{A}}} \nc{\At}{{\wtd{A}}} \nc{\Ah}{{\wht{A}}}
\nc{\Bb}{{\wbar{B}}} \nc{\Bt}{{\wtd{B}}} \nc{\Bh}{{\wht{B}}}
\nc{\Cb}{{\wbar{C}}} \nc{\Ct}{{\wtd{C}}} \nc{\Ch}{{\wht{C}}}
\nc{\Db}{{\wbar{D}}} \nc{\Dt}{{\wtd{D}}} \nc{\Dh}{{\wht{D}}}
\nc{\Eb}{{\wbar{E}}} \nc{\Et}{{\wtd{E}}} \nc{\Eh}{{\wht{E}}}
\nc{\Fb}{{\wbar{F}}} \nc{\Ft}{{\wtd{F}}} \nc{\Fh}{{\wht{F}}}
\nc{\Gb}{{\wbar{G}}} \nc{\Gt}{{\wtd{G}}} \nc{\Gh}{{\wht{G}}}
\nc{\Hb}{{\wbar{H}}} \nc{\Ht}{{\wtd{H}}} \nc{\Hh}{{\wht{H}}}
\nc{\Ib}{{\bar{I}}} \nc{\It}{{\wtd{I}}} \nc{\Ih}{{\wht{I}}}
\nc{\Jb}{{\bar{J}}} \nc{\Jt}{{\wtd{J}}} \nc{\Jh}{{\wht{J}}}
\nc{\Kb}{{\wbar{K}}} \nc{\Kt}{{\wtd{K}}} \nc{\Kh}{{\wht{K}}}
\nc{\Lb}{{\wbar{L}}} \nc{\Lt}{{\wtd{L}}} \nc{\Lh}{{\wht{L}}}
\nc{\Mb}{{\wbar{M}}} \nc{\Mt}{{\wtd{M}}} \nc{\Mh}{{\wht{M}}}
\nc{\Nb}{{\wbar{N}}} \nc{\Nt}{{\wtd{N}}} \nc{\Nh}{{\wht{N}}}
\nc{\Ob}{{\wbar{O}}} \nc{\Ot}{{\wtd{O}}} \nc{\Oh}{{\wht{O}}}
\nc{\Pb}{{\wbar{P}}} \nc{\Pt}{{\wtd{P}}} \nc{\Ph}{{\wht{P}}}
\nc{\Qb}{{\wbar{Q}}} \nc{\Qt}{{\wtd{Q}}} \nc{\Qh}{{\wht{Q}}}
\nc{\Rb}{{\wbar{R}}} \nc{\Rt}{{\wtd{R}}} \nc{\Rh}{{\wht{R}}}
\nc{\Sb}{{\wbar{S}}} \nc{\St}{{\wtd{S}}} \nc{\Sh}{{\wht{S}}}
\nc{\Tb}{{\wbar{T}}} \nc{\Tt}{{\wtd{T}}} \nc{\Th}{{\wht{T}}}
\nc{\Ub}{{\wbar{U}}} \nc{\Ut}{{\wtd{U}}} \nc{\Uh}{{\wht{U}}}
\nc{\Vb}{{\wbar{V}}} \nc{\Vt}{{\wtd{V}}} \nc{\Vh}{{\wht{V}}}
\nc{\Wb}{{\wbar{W}}} \nc{\Wt}{{\wtd{W}}} \nc{\Wh}{{\wht{W}}}
\nc{\Xb}{{\wbar{X}}} \nc{\Xt}{{\wtd{X}}} \nc{\Xh}{{\wht{X}}}
\nc{\Yb}{{\wbar{Y}}} \nc{\Yt}{{\wtd{Y}}} \nc{\Yh}{{\wht{Y}}}
\nc{\Zb}{{\wbar{Z}}} \nc{\Zt}{{\wtd{Z}}} \nc{\Zh}{{\wht{Z}}}

\nc{\CA}{{\mcl{A}}} \nc{\CAb}{{\wbar{\CA}}} \nc{\CAt}{{\wtd{\CA}}} \nc{\CAh}{{\wht{\CA}}}
\nc{\CB}{{\mcl{B}}} \nc{\CBb}{{\wbar{\CB}}} \nc{\CBt}{{\wtd{\CB}}} \nc{\CBh}{{\wht{\CB}}}
\nc{\CC}{{\mcl{C}}} \nc{\CCb}{{\wbar{\CC}}} \nc{\CCt}{{\wtd{\CC}}} \nc{\CCh}{{\wht{\CC}}}
\nc{\cD}{{\mcl{D}}} \nc{\cDb}{{\wbar{\cD}}} \nc{\cDt}{{\wtd{\cC}}} \nc{\cDh}{{\wht{\cD}}}
\nc{\CE}{{\mcl{E}}} \nc{\CEb}{{\wbar{\CE}}} \nc{\CEt}{{\wtd{\CE}}} \nc{\CEh}{{\wht{\CE}}}
\nc{\CF}{{\mcl{F}}} \nc{\CFb}{{\wbar{\CF}}} \nc{\CFt}{{\wtd{\CF}}} \nc{\CFh}{{\wht{\CF}}}
\nc{\CG}{{\mcl{G}}} \nc{\CGb}{{\wbar{\CG}}} \nc{\CGt}{{\wtd{\CG}}} \nc{\CGh}{{\wht{\CG}}}
\nc{\CH}{{\mcl{H}}} \nc{\CHb}{{\wbar{\CH}}} \nc{\CHt}{{\wtd{\CH}}} \nc{\CHh}{{\wht{\CH}}}
\nc{\CI}{{\mcl{I}}} \nc{\CIb}{{\wbar{\CI}}} \nc{\CIt}{{\wtd{\CI}}} \nc{\CIh}{{\wht{\CI}}}
\nc{\CJ}{{\mcl{J}}} \nc{\CJb}{{\wbar{\CJ}}} \nc{\CJt}{{\wtd{\CJ}}} \nc{\CJh}{{\wht{\CJ}}}
\nc{\CK}{{\mcl{K}}} \nc{\CKb}{{\wbar{\CK}}} \nc{\CKt}{{\wtd{\CK}}} \nc{\CKh}{{\wht{\CK}}}
\nc{\CL}{{\mcl{L}}} \nc{\CLb}{{\wbar{\CL}}} \nc{\CLt}{{\wtd{\CL}}} \nc{\CLh}{{\wht{\CL}}}
\nc{\CM}{{\mcl{M}}} \nc{\CMb}{{\wbar{\CM}}} \nc{\CMt}{{\wtd{\CM}}} \nc{\CMh}{{\wht{\CM}}}
\nc{\CN}{{\mcl{N}}} \nc{\CNb}{{\wbar{\CN}}} \nc{\CNt}{{\wtd{\CN}}} \nc{\CNh}{{\wht{\CN}}}
\nc{\CO}{{\mcl{O}}} \nc{\COb}{{\wbar{\CO}}} \nc{\COt}{{\wtd{\CO}}} \nc{\COh}{{\wht{\CO}}}
\nc{\CP}{{\mcl{P}}} \nc{\CPb}{{\wbar{\CP}}} \nc{\CPt}{{\wtd{\CP}}} \nc{\CPh}{{\wht{\CP}}}
\nc{\CQ}{{\mcl{Q}}} \nc{\CQb}{{\wbar{\CQ}}} \nc{\CQt}{{\wtd{\CQ}}} \nc{\CQh}{{\wht{\CQ}}}
\nc{\CR}{{\mcl{R}}} \nc{\CRb}{{\wbar{\CR}}} \nc{\CRt}{{\wtd{\CR}}} \nc{\CRh}{{\wht{\CR}}}
\nc{\CS}{{\mcl{S}}} \nc{\CSb}{{\wbar{\CS}}} \nc{\CSt}{{\wtd{\CS}}} \nc{\CSh}{{\wht{\CS}}}
\nc{\CT}{{\mcl{T}}} \nc{\CTb}{{\wbar{\CT}}} \nc{\CTt}{{\wtd{\CT}}} \nc{\CTh}{{\wht{\CT}}}
\nc{\CU}{{\mcl{U}}} \nc{\CUb}{{\wbar{\CU}}} \nc{\CUt}{{\wtd{\CU}}} \nc{\CUh}{{\wht{\CU}}}
\nc{\CV}{{\mcl{V}}} \nc{\CVb}{{\wbar{\CV}}} \nc{\CVt}{{\wtd{\CV}}} \nc{\CVh}{{\wht{\CV}}}
\nc{\CW}{{\mcl{W}}} \nc{\CWb}{{\wbar{\CW}}} \nc{\CWt}{{\wtd{\CW}}} \nc{\CWh}{{\wht{\CW}}}
\nc{\CX}{{\mcl{X}}} \nc{\CXb}{{\wbar{\CX}}} \nc{\CXt}{{\wtd{\CX}}} \nc{\CXh}{{\wht{\CX}}}
\nc{\CY}{{\mcl{Y}}} \nc{\CYb}{{\wbar{\CY}}} \nc{\CYt}{{\wtd{\CY}}} \nc{\CYh}{{\wht{\CY}}}
\nc{\CZ}{{\mcl{Z}}} \nc{\CZb}{{\wbar{\CZ}}} \nc{\CZt}{{\wtd{\CZ}}} \nc{\CZh}{{\wht{\CZ}}}

\let\eps\epsilon
\let\ups\upsilon
\let\veps\varepsilon
\let\vtht\vartheta

\let\vsgm\varsigma
\let\vphi\varphi
\let\vrho\varrho

\nc{\alphab}{{\bar{\alpha}}} \nc{\alphat}{{\td{\alpha}}} \nc{\alphah}{{\hat{\alpha}}}
\nc{\betab}{{\bar{\beta}}}   \nc{\betat}{{\td{\beta}}}   \nc{\betah}{{\hat{\beta}}} 
\nc{\gammab}{{\bar{\gamma}}} \nc{\gammat}{{\td{\gamma}}} \nc{\gammah}{{\hat{\gamma}}} 
\nc{\deltab}{{\bar{\delta}}} \nc{\deltat}{{\td{\delta}}} \nc{\deltah}{{\hat{\delta}}} 
\nc{\epsilonb}{{\bar{\eps}}} \nc{\epsilont}{{\td{\eps}}} \nc{\epsilonh}{{\hat{\eps}}} 
\nc{\vepsb}{{\bar{\veps}}}   \nc{\vepst}{{\td{\veps}}}   \nc{\vepsh}{{\hat{\veps}}} 
\nc{\zetab}{{\bar{\zeta}}}   \nc{\zetat}{{\td{\zeta}}}   \nc{\zetah}{{\hat{\zeta}}} 
\nc{\etab}{{\bar{\eta}}}     \nc{\etat}{{\td{\eta}}}     \nc{\etah}{{\hat{\eta}}} 
\nc{\thetab}{{\bar{\theta}}} \nc{\thetat}{{\td{\theta}}} \nc{\thetah}{{\hat{\theta}}} 
\nc{\vthetab}{{\bar{\vtht}}} \nc{\vthetat}{{\td{\vtht}}} \nc{\vthetah}{{\hat{\vtht}}} 
\nc{\lambdab}{{\bar{\lambda}}} \nc{\lambdat}{{\td{\lambda}}} \nc{\lambdah}{{\hat{\lambda}}} 
\nc{\iotab}{{\bar{\iota}}}   \nc{\iotat}{{\td{\iota}}}   \nc{\iotah}{{\hat{\iota}}} 
\nc{\kappab}{{\bar{\kappa}}} \nc{\kappat}{{\td{\kappa}}} \nc{\kappah}{{\hat{\kappa}}} 
\nc{\lmdb}{{\bar{\lmd}}}     \nc{\lmdt}{{\td{\lmd}}}     \nc{\lmdh}{{\hat{\lmd}}} 
\nc{\mub}{{\bar{\mu}}}       \nc{\mut}{{\td{\mu}}}       \nc{\muh}{{\hat{\mu}}} 
\nc{\nub}{{\bar{\nu}}}       \nc{\nut}{{\td{\nu}}}       \nc{\nuh}{{\hat{\nu}}} 
\nc{\xib}{{\bar{\xi}}}       \nc{\xit}{{\td{\xi}}}       \nc{\xih}{{\hat{\xi}}} 
\nc{\pib}{{\bar{\pi}}}       \nc{\pit}{{\td{\pi}}}       \nc{\pih}{{\hat{\pi}}} 
\nc{\vpib}{{\bar{\vpi}}}     \nc{\vpit}{{\td{\vpi}}}     \nc{\vpih}{{\hat{\vpi}}} 
\nc{\rhob}{{\bar{\rho}}}     \nc{\rhot}{{\td{\rho}}}     \nc{\rhoh}{{\hat{\rho}}} 
\nc{\vrhob}{{\bar{\vrho}}}   \nc{\vrhot}{{\td{\vrho}}}   \nc{\vrhoh}{{\hat{\vrho}}} 
\nc{\sigmab}{{\bar{\sigma}}} \nc{\sigmat}{{\td{\sigma}}} \nc{\sigmah}{{\hat{\sigma}}} 
\nc{\vsigmab}{{\bar{\vsgm}}} \nc{\vsigmat}{{\td{\vsgm}}} \nc{\vsigmah}{{\hat{\vsgm}}} 
\nc{\taub}{{\bar{\tau}}}     \nc{\taut}{{\td{\tau}}}     \nc{\tauh}{{\hat{\tau}}} 
\nc{\upsb}{{\bar{\ups}}} \nc{\upst}{{\td{\ups}}} \nc{\upsh}{{\hat{\ups}}} 
\nc{\phib}{{\bar{\phi}}}     \nc{\phit}{{\td{\phi}}}     \nc{\phih}{{\hat{\phi}}} 
\nc{\varphib}{{\bar{\vphi}}}   \nc{\varphit}{{\td{\vphi}}}   \nc{\varphih}{{\hat{\vphi}}} 
\nc{\chib}{{\bar{\chi}}}     \nc{\chit}{{\td{\chi}}}     \nc{\chih}{{\hat{\chi}}} 
\nc{\psib}{{\bar{\psi}}}     \nc{\psit}{{\td{\psi}}}     \nc{\psih}{{\hat{\psi}}} 
\nc{\omegab}{{\bar{\omega}}} \nc{\omegat}{{\td{\omega}}} \nc{\omegah}{{\hat{\omega}}} 

\nc{\Gammab}{{\wbar{\Gamma}}}     \nc{\Gammat}{{\wtd{\Gamma}}}     \nc{\Gammah}{{\wht{\Gamma}}}
\nc{\Deltab}{{\wbar{\Delta}}}     \nc{\Deltat}{{\wtd{\Delta}}}     \nc{\Deltah}{{\wht{\Delta}}}
\nc{\Thetab}{{\wbar{\Theta}}}     \nc{\Thetat}{{\wtd{\Theta}}}     \nc{\Thetah}{{\wht{\Theta}}}
\nc{\Lambdab}{{\wbar{\Lambda}}}   \nc{\Lambdat}{{\wtd{\Lambda}}}   \nc{\Lambdah}{{\wht{\Lambda}}}
\nc{\Xib}{{\wbar{\Xi}}}           \nc{\Xit}{{\wtd{\Xi}}}           \nc{\Xih}{{\wht{\Xi}}}
\nc{\Pib}{{\wbar{\Pi}}}           \nc{\Pit}{{\wtd{\Pi}}}           \nc{\Pih}{{\wht{\Pi}}}
\nc{\Sigmab}{{\wbar{\Sigma}}}     \nc{\Sigmat}{{\wtd{\Sigma}}}     \nc{\Sigmah}{{\wht{\Sigma}}}
\nc{\Upsilonb}{{\wbar{\Upsilon}}} \nc{\Upsilont}{{\wtd{\Upsilon}}} \nc{\Upsilonh}{{\wht{\Upsilon}}}
\nc{\Phib}{{\wbar{\Phi}}}         \nc{\Phit}{{\wtd{\Phi}}}         \nc{\Phih}{{\wht{\Phi}}}
\nc{\Psib}{{\wbar{\Psi}}}         \nc{\Psit}{{\wtd{\Psi}}}         \nc{\Psih}{{\wht{\Psi}}}
\nc{\Omegab}{{\wbar{\Omega}}}     \nc{\Omegat}{{\wtd{\Omega}}}     \nc{\Omegah}{{\wht{\Omega}}}

\newcommand{\rmd}{\mathrm{d}}

\title{Quiver gauge theories and integrable lattice models}

\author{Junya Yagi}

\emailAdd{junya.yagi@sissa.it}

\affiliation{International School for Advanced Studies (SISSA), \\
via Bonomea 265, 34136 Trieste, Italy \\
INFN --- Sezione di Trieste, \\
via Valerio 2, 34149 Trieste, Italy}

\abstract{We discuss connections between certain classes of
  supersymmetric quiver gauge theories and integrable lattice models
  from the point of view of topological quantum field theories
  (TQFTs).  The relevant classes include 4d $\CN = 1$ theories known
  as brane box and brane tilling models, 3d $\CN = 2$ and 2d $\CN =
  (2,2)$ theories obtained from them by compactification, and 2d $\CN
  = (0,2)$ theories closely related to these theories.  We argue that
  their supersymmetric indices carry structures of TQFTs equipped with
  line operators, and as a consequence, are equal to the partition
  functions of lattice models.  The integrability of these models
  follows from the existence of extra dimension in the TQFTs, which
  emerges after the theories are embedded in M-theory.  The
  Yang-Baxter equation expresses the invariance of supersymmetric
  indices under Seiberg duality and its lower-dimensional analogs.}


\arxivnumber{}

\begin{document}
\maketitle

\section{Introduction}

The present work is motivated by an intriguing connection discovered
in the past few years between supersymmetric quiver gauge theories and
integrable lattice models in statistical~mechanics.

In~\cite{Bazhanov:2010kz, Bazhanov:2011mz}, Bazhanov and Sergeev
introduced an integrable spin model on a planar lattice, which
generalizes many of previously known integrable lattice models such as
the Kashiwara-Miwa~\cite{Kashiwara:1986tu} and chiral
Potts~\cite{vonGehlen:1984bi, AuYang:1987zc, Baxter:1987eq} models.
Soon after the first paper by Bazhanov and Sergeev appeared,
Spiridonov~\cite{Spiridonov:2010em} gave an interpretation of their
model in terms of $\CN =1$ quiver gauge theories in four dimensions.
The relation to gauge theory was further elucidated by
Yamazaki~\cite{Yamazaki:2012cp, Yamazaki:2013nra}, who realized that
the relevant quiver gauge theories arise naturally from a particular
class of brane configurations in string theory, called brane
tilings~\cite{Franco:2005rj}.  From the gauge theory viewpoint, the
lattice on which the spin model is defined is the quiver diagram, the
partition function is the supersymmetric index, and the Yang-Baxter
equation that guarantees the integrability of the model is a special
instance of Seiberg duality.

This discovery, while quite remarkable, leaves us with a series of
questions: Why are the supersymmetric indices of these theories
captured by a lattice model?  Why is this model integrable?  Are there
structures of integrable lattice models hidden in other theories?

In this paper we answer these questions, combining ideas from two
equally stimulating connections uncovered in recent years.  Of these
connections, one lies between a certain class of quiver gauge theories
and topological quantum field theories (TQFTs), whereas the other
relates TQFTs and integrable lattice models.  Our goal is to connect
quiver gauge theories and integrable lattice models, and TQFTs provide
a bridge between the two~elements.

The first connection in question arises from the M5-brane construction
of 4d $\CN = 2$ theories~\cite{Witten:1997sc, Gaiotto:2009we,
  Gaiotto:2009hg}.  Consider a stack of M5-branes wrapped on $S^1
\times S^3 \times \Sigma$, where $\Sigma$ is a compact Riemann
surface.  In addition, we introduce M5-branes that intersect with
these branes along submanifolds of the form $S^1 \times S^3 \times
\{p_i\}$, with $p_i$ being points on $\Sigma$.  If $\Sigma$ is small
compared to the $S^1$ and $S^3$, this brane system is described at low
energies by an $\CN = 2$ theory on $S^1 \times S^3$.  Often this is a
quiver gauge theory.  The path integral on the geometry $S^1 \times
S^3$ (after analytic continuation to Euclidean spacetime) computes the
supersymmetric index of the theory.  The most important property of
this quantity is that it is invariant under continuous changes of the
parameters of the theory.  Since the geometry of $\Sigma$ encodes such
parameters, it follows that the supersymmetric index is a topological
invariant of $\Sigma$.  In fact, it is equal to a correlation function
$\vev{\prod_i \CO_i(p_i)}$ in a TQFT on $\Sigma$, where $\CO_i$ is a
local operator representing the M5-brane inserted at
$p_i$~\cite{Gadde:2009kb}.  If we consider a protected quantity
different from the index on $S^3$ (such as the lens space
index~\cite{Benini:2011nc}), then we get a different TQFT.

The second connection refers to a general construction of integrable
lattice models from TQFTs of a special kind, due to
Costello~\cite{Costello:2013sla, Costello:2013zra}.  Given a 2d TQFT
equipped with line operators, we can place it on a torus $T^2$ and
wrap line operators around various $1$-cycles that form a lattice.
The topological invariance of the theory implies that the correlation
function for this lattice of line operators coincides with the
partition function of a statistical mechanics model defined on the
same lattice.  Under this correspondence, the Yang-Baxter equation
translates to the statement that the correlator remains the same when
a line operator is moved past the intersection of two other line
operators.  The structure of a TQFT itself is not strong enough to
ensure this property since the move is not topologically trivial; the
system may undergo a phase transition as the moved line hits and
crosses the intersection point.  If, however, the theory has ``extra
dimensions'' along which the line can be shifted, the collision can be
avoided and hence the Yang-Baxter equation holds.  What is more, the
correlator then carries continuous parameters, namely the positions of
the line operators in the extra dimensions.  In the context of
integrable models, continuous parameters on which the Boltzmann weight
depends are called spectral parameters, and their presence is
essential for integrability.  Thus, a 2d TQFT with extra dimensions
produces from line operators a solution of the Yang-Baxter equation
with spectral parameter.

We wish to understand the connection between brane tilings and
integrable lattice models in light of these independent, though
apparently related, developments in the relevant areas.  To this end,
it proves helpful to first study the case of the brane box
construction~\cite{Hanany:1997tb}, which is a precursor of the brane
tiling construction and conceptually simpler.

The brane box construction is similar to the M5-brane construction
described above.  In this construction, we start with a stack of $N$
D5-branes on $S^1 \times S^3 \times T^2$, and put NS5-branes on
submanifolds of the form $S^1 \times S^3 \times C_\alpha$.  Here
$C_\alpha$ are $1$-cycles of $T^2$, making up a lattice.  At low
energies, this brane configuration realizes a 4d $\CN = 1$ quiver
gauge theory, which we refer to as a ``brane box model,'' placed on
$S^1 \times S^3$.  (Actually, $T^2$ can be replaced by any Riemann
surface without breaking the $\CN = 1$ supersymmetry.)  Just as we did
for the M5-brane construction, we can relate this theory to a 2d TQFT
by considering its supersymmetric index.  Adapted to the present
situation, the argument used there shows that the supersymmetric index
is given by a correlation function
$\vev{\prod_\alpha \CL_\alpha(C_\alpha)}$ in a TQFT on $T^2$, where
$\CL_\alpha(C_\alpha)$ is a line operator created by the NS5-brane
wrapped around $C_\alpha$.

According to the construction of lattice models from TQFTs, this
correlator coincides with the partition function of a lattice model.
To establish the integrability of the model, we need extra dimensions
along which line operators can move freely.  An extra dimension indeed
emerges as 11th dimension if we embed the brane system into M-theory
by string dualities.  Hence, we conclude that the supersymmetric index
of a brane box model is equal to the partition function of an
integrable lattice model.  It turns out that the Yang-Baxter equation
reduces to Seiberg duality for $\SU(N)$ SQCD with $2N$ flavors.

Once the connection between the brane box construction and integrable
lattice models is understood, the case of brane tilings is not so
difficult.  In this case we consider deformations of brane box
configurations, in which we let the NS5-branes combine with the
D5-branes over ribbon-shaped neighborhoods of the $1$-cycles
$C_\alpha$ in $T^2$.  Such a deformed brane configuration still yields
a 4d $\CN = 1$ quiver gauge theory, provided that the deformation
meets a certain criterion.  By the same reasoning as above, we deduce
that the supersymmetric index of this theory is given by the
correlation function of ``thickened'' line operators representing the
ribbon neighborhoods, and equal to the partition function of an
integrable lattice model.  In this way the Bazhanov-Sergeev model
arises from brane tilings.  Again, the Yang-Baxter equation boils
down to Seiberg duality, though this time the equation involves a
sequence of four basic duality transformations.

So we have answers to the first two of our questions.  The
supersymmetric indices of theories constructed from brane tilings are
captured by a lattice model since they are given by correlation
functions of line operators in a TQFT, which in turn is a consequence
of the nature of the brane construction and the fact that the index is
a protected quantity.  Furthermore, the integrability of this lattice
model is guaranteed by the hidden extra dimension which emerges after
the brane system is embedded in M-theory.

In sections~\ref{sec:ILM-TQFT}--\ref{sec:BT}, we discuss in greater
detail the connections summarized here among the brane box and brane
tiling constructions, TQFTs with extra dimensions, and integrable
lattice models.  After reviewing in section~\ref{sec:ILM-TQFT} the
construction of integrable lattice models from TQFTs with extra
dimensions, we explain in section~\ref{sec:BB} how it can be applied
to the supersymmetric indices of brane box models.  We identify the
associated integrable lattice model, and show that the Yang-Baxter
equation for this model takes the form of Seiberg duality.  In
section~\ref{sec:BT}, we treat the case of brane tiling models.

Sections~\ref{sec:LD} and \ref{sec:(0,2)} are devoted to answering our
third question, that is, finding more examples of quiver gauge
theories whose supersymmetric indices are captured by integrable
lattice models.

One way to produce more quiver gauge theories is to apply T-duality to
the brane configurations discussed above.  In this manner we get 3d
$\CN = 2$ and 2d $\CN = (2,2)$ theories.  Being related to the 4d
parents by T-duality, the supersymmetric indices of these theories are
also given by the partition functions of some integrable lattice
models.  In section~\ref{sec:LD}, we will see that for these theories,
the Yang-Baxter equation follows from lower-dimensional analogs of
Seiberg duality, namely a variant of Aharony
duality~\cite{Aharony:1997gp} in the 3d case and Hori-Tong
duality~\cite{Hori:2006dk} in the 2d case.

Perhaps more unexpected is that there are 2d $\CN = (0,2)$ quiver
gauge theories whose supersymmetric indices, or elliptic genera,
exhibit integrability.  They have half as many supercharges as the
theories mentioned so far, and cannot be obtained by simple
dimensional reduction from three or four dimensions.

In section~\ref{sec:(0,2)}, we discuss three classes of such $\CN =
(0, 2)$ theories.  Two of them are $\CN = (0,2)$ counterparts of the
classes of $\CN = (2,2)$ theories considered in section~\ref{sec:LD}.
The Yang-Baxter equation for these classes identifies two theories
related by Seiberg-like triality, discovered by Gadde, Gukov and
Putrov~\cite{Gadde:2013lxa}.  The third class consists of $\CN =
(0,2)$ theories constructed from brane cube
configurations~\cite{GarciaCompean:1998kh}, and actually gives rise to
a 3d lattice model.  The integrability condition for this model is
that its Boltzmann weight satisfies Zamolodchikov's tetrahedron
equation~\cite{MR611994, Zamolodchikov:1981kf}, which is the 3d analog
of the Yang-Baxter equation.  Our analysis in this section will be
somewhat incomplete, unfortunately.  For the first two classes, we
will demonstrate the integrability of the associated lattice models,
but not identify the underlying brane constructions or TQFT structures
whose existence is strongly suggested by the integrability.  For the
third class, on the other hand, we will describe the brane
construction and associated lattice model, but not determine the
corresponding solution of the tetrahedron equation.%
\footnote{A different solution of the tetrahedron equation has been
  found recently by Gadde and Yamazaki~\cite{GY}.  Their solution is
  based on $\CN = (0,2)$ SQCD~\cite{Gadde:2013lxa} with all flavor
  nodes having equal ranks.  It remains to be seen whether this one
  may be understood from the perspective adopted in this paper.}

Having established the structures of integrable lattice models in
several classes of quiver gauge theories, we should now ask what we
can do with these structures.  It is likely that knowledge accumulated
in the area of integrable models provides new insights into these
theories or quiver gauge theories in general.  Conversely, tools from
the gauge theory side, such as localization and $1/N$ expansion, may
help further elucidate the physics of integrable~models.

Given the generality of the TQFT construction of integrable lattice
models, we also expect that there are many more applications than
those discussed in this paper.  Below we describe just a few
possibilities.

The NS5-branes in a brane box configuration can be mapped by dualities
to either M5- or M2-branes intersecting with a stack of M5-branes.
These branes represent codimension-$2$ and -$4$ defects in 6d $\CN =
(2, 0)$ superconformal field theory.  From the 4d point of view, they
are domain walls and line operators in an $\CN = 2$ theory.%
\footnote{We can obtain any $\CN = 2$ theory of class
  $\CS$~\cite{Gaiotto:2009we, Gaiotto:2009hg} from the brane box
  configuration \eqref{eq:BB} by replacing $T^2$ with the relevant
  Riemann surface and introducing D5-branes in the $012357$
  directions.}
This observation suggests that integrability plays a key role in the
physics of defects in the 6d theory and 4d $\CN = 2$ theories, and the
ideas contained in this paper may be useful for understanding it.  In
this regard, we point out that the Yang-Baxter equation has made an
unexpected appearance in studies of the moduli spaces of $\CN = 2$
theories compactified on $S^1$~\cite{Gaiotto:2008cd,
  Alexandrov:2010pp}.

Another important appearance of integrability is found in the
Bethe/gauge correspondence~\cite{Nekrasov:2009uh, Nekrasov:2009ui,
  Nekrasov:2009rc} between 2d $\CN = (2,2)$ theories and quantum
integrable systems.  It may be possible to connect that story to ours,
by studying the supersymmetric vacua of $\CN = (2,2)$ theories
associated with brane box and brane tiling configurations.

Lastly, integrability features prominently in the AdS/CFT
correspondence, and the constructions discussed in the present work
may shed light on this aspect; after all, the developments of brane
box and brane tiling techniques were motivated by the AdS/CFT
correspondence.  In fact, brane box models were studied at one-loop
level in the planar limit in~\cite{Sadri:2005gi}, and it was shown
that at this level, their dilatation operator can be identified with
the Hamiltonian of an integrable spin chain.  This fact raises the
hope that the integrability in the AdS/CFT correspondence may be
understood in a framework that extends the one presented in this
paper.

\section{Integrable lattice models from TQFTs with extra dimensions}
\label{sec:ILM-TQFT}

As explained in the introduction, the construction of integrable
lattice models using line operators in TQFTs constitutes an essential
ingredient for our argument.  So let us begin by reviewing this
construction.  Our discussion mainly follows Costello's
paper~\cite{Costello:2013sla}, to which we refer the reader for more
details.

\subsection{Vertex models from TQFTs}

\begin{figure}
  \centering 
  \begin{tikzpicture}
      \draw (0,0) rectangle (3, 2);
      
      \begin{scope}[shift={(0, 0.2)}]
        \draw[r->] (0, 0) node[left] {$r_1$} -- (3, 0);
        \draw[r->] (0, 1) node[left] {$r_2$}  -- (3, 1);
      \end{scope}

      \begin{scope}[shift={(0.2, 0)}]
        \draw[r->] (0, 0) node[below] {$s_1$} -- (0, 2);
        \draw[r->] (1, 0) node[below] {$s_2$} -- (1, 2);
        \draw[r->] (2, 0) node[below] {$s_3$} -- (2, 2);
      \end{scope}

      \begin{scope}[shift={(0, 0.7)}]
        \draw[dotted] (0, 0) -- (3, 0);
        \draw[dotted] (0, 1) -- (3, 1);
      \end{scope}

      \begin{scope}[shift={(0.7, 0)}]
        \draw[dotted] (0, 0) -- (0, 2);
        \draw[dotted] (1, 0) -- (1, 2);
        \draw[dotted] (2, 0) -- (2, 2);
      \end{scope}
    \end{tikzpicture}
    \caption{A $2 \times 3$ lattice of line operators on a torus.  The
      dotted lines divide the lattice into rectangular pieces, each of
      which contains two intersecting segments of line operators.}
  \label{fig:lattice}
\end{figure}
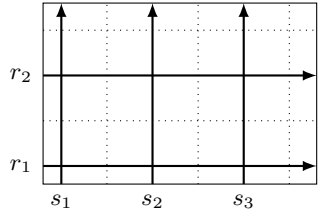

Consider a 2d TQFT equipped with a family of line operators
parametrized by some continuous set, say $\R$ or $\C$.  We denote by
$\CL_r(C)$ a line operator with parameter $r$, supported on an
oriented closed curve $C$.  Let us place this TQFT on a torus $T^2$,
and choose $1$-cycles $A_1$, $\dotsc$, $A_m$, $B_1$, $\dotsc$, $B_n$
that form an $m \times n$ lattice on $T^2$.  The case with $(m, n) =
(2, 3)$ is illustrated in figure~\ref{fig:lattice}.  We are interested
in the correlation function
\begin{equation}
  \label{eq:Z}
  Z(\{r_\alpha\}, \{s_\beta\})
  =
  \biggvev{\prod_{\alpha=1}^m \CL_{r_\alpha}(A_\alpha)
              \prod_{\beta=1}^n \CL_{s_\beta}(B_\beta)}
\end{equation}
of line operators wrapped around these cycles.

We compute this correlation function by breaking up the torus into
smaller rectangular pieces, as shown with dotted lines in
figure~\ref{fig:lattice}.  The idea is that we first perform the path
integral on each of these pieces, and then glue the results together
to reconstruct the path integral on $T^2$.

Each piece in this decomposition is, topologically, a square with two
line operators crossing in the middle:
\begin{equation}
  \label{eq:RM-PI}
  \begin{tikzpicture}[scale=1.5]
    \draw (0,0) rectangle (1,1);
    \draw[r->] (0,0.5) node[left] {$r$} -- (1,0.5);
    \draw[r->] (0.5,0) node[below] {$s$} -- (0.5,1);
  \end{tikzpicture} \quad .
\end{equation}
Using the topological invariance of the theory, we can deform it as
\begin{equation}
  \begin{tikzpicture}[scale=3/4]
    \draw (0,0) -- (1,0);
    \draw (1.5,0) -- (2.5,0);
    \draw (0,2) -- (1,2);
    \draw (1.5,2) -- (2.5,2);
    
    \draw (1,0) .. controls (1.25, 0.75) .. (1.5,0);
    \draw (1,2) .. controls (1.25, 1.25) .. (1.5,2);
    \draw (0,0) .. controls (0.5, 1) .. (0,2);
    \draw (2.5,0) .. controls (2, 1) .. (2.5,2);

    \draw[r->] (0.5,0) node[below] {$r$}
    .. controls (0.5,1) and (2,1) .. (2, 2);
    \draw[r->] (2,0) node[below] {$s$}
    .. controls (2,1) and (0.5,1) .. (0.5, 2);
  \end{tikzpicture} \quad .
\end{equation}
Let $V_r$ be the space of states on a finite interval intersected by
$\CL_r$.  Intuitively, $V_r$ is the Hilbert space of an open string
with a particle attached whose worldline is the line operator $\CL_r$.
In this language, the deformed picture shows an initial state of two
open strings in $V_r \otimes V_s$ evolving into a final state in $V_s
\otimes V_r$ that results from interaction in the middle.  The original
picture may be regarded as a particular case where the curved
boundaries are shrunk to points.  Thus, the path integral on the
square produces a linear map
\begin{equation}
  \RM(r, s)\colon
  V_r \otimes V_s \longto V_s \otimes V_r.
\end{equation}
We call $\check R$ the \emph{R-matrix} of the TQFT with line
operators.

Choosing a basis $\{e_{r, i}\}$ for $V_r$ for all $r$, we can
represent $\RM$ by its matrix elements.  The matrix element $\RM_{i_1
  j_1}^{j_2 i_2}(r, s)$ is the scattering amplitude for a process in
which the initial state $e_{r, i_1} \otimes e_{s, j_1}$ ends up in the
final state $e_{s, j_2} \otimes e_{r, i_2}$.  We represent it
pictorially as
\begin{equation}
  \RM_{i_1 j_1}^{j_2 i_2}(r, s)
   =
  \begin{tikzpicture}
    \draw[r->] (0, 1) node[left] {$r$} -- (2, 1);
    \draw[r->] (1, 0) node[below] {$s$} -- (1, 2);

    \node[below] at (0.5, 1) {$i_1$};
    \node[above] at (1.5, 1) {$i_2$};
    \node[right] at (1, 0.5) {$j_1$};
    \node[left] at (1, 1.5) {$j_2$};
  \end{tikzpicture}
  \quad .
\end{equation}
Then, the correlation function~\eqref{eq:Z} is given by the formula
\begin{equation}
  \label{eq:Z-R}
  Z(\{r_\alpha\}, \{s_\beta\})
  =
  \sum_{\{i_{\alpha, \beta}\}}
  \sum_{\{j_{\beta, \alpha}\}}
  \prod_{\alpha=1}^m \prod_{\beta=1}^n
  \begin{tikzpicture}
    \draw[r->] (0, 1) node[left] {$r_\alpha$} -- (2, 1);
    \draw[r->] (1, 0) node[below] {$s_\beta$} -- (1, 2);

    \node[below] at (0.5, 1) {$i_{\alpha, \beta}$};
    \node[above] at (1.5, 1) {$i_{\alpha, \beta + 1}$};
    \node[right] at (1, 0.5) {$j_{\beta, \alpha}$};
    \node[left] at (1, 1.5) {$j_{\beta, \alpha + 1}$};
  \end{tikzpicture}
  \quad ,
\end{equation}
with the periodic boundary conditions $i_{\alpha, n+1} = i_{\alpha,
  1}$ and $j_{\beta, m+1} = j_{\beta, 1}$.  The right-hand side is
what we get by gluing a collection of pieces like the picture
\eqref{eq:RM-PI} to reconstruct the line operators wrapped on the
torus.

By representing the R-matrix as above, we are treating the open
strings as if their physical degrees of freedom are carried solely by
the particles attached to them, or equivalently, by the edges of the
lattice.  In words, the formula \eqref{eq:Z-R} instructs us to do the
following.  First, choose a state on every edge of the lattice.  Next,
compute the probability amplitude for this configuration of states by
multiplying the corresponding R-matrix elements. Finally, sum over all
possible such configurations to find the answer.

This procedure is, in fact, precisely how the partition function of a
vertex model in statistical mechanics is defined.  In a vertex model,
state variables are assigned to the edges of a lattice, and the
interaction takes place at the vertices.  The total energy $E$ of the
system is the sum of interaction energies at the vertices, so the
Boltzmann weight $e^{-\beta E}$ factorizes into weight factors
associated to the vertices.  These factors are matrix elements of
$\RM$.  In this context, $r$ is called the \emph{spectral parameter}
of the model.

In conclusion, we have found that the correlation function
\eqref{eq:Z} for a lattice of line operators in a 2d TQFT coincides
with the partition function of a vertex model defined on the same
lattice, with the line operator parameter playing the role of a
spectral parameter.

\subsection{Integrability and extra dimensions}

Let us fix the parameters $\{s_\beta\}$ in the lattice of line
operators under consideration.  Viewing the vertical direction of the
torus as the time direction, we define the Hilbert space of the vertex
model to be the tensor product%
\footnote{This is the space of states on the disjoint union of $n$
  intervals, with $\beta$th interval intersected by $\CL_{s_\beta}$.
  The actual Hilbert space of the TQFT is generally smaller than
  $\CH$, since the degrees of freedom associated with the endpoints
  (``Chan-Paton factors'') must match between adjacent intervals.}
\begin{equation}
  \CH = \bigotimes_{\beta=1}^n V_{s_\beta}.
\end{equation}
Furthermore, we introduce the row-to-row \emph{transfer matrix}
\begin{equation}
  T(r)
  =
  \Tr_{V_r} \bigl(\RM(r, s_1) \circ_{V_r} \dotsb \circ_{V_r} \RM(r, s_n)\bigr).
\end{equation}
It is an endomorphism of $\CH$, with matrix elements
\begin{equation}
  T_{j_{1,1}, \dotsc, j_{n,1}}^{j_{1,2}, \dotsc, j_{n,2}}(r)
  =
  \sum_{\{i_\beta\}}
  \prod_{\beta=1}^n
  \begin{tikzpicture}
    \draw[r->] (0, 1) node[left] {$r$} -- (2, 1);
    \draw[r->] (1, 0) node[below] {$s_\beta$} -- (1, 2);

    \node[below] at (0.5, 1) {$i_{\beta}$};
    \node[above] at (1.5, 1) {$i_{\beta + 1}$};
    \node[right] at (1, 0.5) {$j_{\beta, 1}$};
    \node[left] at (1, 1.5) {$j_{\beta, 2}$};
  \end{tikzpicture}
  \quad .
\end{equation}
In terms of the transfer matrix, the partition function \eqref{eq:Z-R}
is written as
\begin{equation}
  \label{eq:Z-T}
  Z(\{r_\alpha\}, \{s_\beta\})
  =
  \Tr_\CH \prod_{\alpha=1}^m T(r_\alpha)
  =
  \sum_{\{j_{\beta, \alpha}\}}
  \prod_{\alpha=1}^m
  T_{j_{1,\alpha}, \dotsc, j_{n,\alpha}}^{j_{1,\alpha+1}, \dotsc, j_{n,\alpha+1}}(r_\alpha).
\end{equation}
In a TQFT, time evolution is trivial unless the state hits some
operator at some point in time.  We may think of $T(r)$ as the time
evolution operator induced by $\CL_r$.

We say that the vertex model on the lattice of line operators is
\emph{integrable} if $T(r)$ is analytic in $r$ and
\begin{equation}
  [T(r),  T(r')] = 0
\end{equation}
for all $r$, $r'$.  The rationale for this terminology is that by
expanding $T(r)$ in powers of $r$, we get a tower of operators that
commute with one another and with the time evolution~operator.

A sufficient condition for integrability is that the R-matrix has
the following two properties.  The first is that $\RM(r, s)$ is an
isomorphism for $r \neq s$, with the inverse being $\RM(s, r)$:
\begin{equation}
  \label{eq:RM*RM=1-eq}
  \RM(r, s) \RM(s, r) = 1.
\end{equation}
The second is that $\RM$ satisfies the \emph{Yang-Baxter equation}:
\begin{equation}
  \label{eq:YBE-eq}
  \RM(s, t) \RM(r, t) \RM(r, s) = \RM(r, s) \RM(r, t) \RM(s, t).
\end{equation}
Each factor on the two sides of the equation is regarded as an
operator on a tensor product of $V_r$, $V_s$ and $V_t$.  For example,
$\RM(r, t)$ on the left-hand side is a map from $V_s \otimes V_r
\otimes V_t$ to~$V_s \otimes V_t \otimes V_r$.

In the TQFT language, the identity \eqref{eq:RM*RM=1-eq} means that two
tangled line operators can be straightened out:
\begin{equation}
  \label{eq:RM*RM=1}
  \begin{tikzpicture}[rounded corners]
    \draw[r->] (0, 0) node[below] {$r$} -- (0, 0.25) -- (1, 0.75) --
    (1, 1.25) -- (0, 1.75) -- (0, 2.1);
    \draw[r->] (1, 0) node[below] {$s$} -- (1, 0.25) -- (0, 0.75) --
    (0, 1.25) -- (1, 1.75) -- (1, 2.1);
  \end{tikzpicture}
  \quad = \quad
  \begin{tikzpicture}
    \draw[r->] (0, 0) node[below] {$r$} -- (0, 2.1);
    \draw[r->] (1, 0) node[below] {$s$} -- (1, 2.1);
  \end{tikzpicture} \quad .
\end{equation}
The Yang-Baxter equation, on the other hand, expresses invariance
under moving a line operator past the intersection of two other line
operators:
\begin{equation}
  \label{eq:YBE}
  \begin{tikzpicture}[rounded corners]
    \draw[r->] (0, 0) node[below] {$r$} -- (0, 0.25) -- (1, 0.75) --
    (1, 1.25) -- (2, 1.75) -- (2, 3.1);
    \draw[r->] (1, 0) node[below] {$s$} -- (1, 0.25) -- (0, 0.75) --
    (0, 2.25) -- (1, 2.75) -- (1, 3.1);
    \draw[r->] (2, 0) node[below] {$t$} -- (2, 1.25) -- (1, 1.75) --
    (1, 2.25) -- (0, 2.75) -- (0, 3.1);
  \end{tikzpicture}
  \quad = \quad
  \begin{tikzpicture}[rounded corners]
    \draw[r->] (0, 0) node[below] {$r$} -- (0, 1.25) -- (1, 1.75) --
    (1, 2.25) -- (2, 2.75) -- (2, 3.1);
    \draw[r->] (1, 0) node[below] {$s$} -- (1, 0.25) -- (2, 0.75) --
    (2, 2.25) -- (1, 2.75) -- (1, 3.1);
    \draw[r->] (2, 0) node[below] {$t$} -- (2, 0.25) -- (1, 0.75) --
    (1, 1.25) -- (0, 1.75) -- (0, 3.1);
  \end{tikzpicture} \quad .
\end{equation}
To see that these properties imply the commutativity of transfer
matrices, consider the following move of line operators:
\begin{equation}
  \begin{tikzpicture}
    \draw[r->, rounded corners] (0, 0) node[left] {$r'$} --
    (0.25, 0) -- (0.75, 1) -- (1, 1) -- (1.5, 0) -- (3.6, 0);
    \draw[r->, rounded corners] (0, 1) node[left] {$r$} --
    (0.25, 1) -- (0.75, 0) -- (1, 0) -- (1.5, 1) -- (3.6, 1);

    \begin{scope}[shift={(2.5, 0)}]
      \fill[white] (0, -0.1) rectangle (0.5, 1.1);
      \draw[thick, dotted] (0, 1) -- (0.5, 1);
      \draw[thick, dotted] (0, 0) -- (0.5, 0);
    \end{scope}

    \draw[r->] (1.75, -0.25) -- (1.75, 1.35);
    \draw[r->] (2.25, -0.25) -- (2.25, 1.35);
    \draw[r->] (3.25, -0.25) -- (3.25, 1.35);
  \end{tikzpicture}
  \quad = \quad
  \begin{tikzpicture}
    \draw[r->, rounded corners] (0, 0) node[left] {$r'$} --
    (0.25, 0) -- (0.75, 1) -- (2.75, 1) -- (3.25, 0) -- (3.6, 0);
    \draw[r->, rounded corners] (0, 1) node[left] {$r$} --
    (0.25, 1) -- (0.75, 0) -- (2.75, 0) -- (3.25, 1) -- (3.6, 1);

    \begin{scope}[shift={(1.75, 0)}]
      \fill[white] (0, -0.1) rectangle (0.5, 1.1);
      \draw[thick, dotted] (0, 1) -- (0.5, 1);
      \draw[thick, dotted] (0, 0) -- (0.5, 0);
    \end{scope}

    \draw[r->] (1, -0.25) -- (1, 1.35);
    \draw[r->] (1.5, -0.25) -- (1.5, 1.35);
    \draw[r->] (2.5, -0.25) -- (2.5, 1.35);
  \end{tikzpicture} \quad .
\end{equation}
Taking the trace over $V_r \otimes V_{r'}$ of both sides, and using
the identity \eqref{eq:RM*RM=1} and the cyclic property of the trace,
we find that the left- and right-hand sides become $T(r) T(r')$ and
$T(r') T(r)$, respectively.

In general, a 2d TQFT does not have the above properties.  Physics may
change discontinuously under the relevant moves due to phase
transitions, as the topology of the line operators does not remain the
same.  Imagine, however, that there are ``extra dimensions'' hidden in
the above pictures.  For instance, the direction perpendicular to the
page may provide one.  The line operators can then sit at different
points in the extra dimensions.  If so, these moves are topologically
trivial and hence the R-matrix satisfies the desired properties.  An
argument of this sort is familiar from studies of knot invariants,
which may be approached either from the perspective of 3d TQFTs or
lattice models~\cite{Witten:1989wf, Witten:1989rw}.

The above observation motivates us to introduce the notion of extra
dimensions to TQFTs.  Consider a $D$-dimensional quantum field theory
formulated on $\Sigma \times M$, where $\Sigma$ is any $d$-manifold
and $M$ is a fixed $(D-d)$-manifold.  We say that the theory is a
\emph{$d$-dimensional TQFT with extra dimensions} if it is topological
on $\Sigma$.%
\footnote{Although such theories might sound exotic, actually many
  have already been studied before.  Typically, they are constructed
  from supersymmetric theories by topological twisting along $\Sigma$.
  2d examples are Costello's theory mentioned below, 4d $\CN = 2$
  superconformal gauge theories with $M$ being any Riemann
  surface~\cite{Kapustin:2006hi}, 5d $\CN = 2$ super Yang-Mills
  theory with $M = S^3$~\cite{Fukuda:2012jr}, and 6d $\CN = (2,0)$
  superconformal field theory with $M = S^1 \times
  S^3$~\cite{Gadde:2009kb}.  3d examples include 5d $\CN = 2$ super
  Yang-Mills theory with $M = \R^2_\eps$~\cite{Dimofte:2010tz,
    Luo:2014sva} and $S^2$~\cite{Yagi:2013fda, Lee:2013ida,
    Cordova:2013cea}, 6d $\CN = (2,0)$ theory with $M = S^1 \times
  S^2$~\cite{Dimofte:2011ju}, $S^3$~\cite{Terashima:2011qi}, and more
  general lens spaces $L(p, q)$~\cite{Dimofte:2014zga}.  4d examples
  are 6d $\CN = (2,0)$ theory with $M$ any Riemann
  surface~\cite{Yagi:2011vd, Yagi:2012xa, Gadde:2013sca}.}
We refer to $M$ as the \emph{internal space} of the theory.  This
definition is very much in the same spirit as the notion of a 2d
conformal field theory ``valued in 4d quantum field
theories''~\cite{Moore:2011ee} developed in connection to the 6d
construction of 4d $\CN = 2$ theories of class
$\CS$~\cite{Gaiotto:2009we, Gaiotto:2009hg}.  In that case, however,
the choice of the 4d spacetime is not part of the definition.  By
contrast, here we demand the theory to be topologically invariant on
$\Sigma$ for a specific choice of $M$.

Now we come to the main point of our discussion.  Suppose that a 2d
TQFT with extra dimensions has line operators.  Place the theory on
$T^2$ and wrap line operators around $1$-cycles $A_\alpha$ and
$B_\beta$.  In the internal space $M$, the $1$-cycles are located at
some points $x_\alpha$ and $y_\beta$, respectively.  The situation is
similar to what we considered at the beginning of this section, but
unlike that case, this time we do \emph{not} assume that the line
operators carry a continuous parameter.  To highlight the difference,
let us consider the extreme case that the theory has only one type of
line operator, $\CL$.  Repeating the same argument, we arrive at the
conclusion that the correlation function
\begin{equation}
  \label{eq:Z-XD}
  Z\bigl(\{x_\alpha\}, \{y_\beta\}\bigr)
  =
  \biggvev{\prod_{\alpha=1}^m \CL(A_\alpha \times \{x_\alpha\})
              \prod_{\beta=1}^n \CL(B_\beta \times \{y_\beta\})}
\end{equation}
is given by the partition function of a vertex model defined on the
lattice of line operators.  As we have just seen, the presence of
extra dimensions guarantees that the R-matrix is invertible and
satisfies the Yang-Baxter equation.

A beautiful insight of Costello~\cite{Costello:2013zra,
  Costello:2013sla} is that even though the line operator lacks a
continuous parameter, the R-matrix still depends on such parameters:
the positions $x_\alpha$, $y_\beta$ of line operators in $M$.  Thus,
the vertex model is integrable if the transfer matrix varies
analytically on $M$.  Based on this idea, it was shown
in~\cite{Costello:2013zra} that a special case of the $6$-vertex model
corresponding to the XXX spin chain, and its generalizations based on
Lie algebras other than $\mathfrak{sl}_2$, arise from a 2d TQFT with
extra dimensions whose internal space $M = \mathbb{CP}^1$.  This TQFT
is obtained from a deformed and topologically twisted version of 4d
$\CN = 1$ super Yang-Mills theory, and the line operators used in the
construction are Wilson lines.  The extra dimensions therefore
elegantly explain not only why the Yang-Baxter equation holds for
these models, but also where the spectral parameter comes from.

Speaking of extra dimensions, it should be noted that although only
line operators were considered in our discussion, we could as well use
higher-dimensional operators that have codimension at least two in
$\Sigma \times M$.  This is because after we wrap them on $1$-cycles
on $\Sigma$, there is still room in $M$ for them to avoid one another.
In our main examples, we will make use of codimension-$2$ defects in a
6d theory.

\subsection{IRF models}

In the construction described above, we obtained a vertex model from
line operators by letting the midpoints of open strings represent all
physical degrees of freedom.  In many cases, however, degrees of
freedom really reside on (and only on) the endpoints.  In such cases
it is more efficient to reformulate the vertex model as an
interaction-round-a-face (IRF) model.

In an IRF model, state variables live on the vertices of a lattice,
and interaction takes place among vertices connected by edges that
surround a face.  We use letters $a$, $b$, $\dotsc$ to denote
state variables.  For a square lattice, the Boltzmann weight for the
interaction is denoted as
\begin{equation}
  \label{eq:W}
  \BW{a}{b}{c}{d}{r}{s}
  =
  \begin{tikzpicture}
    \draw[dr->] (0, 1) node[left] {$r$} -- (2, 1);
    \draw[dr->] (1, 0) node[below] {$s$}-- (1, 2);

    \begin{scope}[shift={(0.5, 0.5)}]
      \node[tnode] (a) at (0, 0) {$a$};
      \node[tnode] (b) at (1, 0) {$b$};
      \node[tnode] (c) at (1, 1) {$c$};
      \node[tnode] (d) at (0, 1) {$d$};
    \end{scope}

    \draw[thick] (a) -- (b) -- (c) -- (d) -- (a);
\end{tikzpicture} \quad .
\end{equation}
The dashed oriented lines are \emph{rapidity lines}, and make up the
dual lattice.  The parameters $r$, $s$ are spectral parameters, also
called \emph{rapidities}.  An IRF model is a vertex model on the dual
lattice, in which the state space assigned to an edge is the direct
product of two spaces and many elements of the R-matrix vanish.

From the TQFT point of view, rapidity lines are line operators, and
the rapidities are their parameters.  The Yang-Baxter
equation~\eqref{eq:YBE} for an IRF model is
\begin{equation}
  \label{eq:YBE-IRF}
  \begin{tikzpicture}[scale=1.5]
    \foreach \x [count=\xi] in {a,...,f}
    {\node[tnode] (\x) at (180+60*\xi:1) {$\x$};}
    \node[tnode] (g) at (0,0) {$g$};
    
    \draw[thick] (a) -- (b) -- (c) -- (d) -- (e) -- (f) -- (a);
    \draw[thick] (a) -- (g);
    \draw[thick] (c) -- (g);
    \draw[thick] (e) -- (g);
    
    \draw[dr->, shift={(120:0.2)}] (210:1.1) node[left] {$s$} -- (30:1.1);
    \draw[dr->, shift={(0.2, 0)}] (-90:1.1) node[below] {$t$} -- (90:1.1);
    \draw[dr->, shift={(-120:0.2)}] (150:1.1) node[left] {$r$} -- (-30:1.1);
  \end{tikzpicture}
  \quad = \quad
  \begin{tikzpicture}[scale=1.5]
    \foreach \x [count=\xi] in {a,...,f}
    {\node[tnode] (\x) at (180+60*\xi:1) {$\x$};}
    \node[tnode] (g) at (0,0) {$g$};

    \draw[thick] (a) -- (b) -- (c) -- (d) -- (e) -- (f) -- (a);
    \draw[thick] (b) -- (g);
    \draw[thick] (d) -- (g);
    \draw[thick] (f) -- (g);

    \draw[dr->, shift={(-60:0.2)}] (210:1.1) node[left] {$s$} -- (30:1.1);
    \draw[dr->, shift={(-0.2, 0)}] (-90:1.1) node[below] {$t$} -- (90:1.1);
    \draw[dr->, shift={(60:0.2)}] (150:1.1) node[left] {$r$} -- (-30:1.1);
  \end{tikzpicture}
  \quad .
\end{equation}
In terms of the Boltzmann weight, the equation reads
\begin{multline}
  \sum_g
  \BW{g}{c}{d}{e}{s}{t}
  \BW{a}{b}{c}{g}{r}{t}
  \BW{f}{a}{g}{e}{r}{s}
  \\
  =
  \sum_g
  \BW{g}{b}{c}{d}{r}{s}
  \BW{f}{g}{d}{e}{r}{t}
  \BW{a}{b}{g}{f}{s}{t}
  .
\end{multline}

\subsection{3d TQFTs and the tetrahedron equation}

As a generalization, we can construct higher-dimensional lattice
models using codimension-$1$ defects in a $d$-dimensional TQFT with $d
> 2$.  Placing the theory on a $d$-torus $T^d$ and putting these
defects on $(d-1)$-cycles that form a $d$-dimensional lattice, we get
a vertex model on this lattice.  If the theory has extra dimensions,
the R-matrix of the model satisfies a $d$-dimensional analog of the
Yang-Baxter equation.

To be concrete, let us take $d = 3$.  In this case, codimension-$1$
defects are surface operators, and three of them intersect at a vertex
of the lattice.  A cubic neighborhood of a vertex looks like
\begin{equation}
  \begin{tikzpicture}[scale=1.5]
    \coordinate (a) at (0,0);
    \coordinate (e) at (0,-1);
    \coordinate (f) at (-1,0);
    \coordinate (d) at (-1, -1);

    \begin{scope}[shift=(-150:0.7)]
      \coordinate (g) at (0,0);
      \coordinate (c) at (0,-1);
      \coordinate (b) at (-1,0);
      \coordinate (h) at (-1, -1);
    \end{scope}

    \node[above left=-2pt] at ($0.5*(f) + 0.5*(b)$) {$r$};
    \node[above] at ($0.5*(f) + 0.5*(a)$) {$s$};
    \node[left] at ($0.5*(b) + 0.5*(h)$) {$t$};

    \draw (a) to (e);
    \draw[dashed] (e) -- (d);
    \draw[dashed] (d) -- (f);
    \draw (f) -- (a);
    \draw (g) -- (c) -- (h) -- (b) -- (g);
    \draw (a) -- (g);
    \draw (f) -- (b);
    \draw (e) -- (c);
    \draw[dashed] (d) -- (h);

    \draw[densely dotted] ($0.5*(a) + 0.5*(g)$) rectangle ($0.5*(d) + 0.5*(h)$);
    \draw[densely dotted] ($0.5*(b) + 0.5*(h)$) -- ++(1, 0) -- ++(30:0.7) --
    ++(-1, 0) -- cycle;
    \draw[densely dotted] ($0.5*(c) + 0.5*(h)$) -- ++(30:0.7) --
    ++(0, 1) -- ++(-150:0.7) -- cycle;

    \draw[r->] ($0.5*(e) + 0.5*(h)$) -- ++(0, 1);
    \draw[r->] ($0.5*(f) + 0.5*(h)$) -- ++(1, 0);
    \draw[r->] ($0.5*(g) + 0.5*(h)$) -- ++(30:0.7);
  \end{tikzpicture}
  \quad .
\end{equation}
The orientations of the lattice edges specify those of the surface
operators (drawn with dotted lines in the picture).  Viewing this
picture as the worldvolumes of three faces of the cube traveling in a
diagonal direction, we see that the R-matrix $R(r, s, t)$ produced by
the path integral on this cube is an endomorphism of $V_{r,s} \otimes
V_{s,t} \otimes V_{t,r}$, where $V_{r,s}$ is the Hilbert spaces on a
square intersected by two surface operators with parameters $r$ and
$s$.  Pictorially, the R-matrix is represented as
\begin{equation}
  R(r, s, t)
  =
  \begin{tikzpicture}[scale=1.5]
    \coordinate (a) at (0,0);
    \coordinate (e) at (0,-1);
    \coordinate (f) at (-1,0);
    \coordinate (d) at (-1, -1);

    \begin{scope}[shift=(-150:0.7)]
      \coordinate (g) at (0,0);
      \coordinate (c) at (0,-1);
      \coordinate (b) at (-1,0);
      \coordinate (h) at (-1, -1);
    \end{scope}

    \draw[r->] ($0.5*(e) + 0.5*(h)$) node[below] {$(r, s)$} -- ++(0, 1);
    \draw[r->] ($0.5*(f) + 0.5*(h)$) node[left] {$(t, r)$} -- ++(1, 0);
    \draw[r->] ($0.5*(g) + 0.5*(h)$) node[below left=-2pt] {$(s, t)$} -- ++(30:0.7);
  \end{tikzpicture}
  \quad .
\end{equation}
To avoid clutter, in what follows we will suppress spectral parameters
in this kind \linebreak of~pictures.

The 3d analog of the Yang-Baxter equation is Zamolodchikov's
tetrahedron equation~\cite{MR611994, Zamolodchikov:1981kf}, obtained
by moving one of four surface operators that form a tetrahedron:
\begin{equation}
  \begin{tikzpicture}[thick, scale=3/4]
    \coordinate (a) at (0,0);
    \coordinate (b) at (30:2);
    \coordinate (c) at (-15:2);
    \coordinate (d) at (-45:1.4);

    \draw[r->] ($(a)!-0.5\qsep!(b)$) -- ($(b)!-0.5\qsep!(a)$);
    \draw ($(a)!-0.5\qsep!(c)$) -- (a);
    \draw[r->] (c) --($(c)!-0.5\qsep!(a)$);
    \draw[r<-] ($(a)!-0.5\qsep!(d)$) -- ($(d)!-0.5\qsep!(a)$);
    \draw[r<-] ($(b)!-0.5\qsep!(c)$) -- ($(c)!-0.5\qsep!(b)$);
    \draw[r<-] ($(b)!-0.5\qsep!(d)$) -- ($(d)!-0.5\qsep!(b)$);
    \draw[r<-] ($(c)!-0.5\qsep!(d)$) -- ($(d)!-0.5\qsep!(c)$);

    \draw[dashed] (a) -- (c);
  \end{tikzpicture}
  \quad = \quad
  \begin{tikzpicture}[thick, scale=3/4]
    \coordinate (a) at (0,0);
    \coordinate (b) at (210:2);
    \coordinate (c) at (165:2);
    \coordinate (d) at (135:1.4);

    \draw[r<-] ($(a)!-0.5\qsep!(b)$) -- ($(b)!-0.5\qsep!(a)$);
    \draw[r<-] ($(a)!-0.5\qsep!(c)$) -- ($(c)!-0.5\qsep!(a)$);
    \draw[r->] ($(a)!-0.5\qsep!(d)$) -- ($(d)!-0.5\qsep!(a)$);
    \draw[r->] ($(b)!-0.5\qsep!(c)$) -- ($(c)!-0.5\qsep!(b)$);
    \draw ($(b)!-0.5\qsep!(d)$) -- (b);
    \draw[r->] (d) -- ($(d)!-0.5\qsep!(b)$);
    \draw[r->] ($(c)!-0.5\qsep!(d)$) -- ($(d)!-0.5\qsep!(c)$);

    \draw[dashed] (b) -- (d);
  \end{tikzpicture}
  \quad .
\end{equation}
The tetrahedron equation implies the commutativity of layer-to-layer
transfer matrices at different values of the spectral parameter.

When the state variables may be considered as living on the cubes
surrounded by surface operators, a vertex model can be reformulated as
an interaction-round-a-cube (IRC) model~\cite{MR838340}.  The
Boltzmann weight of an IRC model is denoted as
\begin{equation}
  W(a|e, f, g|b, c, d|h)
  =
  \begin{tikzpicture}[thick]
    \node[wnode] (a) at (0,0) {$a$};
    \node[wnode] (e) at (0,-1) {$e$};
    \node[wnode] (f) at (-1,0) {$f$};
    \node[wnode] (d) at (-1, -1) {$d$};

    \begin{scope}[shift=(-150:0.7)]
      \node[wnode] (g) at (0,0) {$g$};
      \node[wnode] (c) at (0,-1) {$c$};
      \node[wnode] (b) at (-1,0) {$b$};
      \node[wnode] (h) at (-1, -1) {$h$};
    \end{scope}


    \draw (a) to (e);
    \draw[dashed] (e) -- (d);
    \draw[dashed] (d) -- (f);
    \draw (f) -- (a);
    \draw (g) -- (c) -- (h) -- (b) -- (g);
    \draw (a) -- (g);
    \draw (f) -- (b);
    \draw (e) -- (c);
    \draw[dashed] (d) -- (h);
  \end{tikzpicture}
  \quad .
\end{equation}
The tetrahedron equation for an IRC model is
\begin{multline}
  \label{eq:TE-IRC}
  \begin{split}
    &\sum_d W(a_4|c_1, c_3, c_2|b_3, b_2, b_1|d)
    W(c_1|a_3, b_1, b_2|d, c_6, c_4|b_4) \\
    &\quad \times
    W(b_1|c_4, c_3, d|b_3, b_4, a_2|c_5)
    W(d|b_4, b_3, b_2|c_2, c_6, c_5|a_1)
  \end{split}
  \\
  \begin{split}
    = &\sum_d W(b_1|c_4, c_3, c_1|a_4, a_3, a_2|d)
       W(c_1|a_3, a_4, b_2|c_2, c_6, d|a_1) \\
       &\quad \times
       W(a_4|d, c_3, c_2|b_3, a_1, a_2|c_5)
       W(d|a_3, a_2, a_1|c_5, c_6, c_4|b_4).
  \end{split}
\end{multline}
Graphically, it takes the form of equivalence between two ways of
dividing a dodecahedron \pagebreak \linebreak into four hexahedra:
\begin{equation}
  \begin{tikzpicture}[rotate=-20, thick, scale=3/4]
    \node[wnode] (d) at (0,0) {$d$};

    \node[wnode] (a3) at (.5,2) {$a_3$};
    \node[wnode] (a2) at (-1.5,0) {$a_2$};
    \node[wnode] (a1) at (1.7,-1) {$a_1$};
    \node[wnode] (a4) at (-.7,-1) {$a_4$};
    
    \node[wnode] (c4) at ($(a3) + (a2)$) {$c_4$};
    \node[wnode] (c5) at ($(a2) + (a1)$) {$c_5$};
    \node[wnode] (c6) at ($(a3) + (a1)$) {$c_6$};
    \node[wnode] (c1) at ($(a3) + (a4)$) {$c_1$};
    \node[wnode] (c2) at ($(a1) + (a4)$) {$c_2$};
    \node[wnode] (c3) at ($(a2) + (a4)$) {$c_3$};
    
    \node[wnode] (b3) at ($(a4) + (c5)$) {$b_3$};
    \node[wnode] (b2) at ($(a3) + (c2)$) {$b_2$};
    \node[wnode] (b1) at ($(a3) + (c3)$) {$b_1$};
    \node[wnode] (b4) at ($(a3) + (c5)$) {$b_4$};
    
    \draw[dashed] (b3) -- (c5);
    \draw (b3) -- (c3);
    \draw (b3) -- (c2);
    \draw (b1) -- (c4);
    \draw (b1) -- (c1);
    \draw (b1) -- (c3);
    \draw (b2) -- (c6);
    \draw (b2) -- (c1);
    \draw (b2) -- (c2);
    \draw[dashed] (b4) -- (c6);
    \draw[dashed] (b4) -- (c5);
    \draw[dashed] (b4) -- (c4);
    \draw (a3) -- (c6);
    \draw (a3) -- (c4);
    \draw (a3) -- (c1);
    \draw (a1) -- (c6);
    \draw[dashed] (a1) -- (c5);
    \draw (a1) -- (c2);
    \draw[dashed] (a2) -- (c5);
    \draw[dashed] (a2) -- (c4);
    \draw[dashed] (a2) -- (c3);
    \draw (a4) -- (c1);
    \draw (a4) -- (c3);
    \draw (a4) -- (c2);
    \draw[dashed] (b3) -- (d);
    \draw[dashed] (b1) -- (d);
    \draw[dashed] (b2) -- (d);
    \draw[dashed] (b4) -- (d);
  \end{tikzpicture}
  \quad = \quad
  \begin{tikzpicture}[rotate=-20, thick, scale=3/4]
    \node[wnode] (d) at (0,0) {$d$};

    \node[wnode] (a3) at (.5,2) {$a_3$};
    \node[wnode] (a2) at (-1.5,0) {$a_2$};
    \node[wnode] (a1) at (1.7,-1) {$a_1$};
    \node[wnode] (a4) at (-.7,-1) {$a_4$};
    
    \node[wnode] (c4) at ($(a3) + (a2)$) {$c_4$};
    \node[wnode] (c5) at ($(a2) + (a1)$) {$c_5$};
    \node[wnode] (c6) at ($(a3) + (a1)$) {$c_6$};
    \node[wnode] (c1) at ($(a3) + (a4)$) {$c_1$};
    \node[wnode] (c2) at ($(a1) + (a4)$) {$c_2$};
    \node[wnode] (c3) at ($(a2) + (a4)$) {$c_3$};
    
    \node[wnode] (b3) at ($(a4) + (c5)$) {$b_3$};
    \node[wnode] (b2) at ($(a3) + (c2)$) {$b_2$};
    \node[wnode] (b1) at ($(a3) + (c3)$) {$b_1$};
    \node[wnode] (b4) at ($(a3) + (c5)$) {$b_4$};
    
    \draw[dashed] (b3) -- (c5);
    \draw (b3) -- (c3);
    \draw (b3) -- (c2);
    \draw (b1) -- (c4);
    \draw (b1) -- (c1);
    \draw (b1) -- (c3);
    \draw (b2) -- (c6);
    \draw (b2) -- (c1);
    \draw (b2) -- (c2);
    \draw[dashed] (b4) -- (c6);
    \draw[dashed] (b4) -- (c5);
    \draw[dashed] (b4) -- (c4);
    \draw (a3) -- (c6);
    \draw (a3) -- (c4);
    \draw (a3) -- (c1);
    \draw (a1) -- (c6);
    \draw[dashed] (a1) -- (c5);
    \draw (a1) -- (c2);
    \draw[dashed] (a2) -- (c5);
    \draw[dashed] (a2) -- (c4);
    \draw[dashed] (a2) -- (c3);
    \draw (a4) -- (c1);
    \draw (a4) -- (c3);
    \draw (a4) -- (c2);
    \draw[dashed] (a3) -- (d);
    \draw[dashed] (a1) -- (d);
    \draw[dashed] (a2) -- (d);
    \draw[dashed] (a4) -- (d);
  \end{tikzpicture}
  \quad .
\end{equation}

\section{\texorpdfstring{4d $\boldsymbol{\CN = 1}$ quiver gauge
    theories}{4d N=1 quiver gauge theories}}
\label{sec:BB}

Now we wish to apply the above construction to 4d $\CN = 1$ quiver
gauge theories realized by brane box configurations.  In this section,
we explain how the elements that entered the construction --- TQFT,
lattice of line operators, and extra dimensions --- arise nicely in the
supersymmetric indices of brane box models.  Furthermore, we determine
the associated integrable lattice model, and demonstrate that the
Yang-Baxter equation for this model is equivalent to the equality
between the supersymmetric index of $\SU(N)$ SQCD with $2N$ flavors
and that of its Seiberg dual.

\subsection{Brane box models}

Consider Type IIB superstring theory, and suppose that we have $N$
coincident D5-branes extending in the directions $012346$, and a
number of NS5-branes in $012345$ and $012367$:
\begin{equation}
  \label{eq:BB}
  \begin{tabular}{|l|cccccccccc|}
    \hline
    & 0 & 1 & 2 & 3 & 4 & 5 & 6 & 7 & 8 & 9
    \\ \hline
    D5 & $\times$ & $\times$ & $\times$ & $\times$ & $\times$ & & $\times$
    &&&
    \\
    NS5 & $\times$ & $\times$ & $\times$ & $\times$ & $\times$ & $\times$
    &&&&
    \\
    NS5 & $\times$ & $\times$ & $\times$ & $\times$ & & & $\times$ & $\times$
    &&
    \\ \hline
  \end{tabular}
\end{equation}
These branes are all located at the same point in the $89$-plane.  So
each NS5-brane intersects with the D5-branes along a straight line
parallel to the 4- or 6-axis.

Macroscopically, this brane system realizes a supersymmetric quiver
gauge theory, much the same way as the brane constructions of 3d $\CN
= 4$~\cite{Hanany:1996ie} and 4d $\CN = 2$~\cite{Witten:1997sc} gauge
theories.  In those constructions, D3- or D4-branes are suspended
between NS5-branes placed at finite intervals in the 6-direction.  The
difference here is that the D5-branes are divided by NS5-branes into
finite segments in the 4-direction as well.  The resulting theory
therefore has less spacetime dimensions and supersymmetry, but a more
complicated quiver: it is a 4d $\CN = 1$ gauge theory described by a
planar quiver diagram.  Brane box models, first studied
in~\cite{Hanany:1997tb}, are theories constructed from brane
configurations of this kind.

\begin{figure}
  \centering 
  \begin{subfigure}[b]{5.18\qsep}
    \centering 
    \begin{tikzpicture}[scale=1.2]
    \begin{scope}[scale=1/1.2, shift={(-1.35, 1.58)}]
      \draw[q->] (0,0) -- (0.5, 0) node[right, font =\scriptsize] {$x^4$};
      \draw[q->] (0,0) -- (0, 0.5) node[above, font =\scriptsize] {$x^6$};
    \end{scope}

      \draw (0,0) rectangle (3, 2);
      
      \begin{scope}[shift={(0.15, 0)}]
        \draw[dr->] (0, 0) -- (0, 2);
        \draw[dr->] (1, 0) -- (1, 2);
        \draw[dr->] (2, 0) -- (2, 2);
      \end{scope}
      
      \begin{scope}[shift={(0, 0.15)}]
        \draw[dr->] (0, 0) -- (3, 0);
        \draw[dr->] (0, 1) -- (3, 1);
      \end{scope}

      \begin{scope}[shift={(0.15, 0.15)}]
        \node (a) at (0.5, 0.5) {$a$};
        \node (b) at (1.5, 0.5) {$b$};
        \node (c) at (2.5, 0.5) {$c$};
        \node (d) at (0.5, 1.5) {$d$};
        \node (e) at (1.5, 1.5) {$e$};
        \node (f) at (2.5, 1.5) {$f$};
      \end{scope}
    \end{tikzpicture}
    \caption{}
    \label{fig:brane-box-ex-a}
  \end{subfigure}%
  \qquad\qquad 
  \begin{subfigure}[b]{3.6\qsep}
    \centering
    \begin{tikzpicture}[scale=1.2]
      \draw (0,0) rectangle (3, 2);
      
      \begin{scope}[shift={(0.15, 0.15)}]
        \node[wnode] (a) at (0.5, 0.5) {$a$};
        \node[wnode] (b) at (1.5, 0.5) {$b$};
        \node[wnode] (c) at (2.5, 0.5) {$c$};
        \node[wnode] (d) at (0.5, 1.5) {$d$};
        \node[wnode] (e) at (1.5, 1.5) {$e$};
        \node[wnode] (f) at (2.5, 1.5) {$f$};
      \end{scope}
      
      \draw[q->] (a) -- (b);
      \draw[q->] (b) -- (c);
      \draw[q->] (d) -- (e);
      \draw[q->] (e) -- (f);
      
      \draw[q->] (a) -- (d);
      \draw[q->] (b) -- (e);
      \draw[q->] (c) -- (f);
      
      \draw[q->] (e) -- (a);
      \draw[q->] (f) -- (b);
      
      \draw (a) -- (0, 0);
      \draw (b) -- (1, 0);
      \draw (c) -- (2, 0);
      \draw (d) -- (0, 1);
      
      \draw[q->] (1, 2) -- (d);
      \draw[q->] (2, 2) -- (e);
      \draw[q->] (3, 2) -- (f);
      \draw[q->] (3, 1) -- (c);
      
      \begin{scope}[shift={(0, 0.15)}]
        \draw[q->] (0, 0.5) -- (a);
        \draw (c) -- (3, 0.5);
        \draw[q->] (0, 1.5) -- (d);
        \draw (f) -- (3, 1.5);
      \end{scope}
      
      \begin{scope}[shift={(0.15, 0)}]
        \draw[q->] (0.5, 0) -- (a);
        \draw (d) -- (0.5, 2);
        \draw[q->] (1.5, 0) -- (b);
        \draw (e) -- (1.5, 2);
        \draw[q->] (2.5, 0) -- (c);
        \draw (f) -- (2.5, 2);
      \end{scope}
    \end{tikzpicture}
    \caption{}
    \label{fig:brane-box-ex-b}
  \end{subfigure}
  \caption{The $2 \times 3$ brane box model.  (a) The configuration of
    NS5-brane intersections on the $46$-torus.  (b) The corresponding
    quiver diagram.}
  \label{fig:brane-box-ex}
\end{figure}
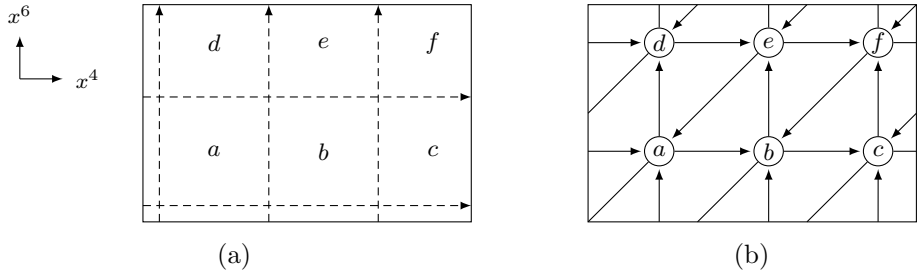

To avoid complications associated with semi-infinite branes, let us
compactify the $46$-plane to a torus $T^2$.  Then, the intersections
of the NS5-branes with the D5-branes form a lattice on the $46$-torus.
We draw the intersections by oriented dashed lines and call them
rapidity lines, anticipating connection with integrable lattice
models.  We choose the orientations for horizontal lines in such a way
that they are parallel (and not antiparallel) to one another, and
similarly for vertical lines.  See figure~\ref{fig:brane-box-ex-a} for
an example of a brane box model, in which the rapidity lines make up a
$2 \times 3$ lattice.

The rule for determining the quiver is as follows.  We label the faces
of the lattice by letters $a$, $b$, $\dotsc$.  To the $a$th face is
associated a vector multiplet for gauge group $\SU(N)_a$, which
originates from open strings whose ends are attached on this part of
the D5-branes.  Since the rank of the gauge group is fixed throughout
our discussion, in our quiver diagram we represent the vector
multiplet simply by a circle node labeled $a$:
\begin{equation}
  \begin{tikzpicture}
    \node[wnode] {$a$};
  \end{tikzpicture}
\end{equation}
From two faces $a$ and $b$ adjacent to each other or sharing a vertex,
we get a chiral multiplet in a bifundamental representation of
$\SU(N)_a \times \SU(N)_b$.  This is represented by an arrow between
the two nodes:
\begin{equation}
  \label{arrow-ab}
  \begin{tikzpicture}
    \node[wnode] (a) at (0,0) {$a$};
    \node[wnode] (b) at (1,0) {$b$};
    \draw[q->] (a) -- (b);
  \end{tikzpicture}
\end{equation}
is a chiral multiplet in the representation $(\overline{\mathbf{N}},
\mathbf{N})$ of $\SU(N)_a \times \SU(N)_b$, coupled to the vector
multiplets.  If four faces $a$, $b$, $c$, $d$ are placed around a
vertex in the counterclockwise order, the corresponding part of the
quiver is given by
\begin{equation}
  \begin{tikzpicture}
    \draw[dr->] (1, 0) -- (1, 2);
    \draw[dr->] (0, 1) -- (2, 1);
    
    \begin{scope}[shift={(0.5, 0.5)}]
      \node[wnode] (a) at (0, 0) {$a$};
      \node[wnode] (b) at (1, 0) {$b$};
      \node[wnode] (c) at (1, 1) {$c$};
      \node[wnode] (d) at (0, 1) {$d$};
    \end{scope}

    \draw[q->] (a) -- (b);
    \draw[q->] (d) -- (c);
    \draw[q->] (a) -- (d);
    \draw[q->] (b) -- (c);
    \draw[q->] (c) -- (a);
  \end{tikzpicture} \quad .
\end{equation}
What makes brane box models physically interesting is that they are
chiral: between nodes, arrows only point in one direction.  The quiver
diagram for the $2 \times 3$ brane box model is shown in
figure~\ref{fig:brane-box-ex-b}.

The same diagram also encodes the superpotential of the theory.  The
quiver consists of a collection of triangles that cover the
$46$-torus.  For each triangle formed by three arrows, there is a
cubic superpotential term given by the trace of the product of the
corresponding bifundamental multiplets, with sign determined by the
orientation of the arrows.  The precise form of the superpotential
will not be important for our discussion; the quantities we will
consider are largely independent of this information.

\begin{figure}
  \centering
  \begin{tikzpicture}[thick]
    \begin{scope}[thin, shift={(-1.63, 1.48)}]
      \draw[q->] (0,0) -- (0.5, 0) node[right, font =\scriptsize] {$x^6$};
      \draw[q->] (0,0) -- (0, 0.5) node[above, font =\scriptsize] {$x^5$};    
    \end{scope}

    \draw (1, 0) -- (1, 2) node[above] {NS5};
    \draw (2, 0) -- (2, 2);
    \draw (3, 0) -- (3, 2);
    
    \begin{scope}[shift={(0, 0.5)}]
      \draw (0, 0) node[left] {$N$ D5} -- (1, 0);
    \end{scope}

    \begin{scope}[shift={(1, 1.5)}]
      \draw (0, 0) -- (1, 0);
    \end{scope}

    \begin{scope}[shift={(2, 0.7)}]
      \draw (0, 0) -- (1, 0);
    \end{scope}

    \begin{scope}[shift={(3, 1.2)}]
      \draw (0, 0) -- (1, 0);
    \end{scope}
  \end{tikzpicture}
  \caption{Sliding D5-brane segments along NS5-branes.  Illustrated
    here is an ideal situation with $g_s = 0$.}
  \label{fig:D5-NS5}
\end{figure}
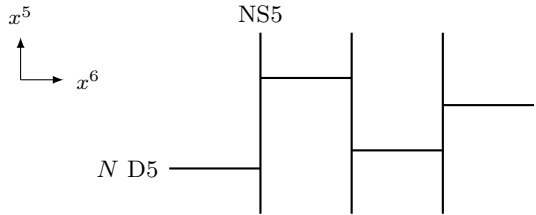

The gauge group of a brane box model is the product of $\SU(N)$s and
not $\U(N)$s.  To understand why, let us suppose for simplicity that
there are only NS5-branes spanning the $012345$ directions.  In this
case we get a 5d $\CN = 1$ theory.  We are taking the string coupling
$g_s$ to be small, which means that the NS5-branes have tension much
larger than that of the D5-branes and can be treated as rigid objects.
These heavy NS5-branes ``chop'' the D5-branes into finite segments in
the $6$-direction.  Imagine sliding these segments slightly in the
$5$-direction.  If $g_s$ were strictly zero, the NS5-branes would be
infinitely heavier than the D5-branes and not affected by such a
deformation at all.  Then we would have a situation sketched in
figure~\ref{fig:D5-NS5}.

\begin{figure}
  \centering
  \begin{subfigure}{3.41\qsep}
    \centering
    \begin{tikzpicture}[thick]
      \draw (1, 0) -- (1, 0.5);
      \draw (1.5, 1.5) -- (1.5, 2) node[above] {NS5};

      \draw (1, 0.5) -- node[below right=-2pt] {$(N,1)$} (1.5, 1.5);
      
      \draw (0, 0.5) node[left] {$N$ D5} -- (1, 0.5);
      \draw (1.5, 1.5) -- (2.5, 1.5);
    \end{tikzpicture}
    \caption{}
    \label{fig:5BW-a}
  \end{subfigure}%
  \qquad\qquad 
  \begin{subfigure}{3.41\qsep}
    \centering
    \begin{tikzpicture}[thick]
      \draw (1, 1.5) -- (1, 2) node[above] {NS5};
      \draw (1.5, 0.5) -- (1.5, 0);
      \draw (1, 1.5) -- node[below left=-2pt] {$(N, -1)$} (1.5, 0.5);
      
      \draw (0, 1.5) node[left] {$N$ D5} -- (1, 1.5);
      \draw (1.5, 0.5) -- (2.5, 0.5);
    \end{tikzpicture}
    \caption{}
    \label{fig:5BW-b}
  \end{subfigure}
  \caption{For $g_s > 0$, taking apart two D5-brane segments creates
    either (a) a tilted $(N, 1)$ 5-brane or (b) a tilted $(N, -1)$
    5-brane.}
  \label{fig:5BW}
\end{figure}
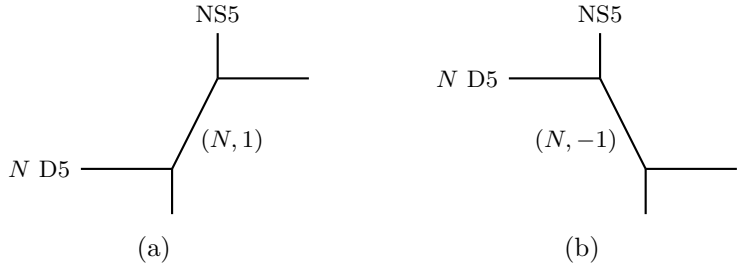

In reality, $g_s$ is small but nonzero, and the deformation does
affect the NS5-branes.  What actually happens is that when a stack of
$N$ D5-branes hit on an NS5-brane, these branes combine to form a
bound state --- an $(N, 1)$ or $(N, -1)$ 5-brane, depending on the
relative positions of the branes --- which then goes off diagonally, as
shown in figure~\ref{fig:5BW}.  As a result, moving the D5-brane
segments in the $5$-direction entails shifts of semi-infinite
NS5-branes in the 6-direction, and this costs an infinite amount of
energy.  The $x^5$-coordinates of the D5-brane segments are therefore
frozen.  As the diagonal $\U(1)$ gauge field on each segment is in the
same 5d multiplet as the scalar $x^5$, it is frozen as well.  Thus,
the gauge group of the 5d theory is the product of $\SU(N)$s, times a
$\U(1)$ factor corresponding to the overall center of mass coordinate
of the D5-branes.  The $\U(1)$ factor is decoupled from the rest of
the dynamics, so we ignore it.

\subsection{R-symmetry and flavor symmetries}

The brane system~\eqref{eq:BB} is invariant under rotations in the
$89$-plane.  The rotational symmetry appears in the effective 4d
theory as an R-symmetry $\U(1)_R$.  However, the R-charges of the
bifundmanetal multiplets are not determined uniquely by the brane
configuration, because of the possibility of shifting them by $\U(1)$
flavor charges.  This ambiguity plays a crucial role in the connection
to integrable lattice models, so let us look at it closely.

We denote by $R_{ab}$ the R-charge of the bifundamental multiplet
represented by an arrow going from node $a$ to $b$.  There are two
conditions for a set of R-charges $\{R_{ab}\}$ to be realized in a
brane box model.  The first is that the superpotential must have
R-charge $2$ so that it preserves $\U(1)_R$.  Thus, for every triangle
\begin{equation}
  \begin{tikzpicture}
    \node[wnode] (a) at (0,0) {$a$};
    \node[wnode] (b) at (1,0) {$b$};
    \node[wnode] (c) at (1,1) {$c$};
    
    \draw[q->] (a) -- (b);
    \draw[q->] (b) -- (c);
    \draw[q->] (c) -- (a);
  \end{tikzpicture}
  \quad \text{or} \quad
  \begin{tikzpicture}
    \node[wnode] (a) at (0,0) {$a$};
    \node[wnode] (b) at (0,1) {$b$};
    \node[wnode] (c) at (1,1) {$c$};
    
    \draw[q->] (a) -- (b);
    \draw[q->] (b) -- (c);
    \draw[q->] (c) -- (a);
  \end{tikzpicture}
\end{equation}
we must have
\begin{equation}
  \label{eq:TC}
  R_{ab} + R_{bc} + R_{ca} = 2.
\end{equation}
The second is that $\U(1)_R$ must be nonanomalous.  This requirement
leads to the condition that at every node, the R-charges of the
bifundamental multiplets starting from or ending on that node add up
to the number of such multiplets minus two.  In our case, for every
\begin{equation}
  \begin{tikzpicture}
    \node[wnode] (a) at (0,0) {$a$};
    \node[wnode] (b) at (1,0) {$b$};
    \node[wnode] (c) at (1,1) {$c$};
    \node[wnode] (d) at (0,1) {$d$};
    \node[wnode] (e) at (-1,0) {$e$};
    \node[wnode] (f) at (-1,-1) {$f$};
    \node[wnode] (g) at (0,-1) {$g$};
    
    \draw[q->] (a) -- (b);
    \draw[q->] (c) -- (a);
    \draw[q->] (a) -- (d);
    \draw[q->] (e) -- (a);
    \draw[q->] (a) -- (f);
    \draw[q->] (g) -- (a);
    \draw[q->] (b) -- (c);
    \draw[q->] (f) -- (e);
    \draw[q->] (d) -- (c);
    \draw[q->] (f) -- (g);
    \draw[q->] (b) -- (g);
    \draw[q->] (d) -- (e);
  \end{tikzpicture}
\end{equation}
we must have
\begin{equation}
\label{eq:U(1)R-AC}
  R_{ab} + R_{ca} + R_{ad} + R_{ea} + R_{af} + R_{ga} = 4.
\end{equation}
The second condition is equivalent to the vanishing of the one-loop
$\beta$-functions for the gauge couplings, and one expects that the
theory flows to a nontrivial infrared fixed point when this condition is
satisfied.

What is the dimension of the space of solutions to the constraint
equations?  The quiver of a brane box model provides a triangulation
of $T^2$.  The number of R-charges to be assigned is equal to the
number of edges $E$, whereas the number of constraints is equal to the
number of faces $F$ plus that of vertices $V$.  Thus, naive counting
would suggest that the dimension is given by minus the Euler
characteristic $\chi = V - E + F$, which vanishes for $T^2$.  This is
not the case, though.  For example, we can assign R-charge $R = 1 - r$
to the horizontal arrows, $R = 1 + r$ to the vertical ones and $R = 0$
to the diagonal ones, with $r$ being a free parameter.

The reason for the mismatch is that the constraints are not all
independent.  Indeed, using the triangle condition~\eqref{eq:TC}, the
anomaly cancellation condition~\eqref{eq:U(1)R-AC} can be replaced~by
\begin{equation}
  R_{ad} - R_{bc} + R_{ga} - R_{fe} = 0.
\end{equation}
The sum of the left-hand side, with $a$ running over all nodes in the
same row of the quiver, is zero due to the periodicity of $T^2$.  So
there is one relation for each row.  Similarly, there is one relation
for each column.  For an $m \times n$ lattice, the dimension of the
space of solutions is therefore at least $m + n$, the total number of
rapidity lines.

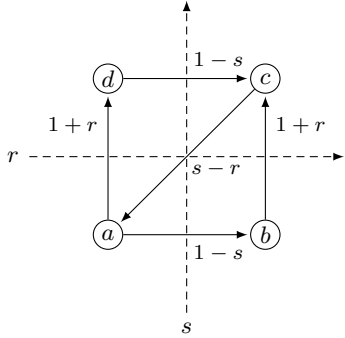
\begin{figure}
  \centering
  \begin{tikzpicture}[scale=1.5*1.15]
    \draw[dr->] (0, 1) node[left] {$r$} -- (2, 1);
    \draw[dr->] (1, 0) node[below] {$s$}-- (1, 2);
    
    \begin{scope}[shift={(0.5, 0.5)}]
      \node[wnode] (a) at (0, 0) {$a$};
      \node[wnode] (b) at (1, 0) {$b$};
      \node[wnode] (c) at (1, 1) {$c$};
      \node[wnode] (d) at (0, 1) {$d$};
    \end{scope}
      
    \draw[q->] (a) -- node[below, near end] {$1 - s$} (b);
    \draw[q->] (d) -- node[above, near end] {$1 - s$} (c);
    \draw[q->] (a) -- node[left, near end] {$1 + r$} (d);
    \draw[q->] (b) -- node[right, near end] {$1 + r$} (c);
    \draw[q->] (c) -- node[below right=-2pt] {$s - r$} (a);
  \end{tikzpicture}
  \caption{The R-charge parametrization for brane box models.}
  \label{fig:RCA}
\end{figure}

These $m + n$ degrees of freedom in the R-charge assignment may be
thought of as degrees of freedom associated to the rapidity lines.
Concretely, we can parametrize the R-charges as follows.  To each
rapidity line we assign a real parameter, and call it the rapidity or
spectral parameter of the rapidity line.  Then, an arrow has $R = 1 -
r$ if it is horizontal and $R = 1 + r$ if vertical, where $r$ is the
spectral parameter of the rapidity line intersecting with that arrow.
The R-charges of the diagonal arrows are determined by the
condition~\eqref{eq:TC}.  The rule is summarized in
figure~\ref{fig:RCA}.  One readily sees that with this rule, the
condition~\eqref{eq:U(1)R-AC} is also satisfied.

In general, a $\U(1)$ R-symmetry can be shifted by $\U(1)$ flavor
symmetries, that is, global $\U(1)$ symmetries that commute with the
supersymmetry algebra.  In view of this fact, the degrees of freedom
in the R-charge assignment just found suggests that there are $m + n$
$\U(1)$ flavor symmetries, each associated to a single rapidity line.

\enlargethispage{6pt}

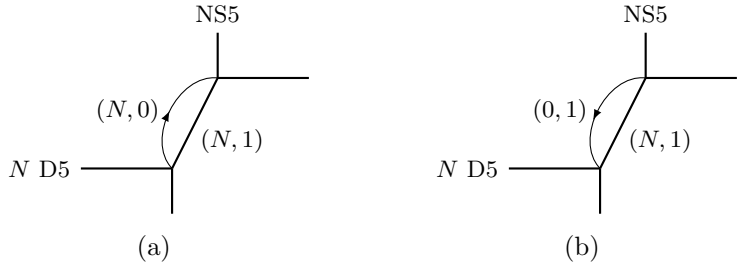
\begin{figure}
  \centering
  \begin{subfigure}{3.41\qsep}
    \centering
    \begin{tikzpicture}[thick]
      \draw (1, 0) -- (1, 0.5);
      \draw (1, 0.5) -- node[below right=-2pt] {$(N,1)$} (1.5, 1.5);
      \draw (1.5, 1.5) -- (1.5, 2) node[above] {NS5};
      
      \begin{scope}[shift={(0, 0.5)}]
        \draw (0, 0) node[left] {$N$ D5} -- (1, 0);
      \end{scope}

      \begin{scope}[shift={(1.5, 1.5)}]
        \draw (0, 0) -- (1, 0);
      \end{scope}

      \draw[->-=0.5, thin] (1, 0.5) to[bend left=70] node[left]
      {$(N, 0)$} (1.5, 1.5);
    \end{tikzpicture}
    \caption{}
    \label{fig:N0=01-a}
  \end{subfigure}%
  \qquad\qquad 
  \begin{subfigure}{3.41\qsep}
    \centering
    \begin{tikzpicture}[thick]
      \draw (1, 0) -- (1, 0.5);
      \draw (1, 0.5) -- node[below right=-2pt] {$(N,1)$} (1.5, 1.5);
      \draw (1.5, 1.5) -- (1.5, 2) node[above] {NS5};
      
      \begin{scope}[shift={(0, 0.5)}]
        \draw (0, 0) node[left] {$N$ D5} -- (1, 0);
      \end{scope}

      \begin{scope}[shift={(1.5, 1.5)}]
        \draw (0, 0) -- (1, 0);
      \end{scope}

      \draw[-<-=0.5, thin] (1, 0.5) to[bend left=70] node[left]
      {$(0, 1)$} (1.5, 1.5);
    \end{tikzpicture}
    \caption{}
    \label{fig:N0=01-b}
  \end{subfigure}
  \caption{(a) An $(N, 0)$-string stretched between the D5-NS5-$(N,
    1)$ junctions.  (b) The same string may be regarded as a $(0,
    1)$-string with the opposite orientation.}
  \label{fig:N0=01}
\end{figure}

The origin of these global symmetries can be most clearly understood
in the original brane picture.  Let us pick an NS5-brane, say the one
corresponding to the $\alpha$th rapidity line, and imagine, as before,
slightly sliding the D5-brane segments that end on it from left and
right.  This operation splits the NS5-brane into the upper and lower
semi-infinite NS5-branes, joined by a finite segment of $(N,\pm 1)$
5-brane; see figure~\ref{fig:5BW}.  For definiteness, consider the
case with an $(N, 1)$ 5-brane.  Bifundamental multiplets arise from
open strings whose ends are localized at the junctions where
semi-infinite NS5-branes meet D5-branes and the $(N, 1)$ 5-brane.
These strings may be viewed as fundamental strings ($(N, 0)$-strings),
as in figure~\ref{fig:N0=01-a}, or D-strings ($(0, 1)$-strings) with
the opposite orientation,%
\footnote{To determine the orientation, we draw an $(N, 0)$-string
  attached on the D5-branes, and move the endpoints to the
  semi-infinite NS5-branes.  When an endpoint passes a junction, an
  $(N, 1)$-string is created due to charge conservation.  When the
  other endpoint passes the other junction, the $(N, 1)$-string is
  annihilated, leaving a $(0, 1)$-string.  The orientation of the $(0,
  1)$-string can be seen from the final picture.}
as in figure~\ref{fig:N0=01-b}.  As such, they are acted upon by
global $\U(1)$ gauge transformations on the semi-infinite NS5-branes.
In the effective 4d theory that emerges when the split branes are
rejoined, gauge transformations on the upper and lower branes appear
as elements in the same symmetry $\U(1)_\alpha$, just
acting in opposite manners.
%
%
Hence, a $\U(1)$ flavor symmetry is associated to each NS5-brane or
rapidity line.

Equivalently, we may regard $\U(1)_\alpha$ as arising from the $\U(1)$
gauge symmetries on the D5-brane segments, which we have seen are
nondynamical.  From this point of view, $\U(1)_\alpha$ acts on all
D5-brane segments on the ``right'' side of the $\alpha$th N5-brane, or
on all of those on the ``left'' side in the opposite manner.  Of
course, the words ``left'' and ``right'' are not really meaningful
when the $4$- and $6$-direction are periodic, but we can make sense of
this interpretation by, roughly speaking, considering the D5-branes to
be sections of a $\U(1)_\alpha$-bundle on the $46$-torus.  

This point is probably easier to understand if we consider the
situation where there are only NS5-branes in the $012345$ directions,
and look at the mass parameter $m_\alpha$ associated to $\U(1)_\alpha$
in the corresponding 5d $\CN = 1$ theory.  $m_\alpha$ is the value of
the real scalar in the background vector multiplet for $\U(1)_\alpha$,
and proportional to the difference between the $x^5$-coordinates of
the two D5-brane segments that end on the $\alpha$th NS5-brane.  The
periodicity in the 6-direction suggests that the sum of $m_\alpha$
over the rapidity lines must vanish.  However, this constraint can be
modified if we compactify the $6$-direction using a twisted boundary
condition that involves a shift in the $5$-direction.  (The same
twisted boundary condition is used in the brane construction of 4d
$\CN=2$ elliptic models~\cite{Witten:1997sc}.)  This modification
introduces one more parameter, namely the amount of shift, leading to
the same number of mass parameters as the rapidity lines.

We normalize the $\U(1)_\alpha$-charge $F_\alpha$ in such a way that
horizontal and vertical bifundamental arrows intersecting with the
$\alpha$th rapidity line have $F_\alpha = -1$ and $+1$, respectively,
while $F_\beta = 0$ for $\beta \neq \alpha$.  The charges of the
diagonal arrows are found by demanding that the superpotential have
$F_\alpha = 0$ for all $\alpha$.  Using $F_\alpha$, our R-charge
parametrization can be concisely written as
\begin{equation}
  \label{eq:R-shift}
  R = R_0 + \sum_\alpha r_\alpha F_\alpha,
\end{equation}
where $R_0$ is the R-charge defined by assigning $R_0 = 1$ to the
horizontal and vertical arrows, and $r_\alpha$ is the parameter of the
$\alpha$th rapidity line.  As expected, the degrees of freedom in the
definition of $R$ come from shifts by flavor symmetries.

\begin{figure}
\centering
\begin{tikzpicture}[scale=1.5]
  \draw (0,0) rectangle (2, 2);

  \begin{scope}[shift={(0.15, 0.15)}]
    \node[wnode] (a) at (0.5, 0.5) {$a$};
    \node[wnode] (b) at (1.5, 0.5) {$b$};
    \node[wnode] (c) at (1.5, 1.5) {$c$};
    \node[wnode] (d) at (0.5, 1.5) {$d$};
  \end{scope}

  \draw[q->] (a) --node[below] {$1 - u$} (b);
  \draw[q<-] (c) -- node[sloped, anchor=center, below] {$1 + s'$} (b);
  \draw[q->] (d) -- node[below] {$1 - u'$} (c);
  \draw[q<-] (d) -- node[sloped, anchor=center, below] {$1 + s$} (a);
  \draw[q->] (c) -- (a);
  
  \draw (a) -- (0, 0);
  \draw (b) -- (1, 0);
  \draw (d) -- (0,1);
  
  \draw[q->] (1, 2) -- (d);
  \draw[q->] (2, 2) -- (c);
  \draw[q->] (2, 1) -- (b);

  \begin{scope}[shift={(0, 0.15)}]
    \draw[q->] (0, 0.5) -- node[below] {$1 - t$} (a);
    \draw[q->] (0, 1.5) -- node[below] {$1 - t'$} (d);
    \draw (b) -- (2, 0.5);
    \draw (c) -- (2, 1.5);
  \end{scope}

  \begin{scope}[shift={(0.15, 0)}]
    \draw[q->] (0.5, 0) -- node[sloped, anchor=center, below] {$1 + r$} (a);
    \draw (d) -- (0.5, 2);
    \draw[q->] (1.5, 0) -- node[sloped, anchor=center, below] {$1 + r'$} (b);
    \draw (c) -- (1.5, 2);
  \end{scope}
\end{tikzpicture}
\caption{The general R-charge assignment for the $2 \times 2$ brane
  box model.  Here only the R-charges of the horizontal and vertical
  arrows are indicated, and the variables satisfy the relation $r' -
  r = s - s' = t - t' = u' - u$.}
\label{fig:RCA-2x2}
\end{figure}
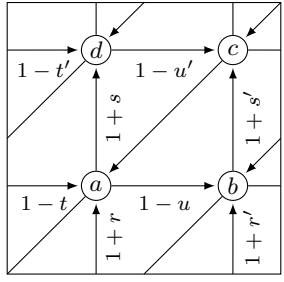

We have seen above that there are at least $m + n$ $\U(1)$ flavor
symmetries, but have not determined the precise number.  Actually, in
some cases there are extra $\U(1)$ symmetries in addition to the ones
discussed above.  For example, for a $2 \times 2$ lattice, the most
general R-charge assignment has five parameters rather than four, as
shown in figure~\ref{fig:RCA-2x2}.  In fact, from the relation between
the brane box and brane tiling constructions explained in
section~\ref{sec:BT}, one can deduce that the number of $\U(1)$ flavor
symmetries is equal to $m + n + \gcd(m,n) -1$.  We will not consider
these extra flavor symmetries.  Alternatively, one may assume that $m$
and $n$ are coprime.

\subsection{Structure of a TQFT with extra dimension}

So far we have described the brane box configuration \eqref{eq:BB}
from the point of view of the effective 4d $\CN = 1$ theory.  To make
contact with the framework discussed in section~\ref{sec:ILM-TQFT},
let us instead think of it as realizing 6d maximally supersymmetric
Yang-Mills theory on the D5-branes, in the presence of
codimension-$1$ defects created by the NS5-branes.  We wish to turn
this 6d theory into a theory that is topological on the $46$-torus,
and regard the defects as line operators with spectral parameter in
the TQFT.  This is achieved by applying an appropriate topological
twist and restricting physical quantities to compute, as we now
explain.

Type IIB superstring theory has $\CN = (2,0)$ supersymmetry in $9+1$
dimensions, generated by Majorana-Weyl spinors $\eps_L$, $\eps_R$
such that
\begin{equation}
  \Gamma_{0123456789} \eps_L = \eps_L, \quad  
  \Gamma_{0123456789} \eps_R = \eps_R.
\end{equation}
Here $\Gamma_M$, $M = 1$, $\dotsc$, $9$ are gamma matrices, and
$\Gamma_{0123456789} = \Gamma_0 \dotsm \Gamma_9$ is the chirality
operator.
The D5-branes preserve supersymmetry generated by $\eps_L$, $\eps_R$
satisfying
\begin{equation}
  \label{D5}
  \eps_L = \Gamma_{012346} \eps_R.
\end{equation}
Likewise, the NS5-branes impose conditions
\begin{equation}
  \eps_L = \Gamma_{012345} \eps_L, \quad  
  \eps_R = -\Gamma_{012345} \eps_R
\end{equation}
and
\begin{equation}
  \eps_L = \Gamma_{012367} \eps_L, \quad  
  \eps_R = -\Gamma_{012367} \eps_R
\end{equation}
on the preserved supersymmetry.  The conditions on $\eps_R$ coming
from the NS5-branes actually follow from those on $\eps_L$ and the
D5-brane condition, hence are redundant.  Since $\Gamma_{012345}$ and
$\Gamma_{012367}$ square to the identity and are traceless, half of
their eigenvalues are $+1$ and the other half are $-1$.  Moreover,
they commute with each other and are simultaneously diagonalizable.
Each NS5-brane condition therefore halves the number of independent
solutions for $\eps_L$.  Given $\eps_L$, the D5-brane condition
determines $\eps_R$.  In total, the 5-branes preserve four
supercharges, which generate the $\CN = 1$ supersymmetry of the 4d
theory we have been studying.

From the NS5-brane conditions, we see $\Gamma_{45} \eps_L =
\Gamma_{67} \eps_L$, or
\begin{equation}
  (\Gamma_{46} + \Gamma_{57}) \eps_L = 0.
\end{equation}
This equations says that if we replace the rotation group
$\SO(2)_{46}$ of the $46$-plane (which we decompactify for a moment)
with the diagonal subgroup $\SO(2)_{46}'$ of $\SO(2)_{46} \times
\SO(2)_{57}$, then the preserved supercharges become scalars under the
new rotation group.  From the same equation we also find that the
NS5-brane conditions on $\eps_L$ can be generalized to
\begin{equation}
  \eps_L
  = \Gamma_{0123} (\Gamma_4 \cos\theta + \Gamma_6 \sin\theta) 
  (\Gamma_5 \cos\theta + \Gamma_7 \sin\theta) \eps_L,
\end{equation}
where $\theta$ is any angle.  This shows that the NS5-branes can be
rotated in the $46$- and $57$-plane by the same angle $\theta$ without
destroying the supercharges, or for that matter, they can intersect
with the $46$-plane along arbitrary closed curves as long as the
slopes are correlated between the $46$- and $57$-plane.

From the point of view of the 6d theory on the D5-branes, the
replacement of $\SO(2)_{46}$ with $\SO(2)_{46}'$ is a topological
twist along the $46$-plane.  After the twisting, eight out of the
sixteen supercharges are scalars on the $46$-plane, and as such, can
be preserved even when the $46$-plane is replaced by a general
$2$-manifold $\Sigma$.  Codimension-$1$ defects can be inserted along
arbitrary closed curves on $\Sigma$ while preserving four of the eight
supercharges.  These defects carry a spectral parameter, as we have
seen from the 4d point of view.

Now, to produce a desired TQFT from this twisted theory, we simply
restrict ourselves to quantities that are independent of the metric on
$\Sigma$, and depend on the codimension-$1$ defects inserted on
$\Sigma$ only through their topology and spectral parameters.  Then,
the twisted theory becomes a TQFT on $\Sigma$, and the defects become
line operators with spectral parameter in this TQFT.

A good example of such a topological quantity is the supersymmetric
index of the corresponding 4d $\CN = 1$
theory~\cite{Romelsberger:2005eg, Kinney:2005ej, Festuccia:2011ws}.
As the theory has the R-symmetry $\U(1)_R$, it can be placed on the
Euclidean spacetime $S^3 \times S^1$ without breaking
supersymmetry~\cite{Sen:1985ph, Romelsberger:2005eg,
  Festuccia:2011ws}.  Under (the double cover of) the isometry group
$\SU(2) \times \SU(2)$ of $S^3$, the four supercharges transform as a
pair of doublets $(\mathbf{2}, \mathbf{1})$.  Let $J_i$, $J'_i$ be the
generators of the $\SU(2)$ factors.  The supersymmetric index is
defined by
\begin{equation}
  \label{eq:I-4d}
  \CI(p, q, \{u_\alpha\}; \{r_\alpha\})
  = \Tr_{S^3} \biggl((-1)^F p^{J_3 + J'_3 + R/2} q^{J_3 - J'_3 + R/2}
  \prod_\alpha u_\alpha^{F_\alpha}\biggr),
\end{equation}
where $p$, $q$, $u_\alpha$ are complex parameters, and the trace is
taken over the Hilbert space on $S^3$.  This quantity is topological
for the following reason.  Bosonic and fermionic states with energy $E
\neq 2J_3 + R$ are paired by the action of a supercharge.  (Here we
are taking the radius of $S^3$ to be $1$.)  Due to the presence of the
factor $(-1)^F$, pairwise cancellations occur and the only
contributions to the trace come from states with $E = 2J_3 + R$.
Under continuous changes of the parameters of the theory (other than
the rapidities $r_\alpha$ on which $R$ implicitly depends), the index
remains invariant since states can be brought into or out of the right
energy level only in boson-fermion pairs.  The index is therefore
protected against such changes, in particular against deformations of
geometric parameters such as the shapes of line operators and the
metric on $\Sigma$.

There are other topological quantities.  For example, we can replace
the $S^3$ in the above definition of the supersymmeric index with $S^2
\times S^1$ or a lens space $L(p, q)$~\cite{Closset:2013vra}.
Unfortunately, though, an explicit expression for the index is not
known except for simple cases such as $L(p, 1)$~\cite{Benini:2011nc}.
We will focus on the $S^3$ index, for which the relevant mathematical
results are available.

So we have obtained from brane box models a 2d TQFT that has a family
of line operators with spectral parameter.  We can now apply the
construction discussed in section~\ref{sec:ILM-TQFT} and get a vertex
model, by placing the TQFT on $T^2$ and making a lattice of line
operators.  However, we cannot tell whether the model is integrable
just from the structure of a TQFT.  For the integrability, it is
necessary to show, in addition, that no phase transitions occur when
two line operator are untangled or a line operator passes through the
intersection of two other line operators.  In Costello's argument,
these properties are guaranteed by the existence of extra dimensions
in which the line operators can miss one another.

We can make a similar argument in the present case.  Let us go back to
the brane picture.  To the configuration \eqref{eq:BB}, we apply
T-duality in the $3$-direction (which we take to be the
$S^1$-direction of the spacetime $S^3 \times S^1$) to convert the
D5-branes to D4-branes, and then lift the resulting Type IIA brane
system to M-theory:
\begin{equation}
  \label{eq:BB-M5}
\begin{tabular}{|l|ccccccccccc|}
  \hline
  & 0 & 1 & 2 & 3 & 4 & 5 & 6 & 7 & 8 & 9 & 10
  \\ \hline
  M5 & $\times$ & $\times$ & $\times$ & 
        & $\times$ & & $\times$ & & & & $\times$
  \\
  M5 & $\times$ & $\times$ & $\times$ & $\times$ & $\times$ & $\times$ &
  &&&& 
 \\
  M5 & $\times$ & $\times$ & $\times$ & $\times$ & & & $\times$ &
  $\times$ &
  &&
  \\ \hline
\end{tabular}
\end{equation}
These dualities transform the 6d theory on the D5-branes into an $\CN
= (2,0)$ theory on a stack of M5-branes, and the codimension-$1$
defects into codimension-$2$ defects.  There is now an extra dimension
along which line operators can move, namely the M-theory circle.  Its
presence ensures the integrability of the vertex model.

The similarity to Costello's argument suggests that the
$x^{10}$-coordinates of line operators are related to the spectral
parameters $r_\alpha$.  This is indeed true, as the following
analysis~shows.

If we trade off the flavor fugacities $u_\alpha$ for new parameters
$s_\alpha$ given by $u_\alpha = (pq)^{s_\alpha/2}$, the
supersymmetric index~\eqref{eq:I-4d} depends on $s_\alpha$ only
through the combination $R + \sum_\alpha s_\alpha F_\alpha$.  Hence,
the effect of turning on nonzero values for $s_\alpha$ is equivalent
to shifting $r_\alpha \to r_\alpha + s_\alpha$, and $s_\alpha$
represent the same degrees of freedom as $r_\alpha$.  We want to
understand how $s_\alpha$ are described in the language of branes, and
keep track of them under the dualities.

As we saw before by considering the deformation depicted in
figure~\ref{fig:N0=01}, $\U(1)_\alpha$ can be interpreted as either
the global gauge symmetry on the lower half of the $\alpha$th
NS5-brane, or that on the upper half acting in the opposite manner.
The conserved current $J_\alpha$ for $\U(1)_\alpha$ is therefore
supported along $C_\alpha$ on the $46$-torus, and couples to the 6d
theory through a term of the form
\begin{equation}
  i\int_{S^3 \times S^1 \times T^2}
  \bigl(A_{\alpha}^- -A_{\alpha}^+\bigr) \wedge \star J_\alpha.
\end{equation}
Here $A_{\alpha}^-$ and $A_{\alpha}^+$ are the gauge fields on the
lower and upper halves of the NS5-brane, restricted to the brane
intersection; they are related there to the $\U(1)$ parts of the gauge
fields on the D5-brane segments by some boundary condition.  We give a
background value to the gauge fields such that $A_\alpha^-
-A_\alpha^+$ is constant along $S^3$ and $C_\alpha$.  As $F_\alpha$ is
the integral of $\star J_\alpha$ over $S^3$ and $T^2$, we find that
the above term becomes
\begin{equation}
  i\int_{S^1} \bigl(A_\alpha^- -A_\alpha^+\bigr) F_\alpha.
\end{equation}
This term enters the path integral as an exponentiated factor, so
$s_\alpha$ is proportional to the difference between the holonomies of
$A_\alpha^-$ and $A_\alpha^+$ around the $3$-circle.

These holonomies are mapped under the T-duality to the background
values of a scalar field on the upper and lower halves of the
corresponding Type IIA NS5-brane.  Upon lifting to M-theory, the
NS5-brane becomes an M5-brane, and the scalar is mapped to the
$x^{10}$-coordinate.  Therefore, the spectral parameter for a line
operator is given by the difference between the $x^{10}$-coordinates
of the upper and lower halves of the M5-brane.

We remark that if the holonomies of $A_\alpha^+$ and $A_\alpha^-$ are
equal, then the whole $\alpha$th NS5-brane is mapped to a single point
on the M-theory circle, and the supersymmetric index does not depend
on its position at all.  This observation is consistent with the fact
that the $\CN = (2, 0)$ theory compactified on $S^3$ has a topological
sector that is equivalent to Chern-Simons theory with complex gauge
group at level $k = 1$~\cite{Cordova:2013cea}.  In the present
setting, the NS5-brane creates a Wilson line in the Chern-Simons
theory on $T^2 \times S^1$, and its precise position on the
$3$-manifold is irrelevant.  When the holonomies are different, the
Wilson line splits into two line operators, and apparently they no
longer belong to the topological sector.  It is clear from our
viewpoint that the same consideration should apply to other indices
obtained by replacing $S^3$ with appropriate $3$-manifolds.  Indeed,
the $\CN = (2,0)$ theory on $S^2 \times S^1$ has a topological sector
equivalent to complex Chern-Simons theory at level $k =
0$~\cite{Yagi:2013fda, Lee:2013ida}, while the lens space $L(k, 1)$
gives complex Chern-Simons theory at level
$k$~\cite{Dimofte:2014zga}.  Similarly, a twisted product of $\R^2$
and $S^1$ leads to analytically continued Chern-Simons theory, which
is the holomorphic sector of complex Chern-Simons
theory~\cite{Witten:2010zr, Witten:2011zz, Beem:2012mb, Luo:2014sva}.

\subsection{Integrable lattice model for brane box models}

Finally, let us identify the integrable lattice model associated with
brane box models.  To do so, we look at the integral formula for the
supersymmetric index.

Recall that the flavor fugacities $u_\alpha$ represent the same
degrees of freedom as the spectral parameters $r_\alpha$.  Since we
are going to keep the dependence on $r_\alpha$, we set $u_\alpha = 1$.
With this understood, the supersymmetric index of a brane box model is
computed as follows~\cite{Romelsberger:2005eg, Dolan:2008qi}.  First,
we assign the factor
\begin{equation}
  \CI_{\text{V}}(z_a)
  = \bigl((p;p)_\infty (q;q)_\infty\bigr)^{N-1}
      \prod_{i \neq j} \frac{1}{\Gamma(z_{a,i}/z_{a,j}; p, q)}
\end{equation}
to the vector multiplet represented by node $a$, and
\begin{equation}
  \CI_{\text{B}}(z_a, z_b; R_{ab})
  = \prod_{i, j} \Gamma\bigl((pq)^{R_{ab}/2} z_{b,j}/z_{a,i}; p, q\bigr)
\end{equation}
to the bifundamental chiral multiplet with $R = R_{ab}$ represented by
an arrow from node~$a$ to~$b$.  Here
\begin{equation}
  \Gamma(z; p, q)
  =
  \prod_{i, j = 0}^\infty
  \frac{1 - z^{-1} p^{i+1} q^{j+1}}{1 - z p^i q^j}
\end{equation}
is the elliptic gamma function, $(z; q)_\infty = \prod_{i=0}^\infty (1
- zq^i)$, and $z_a$ collectively denotes the fugacities $z_{a,i}$, $i
= 1$, $\dotsc$, $N$ for the gauge group $\SU(N)_a$, obeying
$\prod_{i=1}^N z_{a,i} = 1$.  Then, we multiply the factors from all
nodes and arrows, and integrate over the fugacities:
\begin{equation}
\label{eq:I-int-4d}
  \CI
  = \oint
     \prod_{\circled{a}} \frac{1}{N!}
     \prod_{j=1}^{N-1} \frac{\rmd z_{a, j}}{2\pi i z_{a,j}}
     \CI_{\text{V}}(z_a)
     \prod_{\circled{a} \qto \circled{b}} \CI_{\text{B}}(z_a, z_b; R_{ab}).
\end{equation}
The integration for each fugacity is performed over the unit circle
$\{|z_{a,i}| = 1\}$.  

The above procedure clearly shows the statistical mechanical nature of
the supersymmetric index.  In the statistical mechanical
interpretation, $z_{a,i}$ are continuous state variables, and the
factors $\CI_{\text{V}}$ and $\CI_{\text{B}}$ are the Boltzmann
weights for self- and nearest-neighbor interactions, respectively.
The partition function is defined by multiplying the weights and
summing over all possible state configurations, and this is identified
with the supersymmetric index.

Since the state variables $z_{a,i}$ are assigned to the nodes of the
quiver which correspond to the faces of the lattice on the $46$-torus,
it is natural to formulate this model as an IRF model.  We can take
its Boltzmann weight to be%
\footnote{This is not the only choice.  For example, one may replace
  the right-hand side with the more symmetric expression
  \begin{multline}
    \sqrt{\CI_{\text{V}}(z_a) \CI_{\text{B}}(z_a,z_b; 1-s)
      \CI_{\text{B}}(z_b,z_c; 1+r)\CI_{\text{B}}(z_c,z_a; s-r)}
    \\ \times
    \sqrt{\CI_{\text{V}}(z_c) \CI_{\text{B}}(z_c,z_a; s-r)
      \CI_{\text{B}}(z_a,z_d; 1+r)\CI_{\text{B}}(z_d,z_c; 1-s)}
  \end{multline}
  without spoiling the integrability.}
\begin{equation}
  \BW{z_a}{z_b}{z_c}{z_d}{r}{s}
  = \CI_{\text{V}}(z_a) \CI_{\text{B}}(z_a, z_b; 1 - s)
     \CI_{\text{B}}(z_b, z_c; 1 + r)
     \CI_{\text{B}}(z_c, z_a; s - r).
\end{equation}
Graphically, it can be represented as
\begin{equation}
  \label{eq:W-4d}
  \begin{tikzpicture}
    \draw[dr->] (0, 1) node[left] {$r$} -- (2, 1);
    \draw[dr->] (1, 0) node[below] {$s$}-- (1, 2);

    \begin{scope}[shift={(0.5, 0.5)}]
      \node[tnode] (a) at (0,0) {$a$};
      \node[tnode] (b) at (1,0) {$b$};
      \node[tnode] (c) at (1,1) {$c$};
      \node[tnode] (d) at (0,1) {$d$};
    \end{scope}

    \draw[thick] (a) -- (b) -- (c) -- (d) -- (a);
  \end{tikzpicture}
  \quad = \quad
  \begin{tikzpicture}
    \node[dnode] (a) at (0,0) {$a$};
    \node[snode] (b) at (1,0) {$b$};
    \node[snode] (c) at (1,1) {$c$};

    \draw[q->] (a) -- node[below] {$1 - s$} (b);
    \draw[q->] (b) -- node[right] {$1 + r$} (c);
    \draw[q->] (c) -- node[above left=-2pt] {$s - r$} (a);
  \end{tikzpicture} \quad .
\end{equation}
(Notice a slight discrepancy in the notation compared to the original
definition \eqref{eq:W}.  Here $a$, $b$, etc.\ on the left-hand side
really stands for $z_a$, $z_b$, etc.)  A square node
\tikz[anchor=base, baseline]{\node[snode] {$a$};} indicates that the
associated factor $\CI_{\text{V}}(z_a)$ is absent, and a node
\tikz[anchor=base, baseline]{\node[dnode] {$a$};} drawn with a dashed
line means that the factor is present but integration is not performed
over $z_a$.  The intersections of NS5-branes with the D5-branes, which
create line operators in the TQFT on the $46$-torus, play the roles of
rapidity lines, as anticipated.

Let us show that this Boltzmann weight satisfies the Yang-Baxter
equation~\eqref{eq:YBE-IRF}.  For this purpose we will need the
identity
\begin{equation}
  \label{eq:IBIB=1}
  \CI_{\text{B}}(z_a, z_b; R_{ab}) \CI_{\text{B}}(z_b, z_a; R_{ba}) = 1
\end{equation}
which holds for $R_{ab} + R_{ba} = 2$.  In effect, it means that we
can erase two arrows going in opposite directions between the same
pair of nodes:
\begin{equation}
  \label{eq:a<->b}
  \begin{tikzpicture}
    \node[snode] (a) at (0,0) {$a$};
    \node[snode] (b) at (1,0) {$b$};
    \draw[q->] (a) to[bend left] (b);
    \draw[q->] (b) to[bend left] (a);
  \end{tikzpicture}
  \quad = \quad
  \begin{tikzpicture}
    \node[snode] (a) at (0,0) {$a$};
    \node[snode] (b) at (1,0) {$b$};
  \end{tikzpicture}
  \quad .
\end{equation}
This property reflects the fact that if one has chiral multipets
$\Phi_1$, $\Phi_2$ in conjugate representations whose R-charges add up
to $2$, one can give them masses by turning on the superpotential term
$m\Phi_1 \Phi_2$.  The supersymmetric index is independent of $m$, and
in the limit $m \to \infty$, these multiplets become infinitely
massive and decouple, leaving no contribution to the index.

Now, plugging the expression~\eqref{eq:W-4d} into the Yang-Baxter
equation~\eqref{eq:YBE-IRF}, we find
\begin{equation}
  \begin{tikzpicture}[scale=1.5]
    \node[dnode] (a) at (240:1){$a$};
    \node[snode] (b) at (300:1) {$b$};
    \node[snode] (c) at (0:1) {$c$};
    \node[snode] (d) at (60:1) {$d$};
    \node[dnode] (f) at (180:1) {$f$};
    \node[wnode] (g) at (0,0) {$g$};

    \draw[q->] (a) -- node[below] {$1 - t$} (b);
    \draw[q->] (b) -- node[right] {$1 + r$} (c);
    \draw[q->] (a) -- node[left] {$1 + r$} (g);
    \draw[q->] (g) -- node[below, near start] {$1 - t$} (c);
    \draw[q->] (c) -- node[right=4pt, near end] {$t - r$} (a);
    
    \draw[q->] (c) -- node[right, near end] {$1 + s$} (d);
    \draw[q->] (d) -- node[right, near end] {$t - s$} (g);
    
    \draw[q->] (f) -- node[left, near end] {$1 - s$} (a);
    \draw[q->] (g) -- node[below] {$s - r$} (f);
  \end{tikzpicture}
  \quad = \quad
  \begin{tikzpicture}[scale=1.5]
    \node[dnode] (a) at (240:1){$a$};
    \node[snode] (b) at (300:1) {$b$};
    \node[snode] (c) at (0:1) {$c$};
    \node[snode] (d) at (60:1) {$d$};
    \node[dnode] (f) at (180:1) {$f$};
    \node[wnode] (g) at (0,0) {$g$};
    
    \draw[q->] (a) -- node[below] {$1 - t$} (b);
    \draw[q->] (b) to[bend  left] node[left, near start] {$1 + s$}(g);
    \draw[q->] (g) -- node[left, near start] {$t - s$} (a);
    
    \draw[q->] (g) to[bend left] node[right, near start] {$1 - s$} (b);
    \draw[q->] (b) -- node[right] {$1 + r$} (c);
    \draw[q->] (c) -- node[above] {$s - r$} (g);
    
    \draw[q->] (f) -- node[above, near end] {$1 - t$} (g);
    \draw[q->] (g) -- node[right] {$1 + r$} (d);
    \draw[q->] (d) -- node[above=4pt] {$t - r$} (f);
  \end{tikzpicture}
  \quad .
\end{equation}
Some nodes and arrows cancel out between the two sides, while the
arrows between nodes $b$ and $g$ on the right-hand side cancel thanks
to the identity \eqref{eq:a<->b}.  Moving the arrows that are not
connected to node $g$ from the left- to right-hand side using the same
identity, we end up with the equality
\begin{equation}
  \label{eq:Seiberg}
  \begin{tikzpicture}[scale=1.5]
    \node[snode] (a) at (240:1){$a$};
    \node[snode] (c) at (0:1) {$c$};
    \node[snode] (d) at (60:1) {$d$};
    \node[snode] (f) at (180:1) {$f$};
    \node[wnode] (g) at (0,0) {$g$};

    \draw[q->] (a) -- node[left] {$1 + r$} (g);
    \draw[q->] (g) -- node[below, near start] {$1 - t$} (c);
    \draw[q->] (d) -- node[right, near end] {$t - s$} (g);    
    \draw[q->] (g) -- node[below] {$s - r$} (f);
  \end{tikzpicture}
  \quad = \quad
  \begin{tikzpicture}[scale=1.5]
    \node[snode] (a) at (240:1){$a$};
    \node[snode] (c) at (0:1) {$c$};
    \node[snode] (d) at (60:1) {$d$};
    \node[snode] (f) at (180:1) {$f$};
    \node[wnode] (g) at (0,0) {$g$};
    
    \draw[q->] (g) -- node[left, near start] {$t - s$} (a);
    \draw[q->] (c) -- node[above] {$s - r$} (g);
    \draw[q->] (f) -- node[above, near end] {$1 - t$} (g);
    \draw[q->] (g) -- node[right] {$1 + r$} (d);
    \draw[q->] (d) -- node[above=4pt] {$t - r$} (f);

    \draw[q<-] (c) -- node[right=4pt, near end] {$2 - t + r$} (a);
    \draw[q<-] (c) -- node[right, near end] {$1 - s$} (d);
    \draw[q<-] (f) -- node[left, near end] {$1 + s$} (a);
  \end{tikzpicture}
  \quad .
\end{equation}
Interestingly, this equality expresses precisely the invariance of the
supersymmetric index under Seiberg duality.

Seiberg duality says that for $N_f > N_c + 1$, an $\CN = 1$ $\SU(N_c)$
gauge theory with $N_f$ flavors $(Q, \Qt)$ and no superpotential has a
dual description in the infrared as an $\CN = 1$ $\SU(N_f - N_c)$
gauge theory with $N_f$ flavors $(q, \qt)$ and $N_f^2$ singlets $M$,
coupled through a superpotential $W = \qt M q$.  For both theories,
the flavor symmetry is $\SU(N_f)_L \times \SU(N_f)_R \times \U(1)_B$
and the R-symmetry is $\U(1)_R$.  Under these symmetries and the gauge
group $G = \SU(N_c)$ or $\SU(N_f - N_c)$, the chiral multiplets
transform as follows:
\begin{equation}
\label{eq:SD-4d}
\begin{tabular}{|l|ccccc|}
\hline
        & $G$ & $\SU(N_f)_L$ & $\SU(N_f)_R$
        & $\U(1)_B$ & $\U(1)_R$
\\ \hline
$Q$    & $\square$ & $\overline\square$ & $\mathbf{1}$
           & $1$ & $1 - N_c/N_f$
\\
$\Qt$ & $\overline\square$ & $\mathbf{1}$ & $\square$
          & $-1$ & $1 - N_c/N_f$
\\ \hdashline
$q$    & $\square$ & $\overline\square$ & $\mathbf{1}$
          & $-N_c/(N_f - N_c)$ & $N_c/N_f$
\\
$\qt$ & $\overline\square$  & $\mathbf{1}$ & $\square$
          & $N_c/(N_f - N_c)$ & $N_c/N_f$
\\
$M$ & $\mathbf{1}$ & $\square$ & $\overline\square$
        & $0$ & $2(1 - N_c/N_f)$
\\ \hline
\end{tabular} \quad .
\end{equation}
In the case at hand, $N_f = 2N_c = 2N$ and the charged chiral
multiplets all have $R = 1/2$.  This corresponds to the situation with
$(r, s, t) = (-1/2, 0, 1/2)$.  As we stressed, however, we can shift
the R-charge by $\U(1)$ flavor charges.  The quivers that appear in
the equality~\eqref{eq:Seiberg} split the $2N$ flavors into a pair of
$N$ flavors.  This breaks each of $\SU(2N)_L$ and $\SU(2N)_R$ down to
an $\SU(N) \times \SU(N) \times \U(1)$ subgroup.  So in total there
are three $\U(1)$ flavor symmetries that can be used to shift the
R-charge, and they give rise to the three parameters $r$, $s$ and~$t$.

Since the supersymmetric index is protected from continuous changes of
the parameters of the theory, it is invariant under renormalization
group flow and therefore under Seiberg duality.  Indeed, as
observed in~\cite{Dolan:2008qi}, the invariance under Seiberg duality
is equivalent to an identity for the elliptic gamma function proved
in~\cite{MR2630038}.%
\footnote{Historically, the identity was first discovered for $N_c =
  2$ and $N_f = 3$ in \cite{MR1846786}.  Subsequently it was proved
  for $N_c = 2$ and $N_f = 4$ in~\cite{MR2044635} and extended to the
  general case in \cite{MR2630038}.}
Thus, we have demonstrated the integrability of our model.%
\footnote{The invertibility of the R-matrix follows from the identity
\begin{equation}
  \frac{1}{N!} \oint 
  \prod_{j=1}^{N-1} \frac{\rmd z_{c, j}}{2\pi i z_{c,j}}
  \CI_{\text{B}}(z_a, z_c; R) \CI_{\text{V}}(z_c) \CI_{\text{B}}(z_c, z_b; -R) \CI_{\text{V}}(z_b)
  = K \sum_{\sigma \in S_N} \delta(z_a, \sigma(z_b))
\end{equation}
proved in \cite{MR2264067}, where $S_N$ is the symmetric group of
degree $N$ and the delta function $\delta$ is defined with respect to
the measure $\oint \prod_{j=1}^{N-1} \rmd z_{a, j}/2\pi i z_{a, j}$.
The overall factor $K = \prod_{s = \pm1} \Gamma((pq)^{sNR/2}; p, q))$
can be removed by rescaling the R-matrix.}

The appearance of Seiberg duality as an integrability condition might
be unexpected from the point of view of lattice models.  From our
perspective, this is not entirely surprising.  The integrability is
simply a consequence of the topological invariance of the relevant
TQFT with extra dimension that has line operators, and these line
operators are created by NS5-branes.  It is well known that in the
brane construction of 4d $\CN = 1$ gauge theories, Seiberg duality is
realized by rearrangement of NS5-branes~\cite{Elitzur:1997fh,
  Elitzur:1997hc}.

\section{Brane tiling models}
\label{sec:BT}

In the discussion of brane box models in the previous section, we
found it helpful to consider a deformation of the brane configuration
that splits the D5-branes across NS5-branes and creates $(N, \pm 1)$
5-branes, as in figure~\ref{fig:5BW}.  The bifundmantal multiplets
represented by the horizontal and vertical arrows in the quiver
originate from open strings whose ends are attached on the brane
junctions.  The original configuration is recovered in the limit where
the D5-brane segments are joined together and the $(N, \pm 1)$
5-branes disappear, and these multiplets become massless in this
limit.

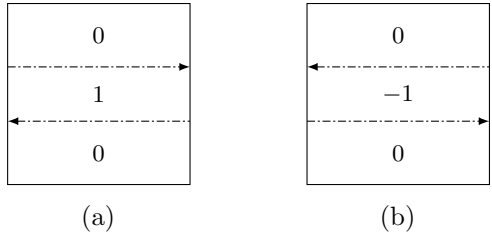
\begin{figure}
  \centering
  \begin{subfigure}[b]{2\qsep}
    \centering
    \begin{tikzpicture}
      \draw (0,0) rectangle (2, 2);
      \draw[ddr<-] (0, 0.7) -- (2, 0.7);
      \draw[ddr->] (0, 1.3) -- (2, 1.3);
      \node at (1, 0.35) {$0$};
      \node at (1, 1.65) {$0$};
      \node at (1, 1) {$1$};
    \end{tikzpicture}
    \caption{}
    \label{fig:5BW-BT-a}
  \end{subfigure}%
  \qquad\qquad 
  \begin{subfigure}[b]{2\qsep}
    \centering 
    \begin{tikzpicture}
      \draw (0,0) rectangle (2, 2);
      \draw[ddr->] (0, 0.7) -- (2, 0.7);
      \draw[ddr<-] (0, 1.3) -- (2, 1.3);
      \node at (1, 0.35) {$0$};
      \node at (1, 1.65) {$0$};
      \node at (1, 1) {$-1$};
    \end{tikzpicture}
    \caption{}
    \label{fig:5BW-BT-b}
  \end{subfigure}
  \caption{The $46$-space representations for the brane configurations
    in figure~\ref{fig:5BW-a} and~\ref{fig:5BW-b}.  The number $q$
    labeling a region indicates that there is an $(N,q)$ 5-brane on
    that region.}
  \label{fig:5BW-BT}
\end{figure}

What happens if we split the D5-branes across every NS5-brane, and
keep the distances between the segments finite?  In such a situation,
the horizontal and vertical multiplets are all massive and absent in
the infrared theory.  As far as renormalization group invariant
quantities such as the supersymmetric index are concerned, these
multiplets can therefore be ignored.  On the other hand, there are now
far more ways in which diagonal multiplets arise.  To see this
clearly, let us represent brane junctions by oriented dash-dot lines,
which we call \emph{zigzag paths}.  The orientation of a zigzag path
is such that if it is going up, then there is an $(N, q)$ 5-brane on
the left and $(N, q+1)$ 5-brane on the right for some $q$.  For
example, the two configurations in figure~\ref{fig:5BW} are
represented on the $46$-torus as in figure~\ref{fig:5BW-BT}.  Diagonal
multiplets arise from open strings localized near intersections of
rapidity lines or zigzag paths.  When the D5-branes split, each
rapidity line splits into two antiparallel zigzag paths.
Correspondingly, each intersection point breaks apart into four
intersection points.  As the number of intersection points increases,
we expect to get more diagonal multiplets.

\enlargethispage{14pt}

The number of $\U(1)$ flavor symmetries also gets larger.  Recall that
these come from global gauge symmetries on the NS5-branes.  In a brane
box model, gauge transformations on the upper and lower halves of an
NS5-brane give the same flavor symmetry in the 4d theory since they
simply act in opposite manners on the horizontal and vertical
multiplets (and the action on the diagonal multiplets is determined by
the requirement that the superpotential be neutral).  After the brane
configuration is deformed, there are no horizontal or vertical
multiplets anymore, so gauge transformations on the two halves can
lead to different flavor symmetries.

We can make these statements more precise when the deformed system
contains only $(N,0)$ and $(N, \pm 1)$ 5-branes.  In fact, in such a
situation the system is still described by an $\CN = 1$ quiver gauge
theory~\cite{Franco:2005rj}.  The quiver is such that an $\SU(N)$ node
is assigned to each $(N,0)$ region, and an arrow connects two nodes on
regions sharing a vertex:
\begin{equation}
  \label{eq:BT-quiver}
  \begin{tikzpicture}
    \draw[ddr->] (0, 1) node[left] {$i$} -- (2, 1);
    \draw[ddr->] (1, 0) node[below] {$j$} -- (1, 2);
    \node at (0.5, 0.5) {$0$};
    \node at (1.5, 1.5) {$0$};
    \node at (1.5, 0.5) {$1$};
    \node at (0.5, 1.5) {$-1$};
  \end{tikzpicture}
  \quad = \quad
  \begin{tikzpicture}
    \draw[ddr->] (0, 1) node[left] {$i$} -- (2, 1);
    \draw[ddr->] (1, 0) node[below] {$j$} -- (1, 2);
    \node[wnode] (a) at (0.5, 0.5) {$$};
    \node[wnode] (b) at (1.5, 1.5) {$$};
    \draw[q->] (a) -- node[below right] {} (b);
  \end{tikzpicture}
  \quad .
\end{equation}
(An $(N, \pm 1)$ 5-brane is dual to a single D5-brane, so it supports
a $\U(1)$ gauge field which is frozen at low energies.)  To the $i$th
zigzag path is associated a flavor symmetry $\U(1)_i$, and arrows
intersecting with this path have nonzero $\U(1)_i$-charge $F_i$.  We
choose the normalization for the $F_i$ such that the arrow in the
above picture has $F_i = +1$ and $F_j = -1$.  It is known that the
$\U(1)_i$ account for all nonanomalous $\U(1)$ flavor symmetries
present in the gauge theory~\cite{Imamura:2006ie}.  They are not
independent, however.  Since each arrow is intersected precisely by
two zigzag paths assigning opposite charges to it, the sum of the
charges~vanishes:
\begin{equation}
  \label{sum-F=0}
  \sum_i F_i = 0.  
\end{equation}
Hence, the number of $\U(1)$ flavor symmetries is equal to the number
of zigzag paths minus~one.

\begin{figure}
  \centering
    \begin{tikzpicture}[scale=1.2]
      \draw (0,0) rectangle (3, 2);
      
      \begin{scope}[shift={(0.2, 0.2)}]
        \node (a) at (0.5, 0.5) {$a$};
        \node (b) at (1.5, 0.5) {$b$};
        \node (c) at (2.5, 0.5) {$c$};
        \node (d) at (0.5, 1.5) {$d$};
        \node (e) at (1.5, 1.5) {$e$};
        \node (f) at (2.5, 1.5) {$f$};
      \end{scope}
      
      \begin{scope}[shift={(0.2, 0)}]
        \draw[ddr<-] (0, 0) -- (0, 2);
        \draw[ddr<-] (1, 0) -- (1, 2);
        \draw[ddr<-] (2, 0) -- (2, 2);
      \end{scope}
      
      \begin{scope}[shift={(0, 0.2)}]
        \draw[ddr<-] (0, 0) -- (3, 0);
        \draw[ddr<-] (0, 1) -- (3, 1);
      \end{scope}
      
      \draw[ddr->] (0.5, 0) -- (2.5, 2);
      \draw[ddr->] (1.5, 0) -- (3, 1.5);
      \draw[ddr->] (2.5, 0) -- (3, 0.5);
      \draw[ddr->] (0, 0.5) -- (1.5, 2);
      \draw[ddr->] (0, 1.5) -- (0.5, 2);
    \end{tikzpicture}
    \caption{The $2 \times 3$ brane box model as a brane tiling
      model.}
  \label{fig:BB-by-BT}
\end{figure}
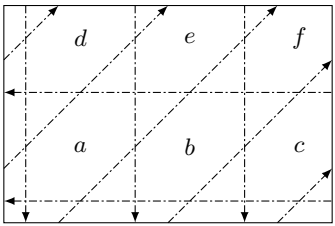

Once we start thinking in terms of zigzag paths, we can construct many
more $\CN = 1$ theories.  We can consider various configurations of
zigzag paths, not necessarily obtained by deforming brane box
configurations.  Such configurations of zigzag paths are called
\emph{brane tilings}.  Provided that only $(N,0)$ and $(N,\pm 1)$
5-branes appear on the graph of zigzag paths, brane tilings yield $\CN
= 1$ quiver gauge theories, whose quivers are determined by the same
rule as above.  Actually, brane box models can also be described as a
brane tiling model; see figure~\ref{fig:BB-by-BT} for the brane tiling
configuration for the $2 \times 3$ brane box model, whose quiver is
shown in figure~\ref{fig:brane-box-ex-b}.  In this sense, brane
tilings generalize the brane box construction.  The reader may consult
the reviews~\cite{Kennaway:2007tq, Yamazaki:2008bt} for more extensive
discussions on brane tilings and their relations to other subjects
such as the AdS/CFT correspondence.

Now, let us try to connect brane tilings to an integrable lattice
model.  Since the only difference compared to the brane box case is
the type of defects inserted in the 6d theory on the D5-branes, much
of the previous argument carries over.  Using this 6d theory, we can
define a 2d TQFT whose correlation function for a configuration of
zigzag paths is given by the supersymmetric index of the corresponding
4d $\CN = 1$ theory.  An extra dimension emerges if we utilize
dualities and embed the 6d theory in M-theory.  Zigzag paths are
mapped to semi-infinite M5-branes, and their $x^{10}$-coordinates
provide spectral parameters.  So we indeed have all ingredients
required for the construction of an integrable lattice model.  It is
reassuring that we have found the relation \eqref{sum-F=0}, which
implies that the supersymmetric index is invariant under an overall
shift of the $x^{10}$-coordinates.

\begin{figure}
  \centering
  \begin{tikzpicture}[rotate=60]
    \node (O') at ($(30:1.5) + (120:{0.5/3*sqrt(3)} )$) {$q$};
    \node at ($(O') + (-120:{0.5*4.5/3*sqrt(3)} )$) {$q + 2$};
    \node at ($(O') + (0:{0.5*4.5/3*sqrt(3)} )$) {$q$};
    \node at ($(O') + (120:{0.5*4.5/3*sqrt(3)} )$) {$q$};
    \node at ($(O') + (60:{0.5*3/3*sqrt(3)} )$) {$q - 1$};
    \node at ($(O') + (180:{0.5*3/3*sqrt(3)} )$) {$q + 1$};
    \node at ($(O') + (-60:{0.5*3/3*sqrt(3)} )$) {$q + 1$};
    
    \draw[ddr->] (0,0) -- ++(30:3);

    \begin{scope}[shift={({0.5*sqrt(3)}, -0.5)}]
      \draw[ddr<-] (0,0) -- ++(90:3);
    \end{scope}

    \begin{scope}[shift={(0, 2)}]
      \draw[ddr->] (0,0) -- ++(-30:3);
    \end{scope}
  \end{tikzpicture}
  \quad = \quad
  \begin{tikzpicture}
    \node (O) at ($(30:1.5) + (120:{0.5/3*sqrt(3)} ) + (0.05, 0)$) {$q + 1$};
    \node at ($(O) + (-120:{0.5*4.5/3*sqrt(3)} ) - (0.05, 0)$) {$q + 1$};
    \node at ($(O) + (0:{0.5*4.5/3*sqrt(3)} )$) {$q + 1$};
    \node at ($(O) + (120:{0.5*4.5/3*sqrt(3)} ) - (0.05, 0)$) {$q - 1$};
    \node at ($(O) + (60:{0.5*3/3*sqrt(3)} )$) {$q$};
    \node at ($(O) + (180:{0.5*3/3*sqrt(3)} )$) {$q$};
    \node at ($(O) + (-60:{0.5*3/3*sqrt(3)} )$) {$q + 2$};
    
    \draw[ddr->] (0,0) -- ++(30:3);

    \begin{scope}[shift={({0.5*sqrt(3)}, -0.5)}]
      \draw[ddr->] (0,0) -- ++(90:3);
    \end{scope}

    \begin{scope}[shift={(0, 2)}]
      \draw[ddr->] (0,0) -- ++(-30:3);
    \end{scope}
  \end{tikzpicture}
  \caption{The Yang-Baxter equation for zigzag paths.}
  \label{fig:YBE-DW}
\end{figure}
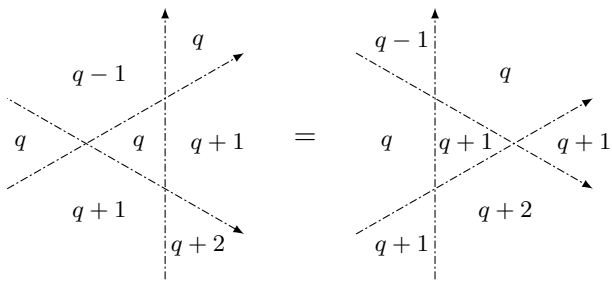

The next step would be to write down an integral expression for the
supersymmetric index and read off the R-matrix.  Here we get into
trouble.  Clearly, we cannot do this for an arbitrary configuration of
zigzag paths; in general, we find $(N, q)$ regions with $|q| > 1$, and
do not know what the corresponding 4d theory is.%
\footnote{An exception is the case with $N = 2$.  In this case, every
  $(N, q)$ 5-brane can be transformed to $(N,0)$ or $(N, 1)$ 5-brane
  by an $\SL(2;\Z)$ transformation.  So there are only two equivalence
  classes of 5-branes, which we can distinguish by painting black or
  white.  This gives a checkerboard pattern on $T^2$.  Presumably, our
  construction applied to this case gives an integrable lattice model
  discovered in~\cite{Bazhanov:2010kz}.  The Yang-Baxter equation for
  this model first appeared in \cite{MR2076912}, and is equivalent to an
  identity proved in \cite{MR1846786}.  See \cite{Derkachov:2012iv,
    Spiridonov:2013zma} for more discussions and further developments
  in this direction.}
Then we are forced to restrict our attention to configurations without
such regions.  Sadly, this restriction is not compatible with
integrability: the Yang-Baxter equation for three zigzag paths always
involves regions with $|q| > 1$, as can be seen from
figure~\ref{fig:YBE-DW}.

The difficulty stems from the fact that zigzag paths are not really
line operators.  They are more properly called \emph{domain walls}
since they separate regions with different physical properties.  This
is a problem we did not encounter for brane box models.  In that case,
rapidity lines were genuine line operators.

Our consideration at the beginning of this section suggests a
resolution to this problem.  We have seen that when a brane box model
is deformed to a brane tiling model, a rapidity line splits into two
antiparallel zigzag paths.  Conversely, we can take two neighboring
antiparallel zigzag paths, and regard them as one thick line (or
``ribbon'') operator.  In this way we can make two kinds of line
operators, namely
\begin{equation}
  \label{eq:LO}
  \begin{tikzpicture}[scale=1.5]
    \draw[r->] (0,0) node[left] {$(r_1, r_2)$} -- (1,0);
  \end{tikzpicture}
  \quad = \quad
  \begin{tikzpicture}[scale=1.5]
    \draw[ddr->] (0,0) node[left] {$r_1$} -- (1,0);
    \draw[ddr->] (1, -0.2) node[right] {$r_2$} -- (0, -0.2);
  \end{tikzpicture}
\end{equation}
and another one for which the orientations of the zigzag paths on the
right-hand side are flipped.  Below we will only use the first one.

If we use this line operator and give the ``background charge'' $q =
-1$ to the TQFT, then regions with $|q| > 1$ never arise:
\begin{equation}
  \begin{tikzpicture}
    \draw[r->] (0, 1) -- (2, 1);
    \draw[r->] (1, 0) -- (1, 2);

    \node at (0.5, 0.5) {$-1$};
    \node at (0.5, 1.5) {$-1$};
    \node at (1.5, 0.5) {$-1$};
    \node at (1.5, 1.5) {$-1$};
  \end{tikzpicture}
  \quad = \quad
  \begin{tikzpicture}
    \draw[ddr->] (2, 0.7) -- (0, 0.7);
    \draw[ddr->] (0, 1.3) -- (2, 1.3);
    \draw[ddr->] (0.7, 0) -- (0.7, 2);
    \draw[ddr->] (1.3, 2) -- (1.3, 0);
  
    \node at (1, 0.35) {$0$};
    \node at (1, 1.65) {$0$};
    \node at (0.35, 1) {$0$};
    \node at (1.65, 1) {$0$};
    \node at (1, 1) {$1$};
    \node at (0.35, 0.35) {$-1$};
    \node at (0.35, 1.65) {$-1$};
    \node at (1.65, 0.35) {$-1$};
    \node at (1.65, 1.65) {$-1$};
  \end{tikzpicture} \quad .
\end{equation}
Thus, a brane tiling constructed with line operators of this type is
always described by a quiver gauge theory, and we can readily write
down an integral expression for its supersymmetric index.  From this
expression we can read off the R-matrix of the associated vertex
model.  Just as in the case for brane box models, we expect that the
Yang-Baxter equation for this R-matrix can be understood as an
equality between the supersymmetric indices of two dual gauge
theories.  Let us see if this is the case.

\begin{figure}
  \centering
  \begin{subfigure}{3.6\qsep}
    \centering
    \begin{tikzpicture}[scale=1.2]
      \draw (0,0) rectangle (3, 2);

      \begin{scope}[shift={(0.25, 0)}]
        \draw[r->] (0, 0) -- (0, 2);
        \draw[r->] (1, 0) -- (1, 2);
        \draw[r->] (2, 0) -- (2, 2);
      \end{scope}

      \begin{scope}[shift={(0, 0.25)}]
        \draw[r->] (0, 0) -- (3, 0);
        \draw[r->] (0, 1) -- (3, 1);
      \end{scope}

      \begin{scope}[shift={(0.25, 0.25)}]
        \node[above] at (0.5, 0) {$i_1$};
        \node[above] at (1.5, 0) {$i_2$};
        \node[above] at (2.5, 0) {$i_3$};
        \node[above] at (0.5, 1) {$j_1$};
        \node[above] at (1.5, 1) {$j_2$};
        \node[above] at (2.5, 1) {$j_3$};

        \node[right] at (0, 0.5) {$k_1$};
        \node[right] at (1, 0.5) {$l_1$};
        \node[right] at (2, 0.5) {$m_1$};
        \node[right] at (0, 1.5) {$k_2$};
        \node[right] at (1, 1.5) {$l_2$};
        \node[right] at (2, 1.5) {$m_2$};
      \end{scope}
    \end{tikzpicture}
    \caption{}
    \label{fig:VM-a}
  \end{subfigure}%
  \qquad\qquad 
  \begin{subfigure}{3.6\qsep}
    \centering 
    \begin{tikzpicture}[scale=1.2]
      \draw (0,0) rectangle (3, 2);

      \begin{scope}[shift={(0.25, 0.25)}]
        \node[wnode] (i1) at (0.5, 0) {$i_1$};
        \node[wnode] (i2) at (1.5, 0) {$i_2$};
        \node[wnode] (i3) at (2.5, 0) {$i_3$};
        \node[wnode] (j1) at (0.5, 1) {$j_1$};
        \node[wnode] (j2) at (1.5, 1) {$j_2$};
        \node[wnode] (j3) at (2.5, 1) {$j_3$};

        \node[wnode] (k1) at (0, 0.5) {$k_1$};
        \node[wnode] (l1) at (1, 0.5) {$l_1$};
        \node[wnode] (m1) at (2, 0.5) {$m_1$};
        \node[wnode] (k2) at (0, 1.5) {$k_2$};
        \node[wnode] (l2) at (1, 1.5) {$l_2$};
        \node[wnode] (m2) at (2, 1.5) {$m_2$};
      \end{scope}

      \draw[q->] (i1) -- (l1);
      \draw (i1) -- ++(-135:{0.25*sqrt(2)} );
      \draw[q<-] (i1) -- ++(-45:{0.25*sqrt(2)} );
      \draw[q->] (i2) -- (m1);
      \draw (i2) -- ++(-135:{0.25*sqrt(2)} );
      \draw[q<-] (i2) -- ++(-45:{0.25*sqrt(2)} );
      \draw (i3) -- ++(-135:{0.25*sqrt(2)} );
      \draw[q<-] (i3) -- ++(-45:{0.25*sqrt(2)} );
      \draw (i3) -- ++(45:{0.25*sqrt(2)} );
      \draw[q->] (k1) -- (i1);
      \draw (k1) -- ++(-135:{0.25*sqrt(2)} );
      \draw (k1) -- ++(135:{0.25*sqrt(2)} );
      \draw[q->] (l1) -- (i2);
      \draw[q->] (l1) -- (j1);
      \draw[q->] (m1) -- (i3);
      \draw[q->] (m1) -- (j2);
      \draw[q->] (j1) -- (k1);
      \draw[q->] (j1) -- (l2);
      \draw[q->] (j2) -- (m2);
      \draw[q->] (j2) -- (l1);
      \draw[q->] (j3) -- (m1);
      \draw (j3) -- ++(45:{0.25*sqrt(2)} );
      \draw (j3) -- ++(-45:{0.25*sqrt(2)} );
      \draw[q->] (k2) -- (j1);
      \draw[q<-] (k2) -- ++(45:{0.25*sqrt(2)} );
      \draw (k2) -- ++(135:{0.25*sqrt(2)} );
      \draw[q<-] (k2) -- ++(-135:{0.25*sqrt(2)} );
      \draw[q->] (l2) -- (j2);
      \draw (m2) -- ++(135:{0.25*sqrt(2)} );
      \draw[q<-] (l2) -- ++(45:{0.25*sqrt(2)} );
      \draw (l2) -- ++(135:{0.25*sqrt(2)} );
      \draw[q->] (m2) -- (j3);
      \draw[q<-] (m2) -- ++(45:{0.25*sqrt(2)} );
    \end{tikzpicture}
    \caption{}
    \label{fig:VM-b}
  \end{subfigure}
  \caption{(a) The $2 \times 3$ lattice model constructed from the
    line operator \eqref{eq:LO}.  (b) The quiver of the corresponding
    gauge theory.}
  \label{fig:VM}
\end{figure}
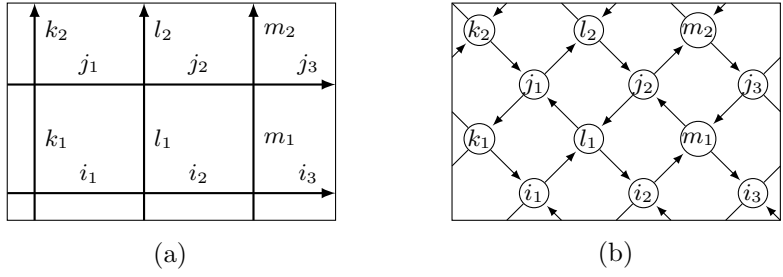

From the quiver rule \eqref{eq:BT-quiver}, we find that the lattice
model in question is a vertex model%
\footnote{The same model can also be described as an IRF model.
  However, the IRF description uses line operators of both types, and
  the Yang-Baxter equation involves twisting of one of the three
  ribbons.  See~\cite{Yamazaki:2013nra} for a discussion from the IRF
  viewpoint.}
with the R-matrix given by
\begin{equation}
  \begin{tikzpicture}
    \draw[thick] (0,1) node[left] {$(r_1, r_2)$}
    -- node[below] {$i_1$} (1,1);
    \draw[r->] (1,1) -- node[above] {$i_2$} (2, 1);

    \draw[thick] (1,0) node[below] {$(s_1, s_2)$}
    -- node[right] {$j_1$} (1,1);
    \draw[r->] (1,1) -- node[left] {$j_2$} (1, 2);
  \end{tikzpicture}
  \quad = \quad
  \begin{tikzpicture}
    \node[dnode] (a) at (0,0) {$i_1$};
    \node[dnode] (b) at (-45:1) {$j_1$};
    \node[snode] (c) at ($(b) + (45:1)$) {$i_2$};
    \node[snode] (d) at (45:1) {$j_2$};

    \draw[q->] (a) -- node[above left=-2pt] {$r_1 - s_1$} (d);
    \draw[q->] (d) -- node[above right=-2pt] {$1 + s_2 - r_1$} (c);
    \draw[q->] (c) -- node[below right=-2pt] {$r_2 - s_2$} (b);
    \draw[q->] (b) -- node[below left=-2pt] {$1 + s_1 - r_2$} (a);
  \end{tikzpicture} \quad .
\end{equation}
Figure~\ref{fig:VM} shows the case when the model is placed on a $2
\times 3$ lattice.  The R-charges of the bifundamental multiplets,
specified by the numbers accompanying the arrows, are chosen to be
consistent with the normalization of the flavor charges and add up to
$2$ around loops.  With this choice, the R-charges satisfy the anomaly
cancellation condition.

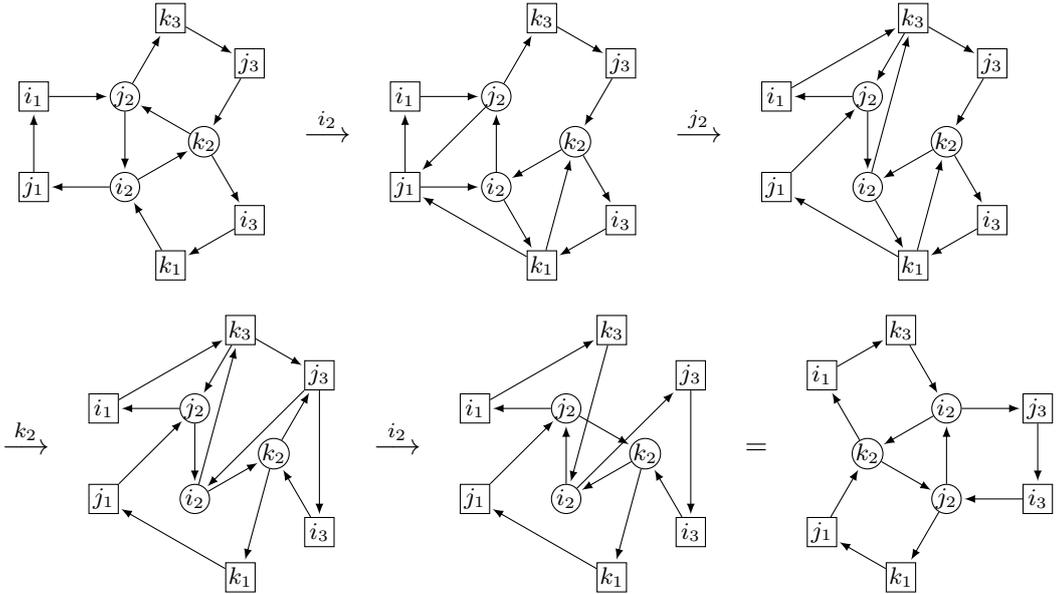
\begin{figure}
  \centering
  \begin{tikzpicture}[rotate=-60]
    \node[wnode] (i2) at (0,0) {$i_2$};
    \node[wnode] (j2) at (150:1) {$j_2$};
    \node[wnode] (k2) at (90:1) {$k_2$};

    \node[snode] (k1) at (0:1) {$k_1$};
    \node[snode] (i3) at ($(k2)+(0:1)$) {$i_3$};

    \node[snode] (j1) at ($(i2) + (-120:1)$) {$j_1$};
    \node[snode] (i1) at ($(j2) + (-120:1)$) {$i_1$};

    \node[snode] (j3) at ($(k2) + (120:1)$) {$j_3$};
    \node[snode] (k3) at ($(j2) + (120:1)$) {$k_3$};

    \draw[q->] (j2) -- (k3);
    \draw[q->] (k3) -- (j3);
    \draw[q->] (j3) -- (k2);
    \draw[q->] (k2) -- (j2);

    \draw[q->] (k2) -- (i3);
    \draw[q->] (i3) -- (k1);
    \draw[q->] (k1) -- (i2);
    \draw[q->] (i2) -- (k2);

    \draw[q->] (i2) -- (j1);
    \draw[q->] (j1) -- (i1);
    \draw[q->] (i1) -- (j2);
    \draw[q->] (j2) -- (i2);
  \end{tikzpicture}
  \quad $\stackrel{i_2}{\longto}$ \quad 
  \begin{tikzpicture}[rotate=-60]
    \node[wnode] (i2) at (0,0) {$i_2$};
    \node[wnode] (j2) at (150:1) {$j_2$};
    \node[wnode] (k2) at (90:1) {$k_2$};

    \node[snode] (k1) at (0:1) {$k_1$};
    \node[snode] (i3) at ($(k2)+(0:1)$) {$i_3$};

    \node[snode] (j1) at ($(i2) + (-120:1)$) {$j_1$};
    \node[snode] (i1) at ($(j2) + (-120:1)$) {$i_1$};

    \node[snode] (j3) at ($(k2) + (120:1)$) {$j_3$};
    \node[snode] (k3) at ($(j2) + (120:1)$) {$k_3$};

    \draw[q->] (j2) -- (k3);
    \draw[q->] (k3) -- (j3);
    \draw[q->] (j3) -- (k2);

    \draw[q->] (k2) -- (i3);
    \draw[q->] (i3) -- (k1);
    \draw[q<-] (k1) -- (i2);
    \draw[q<-] (i2) -- (k2);

    \draw[q<-] (i2) -- (j1);
    \draw[q->] (j1) -- (i1);
    \draw[q->] (i1) -- (j2);
    \draw[q<-] (j2) -- (i2);

    \draw[q->] (j2) -- (j1);
    \draw[q->] (k1) -- (k2);
    \draw[q->] (k1) -- (j1);
  \end{tikzpicture}
  \quad $\stackrel{j_2}{\longto}$ \quad 
  \begin{tikzpicture}[rotate=-60]
    \node[wnode] (i2) at (0,0) {$i_2$};
    \node[wnode] (j2) at (150:1) {$j_2$};
    \node[wnode] (k2) at (90:1) {$k_2$};

    \node[snode] (k1) at (0:1) {$k_1$};
    \node[snode] (i3) at ($(k2)+(0:1)$) {$i_3$};

    \node[snode] (j1) at ($(i2) + (-120:1)$) {$j_1$};
    \node[snode] (i1) at ($(j2) + (-120:1)$) {$i_1$};

    \node[snode] (j3) at ($(k2) + (120:1)$) {$j_3$};
    \node[snode] (k3) at ($(j2) + (120:1)$) {$k_3$};

    \draw[q<-] (j2) -- (k3);
    \draw[q->] (k3) -- (j3);
    \draw[q->] (j3) -- (k2);

    \draw[q->] (k2) -- (i3);
    \draw[q->] (i3) -- (k1);
    \draw[q<-] (k1) -- (i2);
    \draw[q<-] (i2) -- (k2);

    \draw[q<-] (i1) -- (j2);
    \draw[q->] (j2) -- (i2);

    \draw[q<-] (j2) -- (j1);
    \draw[q->] (k1) -- (k2);
    \draw[q->] (k1) -- (j1);

    \draw[q->] (i1) -- (k3);
    \draw[q->] (i2) -- (k3);
  \end{tikzpicture}

  \bigskip

  \quad $\stackrel{k_2}{\longto}$ \quad 
  \begin{tikzpicture}[rotate=-60]
    \node[wnode] (i2) at (0,0) {$i_2$};
    \node[wnode] (j2) at (150:1) {$j_2$};
    \node[wnode] (k2) at (90:1) {$k_2$};

    \node[snode] (k1) at (0:1) {$k_1$};
    \node[snode] (i3) at ($(k2)+(0:1)$) {$i_3$};

    \node[snode] (j1) at ($(i2) + (-120:1)$) {$j_1$};
    \node[snode] (i1) at ($(j2) + (-120:1)$) {$i_1$};

    \node[snode] (j3) at ($(k2) + (120:1)$) {$j_3$};
    \node[snode] (k3) at ($(j2) + (120:1)$) {$k_3$};

    \draw[q<-] (j2) -- (k3);
    \draw[q->] (k3) -- (j3);
    \draw[q<-] (j3) -- (k2);

    \draw[q<-] (k2) -- (i3);
    \draw[q->] (i2) -- (k2);

    \draw[q<-] (i1) -- (j2);
    \draw[q->] (j2) -- (i2);

    \draw[q<-] (j2) -- (j1);
    \draw[q<-] (k1) -- (k2);
    \draw[q->] (k1) -- (j1);

    \draw[q->] (i1) -- (k3);
    \draw[q->] (i2) -- (k3);

    \draw[q->] (j3) -- (i2);
    \draw[q->] (j3) -- (i3);
  \end{tikzpicture}
  \quad $\stackrel{i_2}{\longto}$ \quad 
  \begin{tikzpicture}[rotate=-60]
    \node[wnode] (i2) at (0,0) {$i_2$};
    \node[wnode] (j2) at (150:1) {$j_2$};
    \node[wnode] (k2) at (90:1) {$k_2$};

    \node[snode] (k1) at (0:1) {$k_1$};
    \node[snode] (i3) at ($(k2)+(0:1)$) {$i_3$};

    \node[snode] (j1) at ($(i2) + (-120:1)$) {$j_1$};
    \node[snode] (i1) at ($(j2) + (-120:1)$) {$i_1$};

    \node[snode] (j3) at ($(k2) + (120:1)$) {$j_3$};
    \node[snode] (k3) at ($(j2) + (120:1)$) {$k_3$};

    \draw[q<-] (k2) -- (i3);
    \draw[q<-] (i2) -- (k2);

    \draw[q<-] (i1) -- (j2);
    \draw[q<-] (j2) -- (i2);

    \draw[q<-] (j2) -- (j1);
    \draw[q<-] (k1) -- (k2);
    \draw[q->] (k1) -- (j1);

    \draw[q->] (i1) -- (k3);
    \draw[q<-] (i2) -- (k3);

    \draw[q<-] (j3) -- (i2);
    \draw[q->] (j3) -- (i3);

    \draw[q->] (j2) -- (k2);
  \end{tikzpicture}
  \quad = \quad 
  \begin{tikzpicture}
    \node[wnode] (j2) at (0,0) {$j_2$};
    \node[wnode] (k2) at (150:1) {$k_2$};
    \node[wnode] (i2) at (90:1) {$i_2$};

    \node[snode] (i3) at (0:1) {$i_3$};
    \node[snode] (j3) at ($(i2)+(0:1)$) {$j_3$};

    \node[snode] (k1) at ($(j2) + (-120:1)$) {$k_1$};
    \node[snode] (j1) at ($(k2) + (-120:1)$) {$j_1$};

    \node[snode] (k3) at ($(i2) + (120:1)$) {$k_3$};
    \node[snode] (i1) at ($(k2) + (120:1)$) {$i_1$};

    \draw[q->] (j2) -- (k1);
    \draw[q->] (k1) -- (j1);
    \draw[q->] (j1) -- (k2);
    \draw[q->] (k2) -- (j2);

    \draw[q->] (k2) -- (i1);
    \draw[q->] (i1) -- (k3);
    \draw[q->] (k3) -- (i2);
    \draw[q->] (i2) -- (k2);

    \draw[q->] (i2) -- (j3);
    \draw[q->] (j3) -- (i3);
    \draw[q->] (i3) -- (j2);
    \draw[q->] (j2) -- (i2);
  \end{tikzpicture}
  \caption{A sequence of duality transformations that demonstrates the
    Yang-Baxter equation \eqref{eq:YBE-BT}.  The letter on an arrow
    between two quivers indicates the node at which the transformation
    is applied.  Note that circle nodes can be relabeled freely since
    their fugacities are integrated over.}
  \label{fig:YBE-SD}
\end{figure}

What we have just constructed is the integrable lattice model studied
in~\cite{Bazhanov:2011mz, Yamazaki:2012cp}.  (Its generalization to
the lens space index was considered in~\cite{Yamazaki:2013nra}.)  The
Yang-Baxter equation for the model is
\begin{equation}
  \label{eq:YBE-BT}
  \begin{tikzpicture}[rotate=-60]
    \node[wnode] (i2) at (0,0) {$i_2$};
    \node[wnode] (j2) at (150:1) {$j_2$};
    \node[wnode] (k2) at (90:1) {$k_2$};

    \node[snode] (k1) at (0:1) {$k_1$};
    \node[snode] (i3) at ($(k2)+(0:1)$) {$i_3$};

    \node[snode] (j1) at ($(i2) + (-120:1)$) {$j_1$};
    \node[snode] (i1) at ($(j2) + (-120:1)$) {$i_1$};

    \node[snode] (j3) at ($(k2) + (120:1)$) {$j_3$};
    \node[snode] (k3) at ($(j2) + (120:1)$) {$k_3$};

    \draw[q->] (j2) -- (k3);
    \draw[q->] (k3) -- (j3);
    \draw[q->] (j3) -- (k2);
    \draw[q->] (k2) -- (j2);

    \draw[q->] (k2) -- (i3);
    \draw[q->] (i3) -- (k1);
    \draw[q->] (k1) -- (i2);
    \draw[q->] (i2) -- (k2);

    \draw[q->] (i2) -- (j1);
    \draw[q->] (j1) -- (i1);
    \draw[q->] (i1) -- (j2);
    \draw[q->] (j2) -- (i2);
  \end{tikzpicture}
  \quad = \quad 
  \begin{tikzpicture}
    \node[wnode] (j2) at (0,0) {$j_2$};
    \node[wnode] (k2) at (150:1) {$k_2$};
    \node[wnode] (i2) at (90:1) {$i_2$};

    \node[snode] (i3) at (0:1) {$i_3$};
    \node[snode] (j3) at ($(i2)+(0:1)$) {$j_3$};

    \node[snode] (k1) at ($(j2) + (-120:1)$) {$k_1$};
    \node[snode] (j1) at ($(k2) + (-120:1)$) {$j_1$};

    \node[snode] (k3) at ($(i2) + (120:1)$) {$k_3$};
    \node[snode] (i1) at ($(k2) + (120:1)$) {$i_1$};

    \draw[q->] (j2) -- (k1);
    \draw[q->] (k1) -- (j1);
    \draw[q->] (j1) -- (k2);
    \draw[q->] (k2) -- (j2);

    \draw[q->] (k2) -- (i1);
    \draw[q->] (i1) -- (k3);
    \draw[q->] (k3) -- (i2);
    \draw[q->] (i2) -- (k2);

    \draw[q->] (i2) -- (j3);
    \draw[q->] (j3) -- (i3);
    \draw[q->] (i3) -- (j2);
    \draw[q->] (j2) -- (i2);
  \end{tikzpicture}
  \quad .
\end{equation}
For simplicity we have omitted the R-charges in the quivers.  As
expected, this equation follows from Seiberg duality: applying the
basic duality transformation \eqref{eq:Seiberg} four times, we can
turn the left-hand side into the right-hand side.  This process is
illustrated in figure~\ref{fig:YBE-SD}.

Incidentally, a single Seiberg duality transformation is induced by
the Yang-Baxter move involving two zigzag paths and one ``twisted''
ribbon operator:
\begin{equation}
  \begin{tikzpicture}[rounded corners, scale=1.2]
    \draw[ddr->, shift={(120:0.2)}] (210:1) -- (30:1);
    \draw[ddr->, shift={(-120:0.2)}] (150:1) -- (-30:1);

    \draw[ddr<-, shift={(0.2, 0)}] (-0.15, -1) -- (-0.15, -0.25) -- (0.15, 0.25) -- (0.15,1);
    \draw[ddr->, shift={(0.2, 0)}] (0.15, -1) -- (0.15, -0.25) -- (-0.15, 0.25) -- (-0.15,1);
  \end{tikzpicture}
  \quad \longto \quad
  \begin{tikzpicture}[rounded corners, scale=1.2]
    \draw[ddr->, shift={(-60:0.2)}] (210:1) -- (30:1);
    \draw[ddr->, shift={(60:0.2)}] (150:1) -- (-30:1);

    \draw[ddr<-, shift={(-0.2, 0)}] (-0.15, -1) -- (-0.15, -0.25) -- (0.15, 0.25) -- (0.15,1);
    \draw[ddr->, shift={(-0.2, 0)}] (0.15, -1) -- (0.15, -0.25) -- (-0.15, 0.25) -- (-0.15,1);
  \end{tikzpicture}
  \quad .
\end{equation}
This relation between the Yang-Baxter move and Seiberg duality was
pointed out by Hanany and Vegh~\cite{Hanany:2005ss}.

\section{\texorpdfstring{3d $\boldsymbol{\CN = 2}$ and 2d $\boldsymbol{\CN = (2,2)}$ quiver gauge theories}{3d N=2 and 2d N=(2,2) quiver gauge theories}}
\label{sec:LD}

The results from the previous sections imply, via T-duality, that
there exist similar connections between lower-dimensional quiver gauge
theories and integrable lattice models.  In this section we discuss
integrable lattice models associated with 3d $\CN = 2$ and 2d $\CN =
(2,2)$ theories.  We will focus on those coming from brane box
configurations, though our analysis can be adapted straightforwardly
to the brane tiling case.

\subsection{3d brane box models}

Let us consider the following brane setup in Type IIA superstring
theory, which is related to the 4d brane box model \eqref{eq:BB} by
T-duality along the $2$-direction:
\begin{equation}
\label{eq:BB-3d}
\begin{tabular}{|l|cccccccccc|}
  \hline
  & 0 & 1 & 2 & 3 & 4 & 5 & 6 & 7 & 8 & 9
  \\ \hline
  D4 & $\times$ & $\times$ & & $\times$ & $\times$ & & $\times$
  &&&
  \\
  NS5 & $\times$ & $\times$ & $\times$ & $\times$ & $\times$ & $\times$
  &&&&
  \\
  NS5 & $\times$ & $\times$ & $\times$ & $\times$ & & & $\times$ &
  $\times$
  &&
  \\ \hline
\end{tabular}
\end{equation}
This is a brane box configuration that realizes a 3d $\CN = 2$ quiver
gauge theory.  Being related by T-duality, the theory is essentially
the dimensional reduction of the corresponding 4d $\CN = 1$ theory,
and is described by the same quiver (with all Chern-Simons levels
vanishing).  We expect that the supersymmetric index of this theory is
given by the partition function of an integrable lattice model.

As in the 4d case, the theory can be placed on $S^2 \times S^1$, with
all four supercharges unbroken~\cite{Imamura:2011su, Closset:2012ru}.
We are interested in the supersymmetric index
\begin{equation}
  \label{eq:I-3d}
  \CI(x; \{r_\alpha\}) = \Tr_{S^2} \Bigl((-1)^F x^{R +2 J_3}\Bigr),
\end{equation}
where $J_3$ is the Cartan generator of the isometry group $\SU(2)$ of
$S^2$.  It receives contributions only from states with energy $E = R
+ J_3$.  As usual, we have omitted the fugacities $u_\alpha$ for the
flavor symmetries $\U(1)_\alpha$, whose effects can be taken into
account by keeping the dependence of the R-charge $R$ on the spectral
parameters $r_\alpha$.

The 3d supersymmetric index is computed by an integral formula similar
to the one in the 4d case.  The novelty is that there can be
nontrivial magnetic fluxes through $S^2$.  For each gauge group
$\SU(N)_a$, the flux is specified by a set of integers $m_a =
(m_{a,1}, \dotsc, m_{a,N})$ subject to the condition $\sum_i m_{a,i} =
0$.  Therefore, the formula involves a sum over fluxes as well as an
integration over the fugacities for the gauge groups:
\begin{equation}
\label{eq:I-int-3d}
  \CI
  = \sum_{\{m_a\}} \oint
     \prod_{\circled{a}} \frac{1}{|W_{m_a}|}
     \prod_{j=1}^{N-1} \frac{\rmd z_{a, j}}{2\pi i z_{a,j}}
     \CI_{\text{V}}(z_a, m_a)
     \prod_{\circled{a} \qto \circled{b}} \CI_{\text{B}}\bigl((z_a, m_a), (z_b, m_b); R_{ab}\bigr).
\end{equation}
Here $|W_{m_a}|$ is the order of the Weyl group of the subgroup of $G$
that is left unbroken by $m_a$, and the vector and bifundamental
multiplet factors are given by~\cite{Kim:2009wb, Imamura:2011su,
  Kapustin:2011jm}
\begin{align}
  \CI_{\text{V}}(z_a, m_a)
  &= \prod_{i \neq j} x^{-|m_{a,i} - m_{a,j}|/2}
        \Bigl(1 - \frac{z_{a,i}}{z_{a,j}} x^{|m_{a,i} - m_{a,j}|}\Bigr),
\\
  \begin{split}
    \CI_{\text{B}}\bigl((z_a, m_a), (z_b, m_b); R_{ab}\bigr)
    &= \prod_{i,j} \Bigl(x^{1- R_{ab}}
          \frac{z_{a, i}}{z_{b, j}}\Bigr)^{|m_{a,i} - m_{b,j}|/2}
\\ &\qquad \times
          \frac{(x^{|m_{a,i} - m_{b,j}| + 2 - R_{ab}}
                    z_{a,i}/z_{b,j}; x^2)_\infty}
                  {(x^{|m_{a,i} - m_{b,j}| + R_{ab}}
                    z_{b,j}/z_{a,i}; x^2)_\infty}.
\end{split}
\end{align}
The bifundamental factor $\CI_{\text{B}}$ satisfies the identity
\begin{equation}
  \CI_{\text{B}}\bigl((z_a, m_a), (z_b, m_b); R_{ab}\bigr) 
  \CI_{\text{B}}\bigl((z_b, m_b), (z_a, m_a); R_{ba}\bigr)
  = 1
\end{equation}
for $R_{ab} + R_{ba} = 2$.

We can connect the supersymmetric index to a lattice model by
following the same procedure as before.  We define an IRF model on the
lattice drawn on the 46-torus by the NS5-branes.  The state variables
are $(z_a, m_a)$, and the Bolzmann weight is given by the same picture
\eqref{eq:W-4d}.  By construction, the supersymmetric index of a 3d
brane box model coincides with the partition function of this IRF
model.  The Yang-Baxter equation again boils down to the equality
\eqref{eq:Seiberg} between the supersymmetric indices of two gauge
theories.  This equality, if true, establishes the integrability of
the IRF model.

The equality indeed follows from the $\SU(N)$
version~\cite{Aharony:2013dha, Park:2013wta} of the duality for 3d
$\CN = 2$ gauge theories proposed by Aharony~\cite{Aharony:1997gp}.
There is an interesting twist, though: the theories described by the
quivers that appear in the equality are \emph{not} dual to each other.

The correct duality takes the theory on the left-hand side to a theory
whose matter content consists of the chiral multiplets described by
the quiver on the right-hand side, plus two extra singlets $V_+$,
$V_-$.  More generally, Aharony duality relates two 3d $\CN = 2$ gauge
theories with the following matter contents:
\begin{equation}
\begin{tabular}{|l|cccccc|}
\hline
        & $G$ & $\SU(N_f)_L$ & $\SU(N_f)_R$
        & $\U(1)_B$ & $\U(1)_A$ & $\U(1)_R$
\\ \hline
$Q$    & $\square$ & $\overline\square$ & $\mathbf{1}$
          & $1$ & $1$ & $1 - N_c/N_f$ 
\\
$\Qt$ & $\overline\square$ & $\mathbf{1}$ & $\square$
          & $-1$ & $1$ & $1 - N_c/N_f$
\\ \hdashline
$q$    & $\square$ & $\overline\square$ & $\mathbf{1}$
          & $-N_c/(N_f - N_c)$ & $-1$ & $N_c/N_f$
\\
$\qt$ & $\overline\square$  & $\mathbf{1}$ & $\square$
          & $N_c/(N_f - N_c)$ & $-1$ & $N_c/N_f$
\\
$M$ & $\mathbf{1}$ & $\square$ & $\overline\square$
        & $0$ & $2$ & $2(1 - N_c/N_f)$
\\
$V_\pm$ & $\mathbf{1}$ & $\mathbf{1}$ & $\mathbf{1}$
              & $0$ & $-N_f$ & $1$
\\ \hline
\end{tabular} \quad .
\end{equation}
The gauge group $G$ is $\SU(N_c)$ for the theory with $Q$, $\Qt$, and
$\SU(N_f - N_c)$ for its dual.  Although a proof has not been given
yet, there is strong evidence for the equality of the supersymmetric
indices of these two theories in terms of series expansion in
$x$~\cite{Hwang:2011qt, Hwang:2012jh, Park:2013wta}.%
\footnote{For $N_c = 2$ and $N_f = 3$, $4$, the equality was recently
  proved in \cite{Gahramanov:2015cva} in the absence of the fugacity
  for $\U(1)_A$.}

The above table contains one more global symmetry compared to its 4d
counterpart~\eqref{eq:SD-4d}, namely the axial symmetry $\U(1)_A$
which is anomalous in four dimensions.  In the definition of the
index~\eqref{eq:I-3d}, we did not include fugacities for axial
symmetries.  Thanks to this omission, $V_\pm$ always have $R = 1$ for
whatever shift of $R$ by the relevant $\U(1)$ flavor symmetries, and
their contributions to the index cancel since we can turn on a
superpotential $W = mV_+ V_-$.  If we included these fugacities, the
cancellation would no longer occur.  In that case, the presence of
$V_\pm$ would be crucial for the indices of the two theories to match.

Why does the logic fail if the fugacities for axial symmetries are
turned on?  The point is that when we apply T-duality to get a 3d
brane box model from a 4d one, the $2$-direction is compactified to a
circle, however small its radius may be.  In other words, we are
really compactifying the 4d theory on $S^1$ with finite radius, not
dimensional reducing it.  As emphasized in~\cite{Aharony:2013dha}, the
effective 3d theory obtained by compactification has a superpotential
that contains monopole operators.  This superpotential breaks the
axial symmetries, thereby reproducing the effect of anomalies in the
4d theory.  So it is not that the logic fails.  Rather, we have no
fugacities to turn on to begin with.  Upon inclusion of such
superpotentials on both sides, the extra multiplets disappear from the
duality.

The axial symmetries are recovered if we decompactify the
$2$-direction.  Although it is perfectly fine to do so, the resulting
system is not related to a 4d brane box model by T-duality and its
connection to a lattice model requires a more careful analysis.  It is
not clear whether these additional symmetries admit a natural
interpretation in the lattice~model.

\subsection{2d brane box models}
\label{sec:2d-BB}

Applying further T-duality along the $1$-direction, we obtain the
following brane configuration realizing a 2d $\CN = (2,2)$ quiver
gauge theory:
\begin{equation}
\label{eq:BB-2d}
\begin{tabular}{|l|cccccccccc|}
  \hline
  & 0 & 1 & 2 & 3 & 4 & 5 & 6 & 7 & 8 & 9
  \\ \hline
  D3 & $\times$ & & & $\times$ & $\times$ & & $\times$
  &&&
  \\
  NS5 & $\times$ & $\times$ & $\times$ & $\times$ & $\times$ & $\times$
  &&&&
  \\
  NS5 & $\times$ & $\times$ & $\times$ & $\times$ & & & $\times$ &
  $\times$
  &&
  \\ \hline
\end{tabular}
\end{equation}
This theory is described by the same quiver diagram as in the 3d and
4d brane box cases.  However, each node now represents a $\U(N)$ gauge
group, not $\SU(N)$.  The reason is that unlike the case with D4- or
D5-branes, motion of D3-brane segments along NS5-branes creates
disturbances on the NS5-branes that fall off sufficiently rapidly, and
costs only a finite amount of energy.  (In the 3d brane box case, on
the other hand, the required energy becomes arbitrarily large as we
increase the radius of the $2$-circle.)

The supersymmetric index of a 2d $\CN = (2,2)$ theory is known as the
elliptic genus.  It is given by
\begin{equation}
  \label{eq:EG}
  \CI(q, y; \{r_\alpha\}) = \Tr_{S^1} \Bigl((-1)^F q^{H_L} y^{R/2}\Bigr),
\end{equation}
where $H_L$ is the left-moving Hamiltonian, and $R$ is defined as
\emph{twice} the left-moving R-charge so that a superpotential
preserving the R-symmetry has $R = 2$.  The trace receives
contributions only from states with the right-moving Hamiltonian $H_R
= 0$.  The elliptic genus can be computed as the partition function
with twisted boundary conditions on a torus whose complex parameter
$\tau$ is given by $q = e^{2\pi \tau}$.  As such, it enjoys a nice
modular~property.

An integral formula for the elliptic genus of an $\CN = (2,2)$ gauge
theory was derived in~\cite{Gadde:2013ftv, Benini:2013nda,
  Benini:2013xpa}.  For the theory we are considering, the formula
reads%
\footnote{For concreteness we will follow the treatment
  in~\cite{Benini:2013nda, Benini:2013xpa} where the elliptic genus is
  defined in the $(\text{R}, \text{R})$ sector, though the
  considerations given below equally apply to the other sectors.
  See~\cite{Gadde:2013ftv} for the formula for the $(\text{NS},
  \text{NS})$ sector.}
\begin{equation}
\label{eq:I-int-2d}
  \CI
  = \oint
     \prod_{\circled{a}} \frac{1}{N!}
     \prod_{j=1}^N \frac{\rmd u_{a, j}}{2\pi i}
     \CI_{\text{V}}(u_a)
     \prod_{\circled{a} \qto \circled{b}} \CI_{\text{B}}(u_a, u_b; R_{ab}),
\end{equation}
with the vector and bifundamental multplet factors given by
\begin{align}
  \CI_{\text{V}}(u_a)
  &= \biggl(\frac{2\pi\eta(\tau)^3}{\theta_1(-z|\tau)}\biggr)^{N}
        \prod_{i \neq j}
        \frac{\theta_1(u_{a,i} - u_{a,j}|\tau)}
                {\theta_1(u_{a,i} - u_{a,j} - z|\tau)},
\label{eq:IV-(2,2)}
\\
  \CI_{\text{B}}(u_a, u_b; R_{ab})
  &= \prod_{i, j}
        \frac{\theta_1(u_{b,j} - u_{a,i} + (R_{ab}/2 - 1)z|\tau)}
                {\theta_1(u_{b,j} - u_{a,i} + R_{ab}z/2|\tau)}.
\label{eq:IB-(2,2)}
\end{align}
Here $z$ is related to the fugacity $y$ by $y = e^{2\pi i z}$, and
\begin{align}
  \eta(\tau) &= q^{1/24} \prod_{n=1}^\infty (1 - q^n), \\
  \theta_1(z|\tau)
  &= -i\sum_{n=-\infty}^\infty (-1)^n e^{2\pi iz(n + 1/2)} e^{\pi i\tau(n+1/2)^2}
\end{align}
are the Dedekind eta function and a Jacobi theta function.  Note that
$\theta_1(-z|\tau) = -\theta_1(z|\tau)$.  As a consequence, $\CI_{\text{B}}$
obeys the identity \eqref{eq:IBIB=1}.

The above formula takes the same form as the 4d index formula
\eqref{eq:I-int-4d} under the identification $z_{a,i} = e^{2\pi
  iu_{a,i}}$.  The subtlety lies in the integration contour, as was
elucidated in~\cite{Benini:2013nda, Benini:2013xpa} with a careful
path integral analysis.  Since $\theta_1(z|\tau)$ has a pole at $z =
0$, the integrand has poles at various places in the $u$-space, and
the contour must pick up the residues at some but not all of these
poles.  There are many choices that lead to the correct answer.  For
us, a simple choice is such that for each node $a$, every $u_{a,i}$
encircles the pole located at $u_{a,i} = u_{b,j} - R_{ba} z/2$ for some
$(b, j)$, which comes from the factor $\CI_{\text{B}}(u_b, u_a; R_{ba})$
associated to the incoming arrow
\begin{equation}
  \begin{tikzpicture}
    \node[wnode] (a) at (0,0) {$a$};
    \node[wnode] (b) at (1,0) {$b$};
    \draw[q<-] (a) -- (b);
  \end{tikzpicture} \quad .
\end{equation}
Moreover, $u_{a,i}$ and $u_{a,j}$ for $i \neq j$ must encircle
different poles.  We sum over the contributions from all such
contours.  Alternatively, we may choose the contours to encircle poles
coming from outgoing arrows.  The two choices give the same result, up
to an overall sign.  We refer the reader to~\cite{Benini:2013nda,
  Benini:2013xpa} for more details on the integration contour in
general, and section~4.6 of~\cite{Benini:2013xpa} for details specific
to theories closely related to ours.

As in the higher-dimensional cases, the supersymmetric index of a 2d
brane box model coincides with the partition function of an IRF model.
The Boltzmann weight is given by the formula \eqref{eq:W-4d}, and the
Yang-Baxter equation reduces to the equality~\eqref{eq:Seiberg}.
This time, let us demonstrate this equality explicitly.

The elliptic genus of the theory described by the quiver on the
left-hand side is computed by the contour integral
\begin{multline}
  \CI_{\text{LHS}}
  = \frac{1}{N!} \oint
     \prod_{j=1}^N \frac{\rmd u_{g, j}}{2\pi i}
     \CI_{\text{V}}(u_g)
     \CI_{\text{B}}(u_a, u_g; R_{ag}) \CI_{\text{B}}(u_d, u_g; R_{dg})
     \\ \times 
     \CI_{\text{B}}(u_g, u_c; R_{gc}) \CI_{\text{B}}(u_g, u_f; R_{gf}).
\end{multline}
We choose to evaluate it with contours that encircle $N$ poles from
the factors associated with the incoming arrows.  For instance,
suppose that we are evaluating the integral for contours such that the
unordered set of these poles is $\{u_{a,j} - R_{ag} z/2\}$, that is,
each $u_{g,i}$-integral picks up the residue at $u_{g,i} = u_{a,j} -
R_{ag} z/2$ for some $j$.  The $N!$ different ways to assign the poles
to the $u_{g,i}$ make equal contributions, so we only need to consider
the case when $u_{g,i} = u_{a,i} - R_{ag} z/2$ for all $i$.  Using the
fact that the residue of $1/\theta_1(u|\tau)$ at $u = 0$ is
$1/2\pi\eta(\tau)^3$, we find that the sum of the contributions from
these contours is
\begin{equation}
  \label{eq:contrib}
  \CI_{\text{B}}(u_a, u_c; R_{ag} + R_{gc})
  \CI_{\text{B}}(u_a, u_f; R_{ag} + R_{gf}) 
  \CI_{\text{B}}(u_d, u_a; R_{dg} - R_{ag}).
\end{equation}

On the other hand, the elliptic genus of the right-hand side is given by
\begin{multline}
  \CI_{\text{RHS}}
  = \frac{1}{N!} \oint
     \prod_{j=1}^N \frac{\rmd u_{g, j}}{2\pi i}
     \CI_{\text{V}}(u_g)
     \CI_{\text{B}}(u_g, u_d; R_{gd}) \CI_{\text{B}}(u_g, u_a; R_{ga}) 
     \\ \times
     \CI_{\text{B}}(u_c, u_g; R_{cg}) \CI_{\text{B}}(u_f, u_g; R_{fg}) 
     \\ \times
     \CI_{\text{B}}(u_a, u_c; R_{ac})
     \CI_{\text{B}}(u_a, u_f; R_{af})
     \CI_{\text{B}}(u_d, u_c; R_{dc})
     \CI_{\text{B}}(u_d, u_f; R_{df}).
\end{multline}
For the dual theory, we choose the poles from the outgoing arrows.
The contours picking up the residues at the poles $\{u_{d,i} +
R_{gd}/2\}$ contribute
\begin{multline}
  \CI_{\text{B}}(u_d, u_a; R_{ga} - R_{gd}) 
  \CI_{\text{B}}(u_c, u_d; R_{cg} + R_{gd})
  \CI_{\text{B}}(u_f, u_d; R_{fg} + R_{gd}) 
  \\ \times
  \CI_{\text{B}}(u_a, u_c; R_{ac})
  \CI_{\text{B}}(u_a, u_f; R_{af})
  \CI_{\text{B}}(u_d, u_c; R_{dc}) 
  \CI_{\text{B}}(u_d, u_f; R_{df}).
\end{multline}
After canceling out some factors using the identity \eqref{eq:IBIB=1},
we see that this is equal to the contribution~\eqref{eq:contrib} found
above.

\enlargethispage{6pt}

In a similar fashion, one can show that for each set of poles
$\{\uh_{g,i}\} \subset \{u_{a,i} - R_{ag}z/2\} \cup \{u_{d,i} -
R_{dg}z/2\}$ in the theory on the left-hand side, there is a set of
poles $\{\ub_{g,i}\} \subset \{u_{a,i} + R_{ga}z/2\} \cup \{u_{d,i} +
R_{gd}z/2\}$ in the theory on the right-hand side such that the
contours encircling these poles give the same contributions to the
elliptic genera of the respective theories.  (In the case that the
R-charges vanish, $\{\ub_{g,i}\}$ is the complement of $\{\uh_{g,i}\}$
in $\{u_{a,i}\} \cup \{u_{d,i}\}$.)  This is a one-to-one
correspondence between the choices of poles in the two theories.
Therefore, $\CI_{\text{LHS}} = \CI_{\text{RHS}}$.

The equality just proved is basically a consequence of a variant of
Hori-Tong duality~\cite{Hori:2006dk} proposed
in~\cite{Benini:2012ui}.  However, it is important that the fugacities
$u_a$, $u_c$, $u_d$, $u_f$ are unconstrained here.  By contrast, when
regarded as the elliptic genera of dual theories, the two sides of the
equality must be evaluated with the constraint $\sum_i (u_{a,i} +
u_{c,i} + u_{d,i} + u_{f,i}) = 0$ since the diagonal $\U(1)$ subgroup
of the flavor group $\U(N)_a \times \U(N)_c \times \U(N)_d \times
\U(N)_f$ is gauged.  So strictly speaking, the duality does not imply
the Yang-Baxter equation.  Also, note that the fugacity for the axial
symmetry $\U(1)_A$ is set to zero as in the 3d case.  In general, this
can be turned on in the elliptic genus.

Finally, we remark that the duality does not work if we replace the
gauge group with $\SU(N)$.  This is unlike the higher-dimensional
cases, where one can gauge $\U(1)_B$ to obtain the duality for $\U(N)$
theories from that for $\SU(N)$ theories.  Had we started with the
dimensional reduction of the 3d brane box model and simply discarded
the $\U(1)_A$ fugacity, we would have reached a wrong conclusion.

\section{\texorpdfstring{2d $\boldsymbol{\CN = (0,2)}$ quiver gauge
    theories}{2d N=(0,2) quiver gauge theories}}
\label{sec:(0,2)}

Looking at how the duality relation \eqref{eq:Seiberg} was verified
for 2d brane box models, we notice that the numerator and denominator
of the bifundamental factor $\CI_{\text{B}}$ played rather different
roles.  More specifically, while the denominator determined the pole
structure and hence the possible choices for integration contours, the
numerator only provided cancellations of some factors.  The separation
in their roles hints at the existence of integrability in less
supersymmetric situations where an $\CN = (2,2)$ chiral multiplet
decomposes into two multiplets that correspond to the numerator and
denominator.

$\CN = (0,2)$ supersymmetry provides precisely the required
decomposition: an $\CN = (2,2)$ chiral multiplet consists of an $\CN =
(0,2)$ chiral multiplet and Fermi multiplet, with the former
corresponding to the denominator and the latter to the numerator.  In
this section we introduce three classes of $\CN = (0, 2)$ quiver gauge
theories whose elliptic genera are captured by integrable lattice
models.  The first two classes lead to IRF models much like $\CN =
(2,2)$ brane box and brane tiling models discussed in section
\ref{sec:2d-BB}.  The third class gives rise to an IRC model on 3d
lattices.

\subsection[$\CN=(0,2)$ theories related to brane box
configurations]{\texorpdfstring{$\boldsymbol{\CN = (0, 2)}$ theories
    related to brane box configurations}{N=(0,2) theories related to
    brane box configurations}}

Let us take a quiver from some brane box model, and replace the
vertical arrows with dotted arrows.  (Of course, one may as well
choose to replace the horizontal arrows instead.)  For example,
starting from the quiver for the $2 \times 3$ brane box model shown in
figure~\ref{fig:brane-box-ex-b}, we obtain the diagram in
figure~\ref{fig:BB-(0,2)}.  We interpret the resulting diagram as a
quiver of an $\CN = (0,2)$ gauge theory, letting dotted arrows
represent bifundamental Fermi multiplets.  As usual, circle nodes and
solid arrows represent vector multiplets and bifundamental chiral
multiplets.  By this procedure we can associate an $\CN = (0,2)$
quiver gauge theory to every brane box model.  The reader is referred
to~\cite{Witten:1993yc, Gadde:2013lxa} for basics of $\CN = (0,2)$
gauge theories.

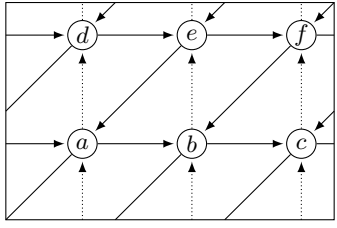
\begin{figure}
  \centering 
  \begin{tikzpicture}[scale=1.2]
    \draw (0,0) rectangle (3, 2);

    \begin{scope}[shift={(0.2, 0.2)}]
      \node[wnode] (a) at (0.5, 0.5) {$a$};
      \node[wnode] (b) at (1.5, 0.5) {$b$};
      \node[wnode] (c) at (2.5, 0.5) {$c$};
      \node[wnode] (d) at (0.5, 1.5) {$d$};
      \node[wnode] (e) at (1.5, 1.5) {$e$};
      \node[wnode] (f) at (2.5, 1.5) {$f$};
    \end{scope}
    
    \draw[q->] (a) -- (b);
    \draw[q->] (b) -- (c);
    \draw[q->] (d) -- (e);
    \draw[q->] (e) -- (f);
    
    \draw[dq->] (a) -- (d);
    \draw[dq->] (b) -- (e);
    \draw[dq->] (c) -- (f);
    
    \draw[q->] (e) -- (a);
    \draw[q->] (f) -- (b);
    
    \draw (a) -- (0, 0);
    \draw (b) -- (1, 0);
    \draw (c) -- (2, 0);
    \draw (d) -- (0, 1);
    
    \draw[q->] (1, 2) -- (d);
    \draw[q->] (2, 2) -- (e);
    \draw[q->] (3, 2) -- (f);
    \draw[q->] (3, 1) -- (c);
    
    \begin{scope}[shift={(0, 0.2)}]
      \draw[q->] (0, 0.5) -- (a);
      \draw (c) -- (3, 0.5);
      \draw[q->] (0, 1.5) -- (d);
      \draw (f) -- (3, 1.5);
    \end{scope}
    
    \begin{scope}[shift={(0.2, 0)}]
      \draw[dq->] (0.5, 0) -- (a);
      \draw[densely dotted] (d) -- (0.5, 2);
      \draw[dq->] (1.5, 0) -- (b);
      \draw[densely dotted] (e) -- (1.5, 2);
      \draw[dq->] (2.5, 0) -- (c);
      \draw[densely dotted] (f) -- (2.5, 2);
    \end{scope}
  \end{tikzpicture}
  \caption{The $\CN = (0,2)$ theory obtained from the $2 \times 3$
    brane box model.}
  \label{fig:BB-(0,2)}
\end{figure}

Although the left-moving R-symmetry is no longer present, the theory
still has the flavor symmetries $\U(1)_\alpha$, at least classically.
So we define the ``R-charge'' in the present case by
\begin{equation}
  R = \sum_\alpha r_\alpha F_\alpha.
\end{equation}
Our claim is that the elliptic genus \eqref{eq:EG} of this theory is
equal to the partition function of an integrable IRF model.

Before we give a proof of this claim, cautionary remarks are in order.
As it is, the theory defined above suffers from mixed anomalies for
the diagonal $\U(1)$ factors of the $\U(N)$ gauge groups.  Similarly,
the flavor symmetry generated by $R$ is anomalous.  These anomalies
must be canceled by introduction of extra multiplets.  For the moment
let us ignore these issues; we will address them later.

The relevant formula for elliptic genera was derived
in~\cite{Benini:2013nda, Benini:2013xpa}.  The only difference
compared to the $\CN = (2,2)$ case is that it involves three kinds of
factors, corresponding to the three kinds of multiplets:
\begin{equation}
\label{eq:I-int-(2,2)}
  \CI
  = \oint
     \prod_{\circled{a}} \frac{1}{N!}
     \prod_{j=1}^N \frac{\rmd u_{a, j}}{2\pi i}
     \CI_{\text{V}}(u_a)
     \prod_{\circled{a} \qto \circled{b}} \CI_{\text{BC}}(u_a, u_b; R_{ab})
     \prod_{\circled{a} \dqto \circled{b}} \CI_{\text{BF}}(u_a, u_b; R_{ab}),
\end{equation}
where the vector, bifundamental chiral and bifundamental Fermi
multiplet factors are%
\footnote{See~\cite{Gadde:2013wq, Gadde:2013lxa} for the formulas for
  the $(\text{NS}, \text{NS})$ sector.}
\begin{align}
  \CI_{\text{V}}(u_a)
  &= \biggl(\frac{2\pi\eta(\tau)^2}{i}\biggr)^N
        \prod_{i \neq j}
        \frac{i\theta_1(u_{a,i} - u_{a,j}|\tau)}
                {\eta(\tau)},
\\
  \CI_{\text{BC}}(u_a, u_b; R_{ab})
  &= \prod_{i, j}
        \frac{i\eta(\tau)}
                {\theta_1(u_{b,j} - u_{a,i} + R_{ab}z/2|\tau)},
\\
  \CI_{\text{BF}}(u_a, u_b; R_{ab})
  &= \prod_{i, j}
        \frac{i\theta_1(u_{b,j} - u_{a,i} + R_{ab}z/2|\tau)}
                {\eta(\tau)}.
\end{align}
Up to an immaterial overall sign, $\CI_{\text{BC}}(u_a, u_b; R_{ab})
\CI_{\text{BF}}(u_a, u_b; R_{ab} - 2)$ is equal to the $\CN = (2,2)$
bifundamental chiral multiplet factor \eqref{eq:IB-(2,2)}.  Likewise,
$\CI_{\text{V}}(u_a) \CI_{\text{BC}}(u_a, u_a; -2)$ is equal to the
$\CN = (2,2)$ vector multiplet factor \eqref{eq:IV-(2,2)}, as is
consistent with the decomposition of an $\CN = (2,2)$ vector multiplet
into an $\CN = (0, 2)$ vector multiplet and adjoint chiral multiplet.
We have
\begin{equation}
  \label{eq:IBC*IBF=1}
  \CI_{\text{BC}}(u_a, u_b; R_{ab}) \CI_{\text{BF}}(u_b, u_a; R_{ba}) = 1
\end{equation}
for $R_{ab} + R_{ba} = 0$, or more graphically,
\begin{equation}
  \begin{tikzpicture}
    \node[snode] (a) at (0,0) {$a$};
    \node[snode] (b) at (1,0) {$b$};
    \draw[q->] (a) to[bend left] (b);
    \draw[dq->] (b) to[bend left] (a);
  \end{tikzpicture}
  \quad = \quad
  \begin{tikzpicture}
    \node[snode] (a) at (0,0) {$a$};
    \node[snode] (b) at (1,0) {$b$};
  \end{tikzpicture}
  \quad .
\end{equation}
The cancellation of arrows corresponds to giving masses to a pair of
chiral multiplet and Fermi multiplet.

For the theory under consideration, the elliptic genus can be computed
as the partition function of an IRF model with Boltzmann weight
\begin{equation}
  \label{eq:W-(0,2)}
  \begin{tikzpicture}
    \draw[dr->] (0, 1) node[left] {$r$} -- (2, 1);
    \draw[dr->] (1, 0) node[below] {$s$}-- (1, 2);

    \begin{scope}[shift={(0.5, 0.5)}]
      \node[tnode] (a) at (0, 0) {$a$};
      \node[tnode] (b) at (1, 0) {$b$};
      \node[tnode] (c) at (1, 1) {$c$};
      \node[tnode] (d) at (0, 1) {$d$};
    \end{scope}

    \draw[thick] (a) -- (b) -- (c) -- (d) -- (a);
  \end{tikzpicture}
  \quad = \quad
  \begin{tikzpicture}
    \node[dnode] (a) at (0, 0) {$a$};
    \node[snode] (b) at (1, 0) {$b$};
    \node[snode] (c) at (1, 1) {$c$};

    \draw[q->] (a) -- node[below] {$-s$} (b);
    \draw[dq->] (b) -- node[right] {$r$} (c);
    \draw[q->] (c) -- node[above left=-2pt] {$s - r$} (a);
  \end{tikzpicture} \quad .
\end{equation}
Plugging this into the Yang-Baxter equation, we find that the
integrability of the model is equivalent to the equality
\begin{equation}
  \label{eq:Seiberg-(0,2)}
  \begin{tikzpicture}[scale=1.5]
    \node[snode] (a) at (240:1){$a$};
    \node[snode] (c) at (0:1) {$c$};
    \node[snode] (d) at (60:1) {$d$};
    \node[snode] (f) at (180:1) {$f$};
    \node[wnode] (g) at (0,0) {$g$};

    \draw[dq->] (a) -- node[left] {$r$} (g);
    \draw[q->] (g) -- node[below, near start] {$-t$} (c);
    \draw[q->] (d) -- node[right, near end] {$t - s$} (g);    
    \draw[q->] (g) -- node[below] {$s - r$} (f);
  \end{tikzpicture}
  \quad = \quad
  \begin{tikzpicture}[scale=1.5]
    \node[snode] (a) at (240:1){$a$};
    \node[snode] (c) at (0:1) {$c$};
    \node[snode] (d) at (60:1) {$d$};
    \node[snode] (f) at (180:1) {$f$};
    \node[wnode] (g) at (0,0) {$g$};
    
    \draw[q->] (g) -- node[left, near start] {$t - s$} (a);
    \draw[q->] (c) -- node[above] {$s - r$} (g);
    \draw[q->] (f) -- node[above, near end] {$-t$} (g);
    \draw[dq->] (g) -- node[right] {$r$} (d);
    \draw[q->] (d) -- node[above=4pt] {$t - r$} (f);

    \draw[dq<-] (c) -- node[right=4pt, near end] {$-t + r$} (a);
    \draw[q<-] (c) -- node[right, near end] {$-s$} (d);
    \draw[dq<-] (f) -- node[left, near end] {$s$} (a);
  \end{tikzpicture}
  \quad .
\end{equation}
Let us demonstrate it.

The left-hand side is given by the contour integral
\begin{multline}
  \CI_{\text{LHS}}
  =
  \frac{1}{N!} \oint
  \prod_{j=1}^N \frac{\rmd u_{g, j}}{2\pi i}
  \CI_{\text{V}}(u_g)
  \CI_{\text{BC}}(u_d, u_g; R_{dg})
  \CI_{\text{BF}}(u_a, u_g; R_{ag}) 
  \\ \times
  \CI_{\text{BC}}(u_g, u_c; R_{gc})
  \CI_{\text{BC}}(u_g, u_f; R_{gf}). 
\end{multline}
If we choose the contours to pick up residues from the incoming arrow,
then the set of relevant poles is $\{u_{d,i} - R_{dg} z/2\}$.  Thus we
have
\begin{equation}
  \label{eq:ILHS}
  \CI_{\text{LHS}}
  =
  \CI_{\text{BF}}(u_a, u_d; R_{ag} - R_{dg})
  \CI_{\text{BC}}(u_d, u_c; R_{dg} + R_{gc})
  \CI_{\text{BC}}(u_d, u_f; R_{dg} + R_{gf}).
\end{equation}

The right-hand side of the equality~\eqref{eq:Seiberg-(0,2)} is
\begin{multline}
  \CI_{\text{RHS}}
  =
  \frac{1}{N!} \oint
  \prod_{j=1}^N \frac{\rmd u_{g, j}}{2\pi i}
  \CI_{\text{V}}(u_g)
  \CI_{\text{BC}}(u_g, u_a; R_{ga}) 
  \CI_{\text{BF}}(u_g, u_d; R_{gd}) 
  \\ \times 
  \CI_{\text{BC}}(u_c, u_g; R_{cg}) 
  \CI_{\text{BC}}(u_f, u_g; R_{fg})
  \\ \times
  \CI_{\text{BF}}(u_a, u_c; R_{ac})
  \CI_{\text{BF}}(u_a, u_f; R_{af})
  \CI_{\text{BC}}(u_d, u_c; R_{dc}) 
  \CI_{\text{BC}}(u_d, u_f; R_{df}).
\end{multline}
The ``complement'' set of poles is $\{u_{a,i} + R_{ga}z/2\}$, and they give
\begin{multline}
  \CI_{\text{RHS}}
  =
  \CI_{\text{BF}}(u_a, u_d; R_{gd} - R_{ga}) 
  \CI_{\text{BC}}(u_c, u_a; R_{cg} + R_{ga}) 
  \CI_{\text{BC}}(u_f, u_a; R_{fg} + R_{ga})
  \\ \times
  \CI_{\text{BF}}(u_a, u_c; R_{ac})
  \CI_{\text{BF}}(u_a, u_f; R_{af})
  \CI_{\text{BC}}(u_d, u_c; R_{dc}) 
  \CI_{\text{BC}}(u_d, u_f; R_{df}).
\end{multline}
Canceling some factors using the identity~\eqref{eq:IBC*IBF=1}, we
find $\CI_{\text{LHS}} = \CI_{\text{RHS}}$.

In fact, the quivers on the two sides of the
equality~\eqref{eq:Seiberg-(0,2)} describe two theories related by the
$(0,2)$ triality transformation~\cite{Gadde:2013lxa}.  The third
theory in the triality is the free theory with quiver
\begin{equation}
  \begin{tikzpicture}
    \node[snode] (a) at (240:1){$a$};
    \node[snode] (c) at (0:1) {$c$};
    \node[snode] (d) at (60:1) {$d$};
    \node[snode] (f) at (180:1) {$f$};

    \draw[dq<-] (d) -- node[right, near end] {$R_{ag} - R_{dg}$} (a);
    \draw[q->] (d) -- node[right] {$R_{dg} + R_{gc}$} (c);
    \draw[q->] (d) -- node[above left=-2pt] {$R_{dg} + R_{gf}$} (f);
  \end{tikzpicture} \quad .
\end{equation}
This is manifest in the expression \eqref{eq:ILHS} of the elliptic
genus.

\begin{figure}
  \centering
  \begin{tikzpicture}[scale=1.5]
    \node[wnode] (a) at (0,0) {$a$};
    \node[wnode] (b) at (1,0) {$b$};
    \node[wnode] (c) at (1,1) {$c$};
    
    \draw[q->] (a) -- (b);
    \draw[dq->] (b) -- (c);
    \draw[q->] (c) -- (a);
    
    \draw[dq->, snake] (b) to[bend left] node[below] {$-NR_{ab} - \eps$} (a);
    \draw[q->, snake] (c) to[bend left] node[right] {$-NR_{cb} + \eps$} (b);
    \draw[dq->, snake] (a) to[bend left] node[above left=-2pt] {$-NR_{ca} - \eps$} (c);
    
    \draw[q->, snake] (a) to[out=180, in=270, loop] node[left=2pt] {$\eps$} (a);
    \draw[q->, snake] (b) to[out=270, in=360, loop] node[right=2pt] {$\eps$} (b);
    \draw[q->, snake] (c) to[out=0, in=90, loop] node[right=2pt] {$\eps$} (c);
  \end{tikzpicture}
  \caption{The R-charges of the extra multiplets for anomaly
    cancellation.}
  \label{fig:extra-m}
\end{figure}
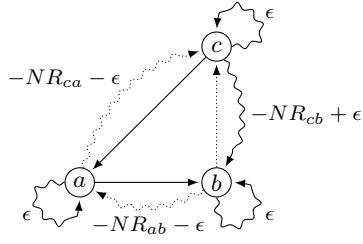

Finally, we come back to the issues of anomalies.  To cancel the mixed
$\U(1)$ gauge anomalies and $R$-anomaly, we add to each solid arrow
\begin{equation}
  \begin{tikzpicture}
    \node[wnode] (a) at (0,0) {$a$};
    \node[wnode] (b) at (1,0) {$b$};
    \draw[q->] (a) -- (b);
  \end{tikzpicture}
\end{equation}
a Fermi multiplet with $R = -NR_{ab} - \eps$ in the representation
$\det_a \otimes \det_b^{-1}$ of $\U(N)_a \times \U(N)_b$, and to
each dotted arrow
\begin{equation}
  \begin{tikzpicture}
    \node[wnode] (a) at (0,0) {$a$};
    \node[wnode] (b) at (1,0) {$b$};
    \draw[dq->] (a) -- (b);
  \end{tikzpicture}
\end{equation}
a chiral multiplet with $R = -NR_{ab} + \eps$ in $\det_a \otimes
\det_b^{-1}$.  Here $\eps$ is a nonzero parameter.  Furthermore, to
each node we introduce a singlet chiral multiplet with $R = \eps$,
though this is not required for anomaly cancellation.  We can
visualize these extra multiplets as wavy arrows, as in
figure~\ref{fig:extra-m}.

\enlargethispage{13pt}

The presence of the extra multiplets modifies the factors associated
to vector and bifundamental multiplets to
\begin{align}
  \CI'_{\text{V}}(u_a)
  &=
  \frac{i\eta(\tau)}{\theta_1(\eps z/2|\tau)}
  \CI_{\text{V}}(u_a),
  \\
  \CI'_{\text{BC}}(u_a, u_b; R_{ab})
  &=
  \frac{i\theta_1(\sum_i u_{a,i} - \sum_i u_{b,i}
          + (-NR_{ab} - \eps)z/2|\tau)}{\eta(\tau)}
  \CI_{\text{BC}}(u_a, u_b; R_{ab}),
  \\
  \CI'_{\text{BF}}(u_a, u_b; R_{ab}) 
  &=
  \frac{i\eta(\tau)}{\theta_1(\sum_i u_{a,i} - \sum_i u_{b,i}
          + (-NR_{ab} + \eps)z/2|\tau)}
  \CI_{\text{BF}}(u_a, u_b; R_{ab}).
\end{align}
Apart from this modification, the computations of the elliptic genera
for the equality \eqref{eq:Seiberg-(0,2)} remain unchanged.  In
particular, we can use the same integration contours, since the extra
multiplets introduce neither solid incoming arrows to the left-hand
side nor solid outgoing arrows to the right-hand side.

\subsection[$\CN=(0,2)$ theories related to brane tiling
configurations]{\texorpdfstring{$\boldsymbol{\CN = (0,2)}$ theories
    related to brane tiling configurations}{N=(0,2) theories related
    to brane tiling configurations}}

Quivers from brane tiling models can also be modified to produce $\CN
= (0,2)$ theories that exhibit integrability.  As in the brane box
case discussed above, we obtain these theories by replacing one of the
arrows in the R-matrix with a dotted arrow:
\begin{equation}
  \begin{tikzpicture}
    \draw[thick] (0,1) node[left] {$(r_1, r_2)$}
    -- node[below] {$i_1$} (1,1);
    \draw[r->] (1,1) -- node[above] {$i_2$} (2, 1);

    \draw[thick] (1,0) node[below] {$(s_1, s_2)$}
    -- node[right] {$j_1$} (1,1);
    \draw[r->] (1,1) -- node[left] {$j_2$} (1, 2);
  \end{tikzpicture}
  \quad = \quad
  \begin{tikzpicture}
    \node[dnode] (i1) at (0,0) {$i_1$};
    \node[dnode] (j1) at (-45:1) {$j_1$};
    \node[snode] (i2) at ($(j1) + (45:1)$) {$i_2$};
    \node[snode] (j2) at (45:1) {$j_2$};

    \draw[q->] (i1) -- node[above left=-2pt] {$r_1 - s_1$} (j2);
    \draw[dq->] (j2) -- node[above right=-2pt] {$s_2 - r_1$} (i2);
    \draw[q->] (i2) -- node[below right=-2pt] {$r_2 - s_2$} (j1);
    \draw[q->] (j1) -- node[below left=-2pt] {$s_1 - r_2$} (i1);
  \end{tikzpicture} \quad .
\end{equation}
Extra multiplets for anomaly cancellations are implicit in this
picture.  Figure \ref{fig:VM-(0,2)} shows the quiver for the $\CN =
(0,2)$ theory obtained from the $2 \times 3$ brane tiling model, whose
quiver is given in figure \ref{fig:VM-b}.

\begin{figure}
  \centering 
  \begin{tikzpicture}[scale=1.2]
    \draw (0,0) rectangle (3, 2);
    
    \begin{scope}[shift={(0.25, 0.25)}]
      \node[wnode] (i1) at (0.5, 0) {$i_1$};
      \node[wnode] (i2) at (1.5, 0) {$i_2$};
      \node[wnode] (i3) at (2.5, 0) {$i_3$};
      \node[wnode] (j1) at (0.5, 1) {$j_1$};
      \node[wnode] (j2) at (1.5, 1) {$j_2$};
      \node[wnode] (j3) at (2.5, 1) {$j_3$};
      
      \node[wnode] (k1) at (0, 0.5) {$k_1$};
      \node[wnode] (l1) at (1, 0.5) {$l_1$};
      \node[wnode] (m1) at (2, 0.5) {$m_1$};
      \node[wnode] (k2) at (0, 1.5) {$k_2$};
      \node[wnode] (l2) at (1, 1.5) {$l_2$};
      \node[wnode] (m2) at (2, 1.5) {$m_2$};
    \end{scope}
    
    \draw[q->] (i1) -- (l1);
    \draw (i1) -- ++(-135:{0.25*sqrt(2)} );
    \draw[q<-] (i1) -- ++(-45:{0.25*sqrt(2)} );
    \draw[q->] (i2) -- (m1);
    \draw (i2) -- ++(-135:{0.25*sqrt(2)} );
    \draw[q<-] (i2) -- ++(-45:{0.25*sqrt(2)} );
    \draw (i3) -- ++(-135:{0.25*sqrt(2)} );
    \draw[q<-] (i3) -- ++(-45:{0.25*sqrt(2)} );
    \draw (i3) -- ++(45:{0.25*sqrt(2)} );
    \draw[dq->] (k1) -- (i1);
    \draw (k1) -- ++(-135:{0.25*sqrt(2)} );
    \draw (k1) -- ++(135:{0.25*sqrt(2)} );
    \draw[dq->] (l1) -- (i2);
    \draw[q->] (l1) -- (j1);
    \draw[dq->] (m1) -- (i3);
    \draw[q->] (m1) -- (j2);
    \draw[q->] (j1) -- (k1);
    \draw[q->] (j1) -- (l2);
    \draw[q->] (j2) -- (m2);
    \draw[q->] (j2) -- (l1);
    \draw[q->] (j3) -- (m1);
    \draw (j3) -- ++(45:{0.25*sqrt(2)} );
    \draw (j3) -- ++(-45:{0.25*sqrt(2)} );
    \draw[dq->] (k2) -- (j1);
    \draw[q<-] (k2) -- ++(45:{0.25*sqrt(2)} );
    \draw (k2) -- ++(135:{0.25*sqrt(2)} );
    \draw[q<-] (k2) -- ++(-135:{0.25*sqrt(2)} );
    \draw[dq->] (l2) -- (j2);
    \draw (m2) -- ++(135:{0.25*sqrt(2)} );
    \draw[q<-] (l2) -- ++(45:{0.25*sqrt(2)} );
    \draw (l2) -- ++(135:{0.25*sqrt(2)} );
    \draw[dq->] (m2) -- (j3);
    \draw[q<-] (m2) -- ++(45:{0.25*sqrt(2)} );
  \end{tikzpicture}
  \caption{The $\CN = (0,2)$ theory obtained from the $2 \times 3$
    brane tiling model.}
  \label{fig:VM-(0,2)}
\end{figure}
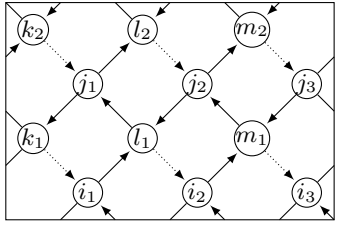

The Yang-Baxter equation for this R-matrix reduces to the equality
\begin{equation}
  \label{eq:YBE-SD-(0,2)}
  \begin{tikzpicture}[rotate=-60]
    \node[wnode] (i2) at (0,0) {$i_2$};
    \node[wnode] (j2) at (150:1) {$j_2$};
    \node[wnode] (k2) at (90:1) {$k_2$};

    \node[snode] (k1) at (0:1) {$k_1$};
    \node[snode] (i3) at ($(k2)+(0:1)$) {$i_3$};

    \node[snode] (j1) at ($(i2) + (-120:1)$) {$j_1$};
    \node[snode] (i1) at ($(j2) + (-120:1)$) {$i_1$};

    \node[snode] (j3) at ($(k2) + (120:1)$) {$j_3$};
    \node[snode] (k3) at ($(j2) + (120:1)$) {$k_3$};

    \draw[q->] (j2) -- (k3);
    \draw[dq->] (k3) -- (j3);
    \draw[q->] (j3) -- (k2);
    \draw[q->] (k2) -- (j2);

    \draw[dq->] (k2) -- (i3);
    \draw[q->] (i3) -- (k1);
    \draw[q->] (k1) -- (i2);
    \draw[q->] (i2) -- (k2);

    \draw[q->] (i2) -- (j1);
    \draw[q->] (j1) -- (i1);
    \draw[q->] (i1) -- (j2);
    \draw[dq->] (j2) -- (i2);
  \end{tikzpicture}
  \quad = \quad
  \begin{tikzpicture}
    \node[wnode] (j2) at (0,0) {$j_2$};
    \node[wnode] (k2) at (150:1) {$k_2$};
    \node[wnode] (i2) at (90:1) {$i_2$};

    \node[snode] (i3) at (0:1) {$i_3$};
    \node[snode] (j3) at ($(i2)+(0:1)$) {$j_3$};

    \node[snode] (k1) at ($(j2) + (-120:1)$) {$k_1$};
    \node[snode] (j1) at ($(k2) + (-120:1)$) {$j_1$};

    \node[snode] (k3) at ($(i2) + (120:1)$) {$k_3$};
    \node[snode] (i1) at ($(k2) + (120:1)$) {$i_1$};

    \draw[q->] (j2) -- (k1);
    \draw[q->] (k1) -- (j1);
    \draw[q->] (j1) -- (k2);
    \draw[dq->] (k2) -- (j2);

    \draw[q->] (k2) -- (i1);
    \draw[q->] (i1) -- (k3);
    \draw[dq->] (k3) -- (i2);
    \draw[q->] (i2) -- (k2);

    \draw[q->] (i2) -- (j3);
    \draw[dq->] (j3) -- (i3);
    \draw[q->] (i3) -- (j2);
    \draw[q->] (j2) -- (i2);
  \end{tikzpicture}
  \quad .
\end{equation}
Application of the $(0,2)$ triality transformation
\eqref{eq:Seiberg-(0,2)} four times (more precisely, one triality
transformation followed by three others in the opposite direction)
turns the left-hand side into the right-hand side.  This is
illustrated in figure \ref{fig:YBE-SD-(0,2)}.

\begin{figure}
  \centering
  \begin{tikzpicture}[rotate=-60]
    \node[wnode] (i2) at (0,0) {$i_2$};
    \node[wnode] (j2) at (150:1) {$j_2$};
    \node[wnode] (k2) at (90:1) {$k_2$};

    \node[snode] (k1) at (0:1) {$k_1$};
    \node[snode] (i3) at ($(k2)+(0:1)$) {$i_3$};

    \node[snode] (j1) at ($(i2) + (-120:1)$) {$j_1$};
    \node[snode] (i1) at ($(j2) + (-120:1)$) {$i_1$};

    \node[snode] (j3) at ($(k2) + (120:1)$) {$j_3$};
    \node[snode] (k3) at ($(j2) + (120:1)$) {$k_3$};

    \draw[q->] (j2) -- (k3);
    \draw[dq->] (k3) -- (j3);
    \draw[q->] (j3) -- (k2);
    \draw[q->] (k2) -- (j2);

    \draw[dq->] (k2) -- (i3);
    \draw[q->] (i3) -- (k1);
    \draw[q->] (k1) -- (i2);
    \draw[q->] (i2) -- (k2);

    \draw[q->] (i2) -- (j1);
    \draw[q->] (j1) -- (i1);
    \draw[q->] (i1) -- (j2);
    \draw[dq->] (j2) -- (i2);
  \end{tikzpicture}
  \quad $\stackrel{i_2}{\longleftarrow}$ \quad 
  \begin{tikzpicture}[rotate=-60]
    \node[wnode] (i2) at (0,0) {$i_2$};
    \node[wnode] (j2) at (150:1) {$j_2$};
    \node[wnode] (k2) at (90:1) {$k_2$};

    \node[snode] (k1) at (0:1) {$k_1$};
    \node[snode] (i3) at ($(k2)+(0:1)$) {$i_3$};

    \node[snode] (j1) at ($(i2) + (-120:1)$) {$j_1$};
    \node[snode] (i1) at ($(j2) + (-120:1)$) {$i_1$};

    \node[snode] (j3) at ($(k2) + (120:1)$) {$j_3$};
    \node[snode] (k3) at ($(j2) + (120:1)$) {$k_3$};

    \draw[q->] (j2) -- (k3);
    \draw[dq->] (k3) -- (j3);
    \draw[q->] (j3) -- (k2);

    \draw[dq->] (k2) -- (i3);
    \draw[q->] (i3) -- (k1);
    \draw[dq<-] (k1) -- (i2);
    \draw[q<-] (i2) -- (k2);

    \draw[q<-] (i2) -- (j1);
    \draw[q->] (j1) -- (i1);
    \draw[q->] (i1) -- (j2);
    \draw[q<-] (j2) -- (i2);

    \draw[dq->] (j2) -- (j1);
    \draw[q->] (k1) -- (k2);
    \draw[q->] (k1) -- (j1);
  \end{tikzpicture}
  \quad $\stackrel{j_2}{\longto}$ \quad 
  \begin{tikzpicture}[rotate=-60]
    \node[wnode] (i2) at (0,0) {$i_2$};
    \node[wnode] (j2) at (150:1) {$j_2$};
    \node[wnode] (k2) at (90:1) {$k_2$};

    \node[snode] (k1) at (0:1) {$k_1$};
    \node[snode] (i3) at ($(k2)+(0:1)$) {$i_3$};

    \node[snode] (j1) at ($(i2) + (-120:1)$) {$j_1$};
    \node[snode] (i1) at ($(j2) + (-120:1)$) {$i_1$};

    \node[snode] (j3) at ($(k2) + (120:1)$) {$j_3$};
    \node[snode] (k3) at ($(j2) + (120:1)$) {$k_3$};

    \draw[dq<-] (j2) -- (k3);
    \draw[dq->] (k3) -- (j3);
    \draw[q->] (j3) -- (k2);

    \draw[dq->] (k2) -- (i3);
    \draw[q->] (i3) -- (k1);
    \draw[dq<-] (k1) -- (i2);
    \draw[q<-] (i2) -- (k2);

    \draw[q<-] (i1) -- (j2);
    \draw[q->] (j2) -- (i2);

    \draw[q<-] (j2) -- (j1);
    \draw[q->] (k1) -- (k2);
    \draw[q->] (k1) -- (j1);

    \draw[q->] (i1) -- (k3);
    \draw[q->] (i2) -- (k3);
  \end{tikzpicture}

  \bigskip

  \quad $\stackrel{k_2}{\longto}$ \quad 
  \begin{tikzpicture}[rotate=-60]
    \node[wnode] (i2) at (0,0) {$i_2$};
    \node[wnode] (j2) at (150:1) {$j_2$};
    \node[wnode] (k2) at (90:1) {$k_2$};

    \node[snode] (k1) at (0:1) {$k_1$};
    \node[snode] (i3) at ($(k2)+(0:1)$) {$i_3$};

    \node[snode] (j1) at ($(i2) + (-120:1)$) {$j_1$};
    \node[snode] (i1) at ($(j2) + (-120:1)$) {$i_1$};

    \node[snode] (j3) at ($(k2) + (120:1)$) {$j_3$};
    \node[snode] (k3) at ($(j2) + (120:1)$) {$k_3$};

    \draw[dq<-] (j2) -- (k3);
    \draw[dq->] (k3) -- (j3);
    \draw[q<-] (j3) -- (k2);

    \draw[q<-] (k2) -- (i3);
    \draw[dq->] (i2) -- (k2);

    \draw[q<-] (i1) -- (j2);
    \draw[q->] (j2) -- (i2);

    \draw[q<-] (j2) -- (j1);
    \draw[q<-] (k1) -- (k2);
    \draw[q->] (k1) -- (j1);

    \draw[q->] (i1) -- (k3);
    \draw[q->] (i2) -- (k3);

    \draw[q->] (j3) -- (i2);
    \draw[dq->] (j3) -- (i3);
  \end{tikzpicture}
  \quad $\stackrel{i_2}{\longto}$ \quad 
  \begin{tikzpicture}[rotate=-60]
    \node[wnode] (i2) at (0,0) {$i_2$};
    \node[wnode] (j2) at (150:1) {$j_2$};
    \node[wnode] (k2) at (90:1) {$k_2$};

    \node[snode] (k1) at (0:1) {$k_1$};
    \node[snode] (i3) at ($(k2)+(0:1)$) {$i_3$};

    \node[snode] (j1) at ($(i2) + (-120:1)$) {$j_1$};
    \node[snode] (i1) at ($(j2) + (-120:1)$) {$i_1$};

    \node[snode] (j3) at ($(k2) + (120:1)$) {$j_3$};
    \node[snode] (k3) at ($(j2) + (120:1)$) {$k_3$};

    \draw[q<-] (k2) -- (i3);
    \draw[q<-] (i2) -- (k2);

    \draw[q<-] (i1) -- (j2);
    \draw[q<-] (j2) -- (i2);

    \draw[q<-] (j2) -- (j1);
    \draw[q<-] (k1) -- (k2);
    \draw[q->] (k1) -- (j1);

    \draw[q->] (i1) -- (k3);
    \draw[dq<-] (i2) -- (k3);

    \draw[q<-] (j3) -- (i2);
    \draw[dq->] (j3) -- (i3);

    \draw[dq->] (j2) -- (k2);
  \end{tikzpicture}
  \quad = \quad 
  \begin{tikzpicture}
    \node[wnode] (j2) at (0,0) {$j_2$};
    \node[wnode] (k2) at (150:1) {$k_2$};
    \node[wnode] (i2) at (90:1) {$i_2$};

    \node[snode] (i3) at (0:1) {$i_3$};
    \node[snode] (j3) at ($(i2)+(0:1)$) {$j_3$};

    \node[snode] (k1) at ($(j2) + (-120:1)$) {$k_1$};
    \node[snode] (j1) at ($(k2) + (-120:1)$) {$j_1$};

    \node[snode] (k3) at ($(i2) + (120:1)$) {$k_3$};
    \node[snode] (i1) at ($(k2) + (120:1)$) {$i_1$};

    \draw[q->] (j2) -- (k1);
    \draw[q->] (k1) -- (j1);
    \draw[q->] (j1) -- (k2);
    \draw[dq->] (k2) -- (j2);

    \draw[q->] (k2) -- (i1);
    \draw[q->] (i1) -- (k3);
    \draw[dq->] (k3) -- (i2);
    \draw[q->] (i2) -- (k2);

    \draw[q->] (i2) -- (j3);
    \draw[dq->] (j3) -- (i3);
    \draw[q->] (i3) -- (j2);
    \draw[q->] (j2) -- (i2);
  \end{tikzpicture}
  \caption{A sequence of $(0,2)$ triality transformations that relates
    the two sides of the Yang-Baxter equation
    \eqref{eq:YBE-SD-(0,2)}.  An arrow between two quivers represents
    a $(0,2)$ triality transformation in the direction from the
    right- to left-hand side of the equality
    \eqref{eq:Seiberg-(0,2)}.}
  \label{fig:YBE-SD-(0,2)}
\end{figure}
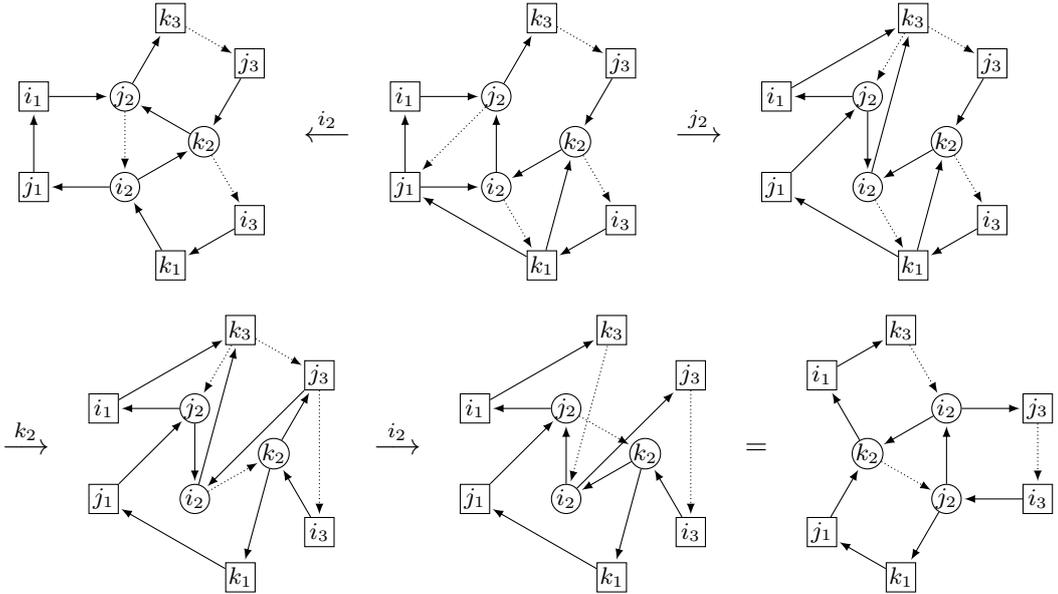

\subsection{Brane cube models}
\label{sec:irc-model}

Lastly, we briefly discuss the class of $\CN = (0,2)$ quiver gauge
theories known as \emph{brane cube
  models}~\cite{GarciaCompean:1998kh}.  These theories are constructed
from brane configurations analogous to brane~boxes:
\begin{equation}
  \label{eq:BB-(0,2)}
  \begin{tabular}{|l|cccccccccc|}
  \hline
  & 0 & 1 & 2 & 3 & 4 & 5 & 6 & 7 & 8 & 9
  \\ \hline
  D4 & $\times$ & $\times$ & $\times$ & & $\times$ & & $\times$
  &&&
  \\
  NS5 & $\times$ & $\times$ & $\times$ & $\times$ & $\times$ & $\times$
  &&&&
  \\
  NS5 & $\times$ & $\times$ & $\times$ & $\times$ & & & $\times$ &
  $\times$
  &&
  \\
  NS5 & $\times$ & $\times$ & & & $\times$ & $\times$ & $\times$ &
  $\times$
  &&
  \\ \hline
\end{tabular}
\end{equation}
Note that brane cube models include 3d brane box models
\eqref{eq:BB-3d} as special cases.  A brane box model is characterized
by the 2d lattice drawn on the $46$-torus by the NS5-branes.  For a
brane cube model, the NS5-branes make a 3d lattice in the $246$-space
which we take to be a $3$-torus $T^3$.

A brane cube configuration can be thought of as describing
intersecting codimension-$1$ defects in the 5d theory on the
D4-branes.  We can topologically twist the theory along the
$246$-space, replacing the holonomy group $\SO(3)_{246}$ with the
diagonal subgroup of $\SO(3)_{246} \times \SO(3)_{357}$.  The twisted
theory has four supercharges that are scalar in the $246$-space, hence
unbroken when this space is a general $3$-manifold $\Sigma$.  These
supercharges generate $\CN = (2,2)$ supersymmetry in the $01$-space.
Half of them, generating an $\CN = (0,2)$ subalgebra, are preserved by
codimension-$1$ defects supported on arbitrary surfaces
inside~$\Sigma$.  Therefore, the twisted theory gives a morphism that
sends a configuration of surfaces in $\Sigma$ to the effective $\CN =
(0,2)$ theory.  Taking the elliptic genus, we get a 3d TQFT on
$\Sigma$ equipped with surface operators.  An extra dimension emerges
upon lifting the system to M-theory, and it follows that the elliptic
genus of a brane cube model is equal to the partition function of a 3d
integrable lattice model.

\enlargethispage{13pt}

Like an IRF model is associated with brane box models, the lattice
model associated with brane cube models is an IRC model.  According to
the quiver rule found in~\cite{GarciaCompean:1998kh}, one way to
describe this model is to set its Boltzmann weight
\begin{equation}
  \begin{tikzpicture}[thick]
    \node[wnode] (a) at (0,0) {$a$};
    \node[wnode] (e) at (0,-1) {$e$};
    \node[wnode] (f) at (-1,0) {$f$};
    \node[wnode] (d) at (-1, -1) {$d$};

    \begin{scope}[shift=(-150:0.7)]
      \node[wnode] (g) at (0,0) {$g$};
      \node[wnode] (c) at (0,-1) {$c$};
      \node[wnode] (b) at (-1,0) {$b$};
      \node[wnode] (h) at (-1, -1) {$h$};
    \end{scope}


    \draw (a) to (e);
    \draw[dashed] (e) -- (d);
    \draw[dashed] (d) -- (f);
    \draw (f) -- (a);
    \draw (g) -- (c) -- (h) -- (b) -- (g);
    \draw (a) -- (g);
    \draw (f) -- (b);
    \draw (e) -- (c);
    \draw[dashed] (d) -- (h);
  \end{tikzpicture}
  \quad = \quad
  \begin{tikzpicture}
    \node[snode] (a) at (0,0) {$a$};
    \node[snode] (d) at (-1, -1) {$d$};

    \draw[q->] (a) -- (h);
    \draw[q->] (h) -- (b);
    \draw[q->] (h) -- (c);
    \draw[q->] (h) -- (d);

    \begin{scope}[shift=(-150:0.7)]
      \node[snode] (c) at (0,-1) {$c$};
      \node[snode] (b) at (-1,0) {$b$};
      \node[dnode] (h) at (-1, -1) {$h$};
    \end{scope}

    \draw[dq->] (a) -- (b);
    \draw[dq->] (a) -- (c);
    \draw[dq->] (a) -- (d);
  \end{tikzpicture}
  \quad .
\end{equation}
This Boltzmann weight, or another one that leads to the same IRC model
on cubic lattices, should solve the tetrahedron
equation~\eqref{eq:TE-IRC}.  It would be interesting to verify this
statement.

\section*{Acknowldgments}

The author would like to thank Sergei Gukov and Edward Witten for
helpful comments, and Nils Carqueville for valuable discussions.  This
work is supported by INFN Postdoctoral Fellowship.

\bibliography{../junya}{}
\bibliographystyle{JHEP}
\end{document}